\documentclass[11pt,a4paper]{article}
\usepackage[T1]{fontenc}
\usepackage{lmodern}
\usepackage[a4paper,margin=25mm]{geometry}
\usepackage{amsmath,amssymb,bm,mathtools}
\usepackage{graphicx,booktabs,array,multirow,adjustbox,float,placeins}
\usepackage{xcolor,microtype,etoolbox}
\usepackage{tikz}
\usetikzlibrary{arrows.meta,fit}
\usepackage[round]{natbib}
\usepackage{xurl}
\usepackage{hyperref}
\definecolor{linkblue}{HTML}{1A5DAB}
\hypersetup{colorlinks=true,linkcolor=linkblue,citecolor=linkblue,urlcolor=linkblue,pdfborder={0 0 0},pdfauthor={Ehsan Roohi},pdftitle={Population structure of the velocity distribution in normal shock waves: Gaussian representations, Mach-number similarity, and prediction}}
\graphicspath{{./}{figures/}}

\newcommand{\doilink}[1]{\href{https://doi.org/#1}{\nolinkurl{doi:#1}}}
\newcommand{\preprintdoilink}[1]{\href{https://doi.org/#1}{\nolinkurl{arXiv DOI:#1}}}
\newcommand{\onlinelink}[1]{\href{#1}{Online}}
\newcommand{\xih}{\hat\xi}
\newcommand{\Tpar}{T_\parallel}
\newcommand{\Tperp}{T_\perp}
\numberwithin{equation}{section}
\newenvironment{bmhead}[1][]{\par\addvspace{8pt}\noindent\textbf{#1}\ }{\par\addvspace{7pt}}
\AtBeginEnvironment{figure}{\centering}

\makeatletter\def\fps@table{!htbp}\def\fps@figure{!htbp}\makeatother
\title{Population structure of the velocity distribution in normal shock waves: Gaussian representations, Mach-number similarity, and prediction}
\author{Ehsan Roohi\\[0.5em]
\small Department of Mechanical and Industrial Engineering\\
\small University of Massachusetts Amherst\\
\small Amherst, Massachusetts 01003, USA\\[0.3em]
\small\href{mailto:roohie@umass.edu}{roohie@umass.edu}}
\date{}
\AtBeginEnvironment{thebibliography}{\raggedright}
\begin{document}
\pagestyle{plain}
\maketitle
\begin{abstract}
Normal shock waves are commonly described as mixtures of upstream and downstream Maxwellians. Discrete-velocity solutions of three relaxation models up to Mach 12 and direct simulation Monte Carlo (DSMC) of hard-sphere and variable-hard-sphere gases up to Mach 25 are used to examine the population structure, Mach-number similarity and upstream precursor of their velocity distributions. Three anisotropic Gaussian populations reproduce the streamwise reduced distributions of the Bhatnagar--Gross--Krook (BGK) model and of DSMC to within 4\% up to Mach 12, and of DSMC to within 6\% at Mach 25, whereas two Maxwellians incur errors up to 50\% (BGK) and 37\% (DSMC). Collision histories identify the hard-sphere beam with unscattered stream molecules, which carry half of the peak viscous stress; the other populations are effective descriptions, not collision classes. Above Mach 5, Rankine--Hugoniot-scaled distributions are nearly Mach-independent, with corrections linear in $1/M^2$, consistent with the scaled kinetic equation. At equal normalised density, Mach 8 and 12 give the heat flux and stress at Mach 25 within 1\%, comparable to the accuracy of the reference; with one moment equation for position, physical-space profiles follow within 5\%. Comparing three prediction methods shows that the accuracy comes from the scaling; the Gaussian form adds compactness. Ahead of the shock, the mass of upstream-moving molecules decays as a stretched exponential with effective exponent 0.68--0.87. Collisions regenerate upstream-moving molecules, which explains a decay slower than free-flight survival; a linear cascade model with a downstream-Maxwellian source and no input from the shock interior reproduces the decay rates within 1.2\%.
\end{abstract}
\section{Introduction}

Rarefied gas simulations contain information at two levels. The first is macroscopic: density, mean velocity, and temperature. The second is kinetic: the asymmetry and anisotropy of the velocity distribution function (VDF), heat flux, stress, skewness, and fourth-order content. In a transitional or rarefied flow, agreement in $\rho$, $\bm u$, and $T$ is not sufficient, because distributions with nearly identical primary fields can transport different momentum and energy. Odd, mixed, and high-order velocity moments measure the departure from local equilibrium and expose errors that are nearly invisible in conserved variables~\citep{Bird1994,Cercignani2000,Sone2007,RoohiStefanov2016,AkhlaghiRoohiStefanov2023}.

The normal shock is the simplest flow in which this distinction matters. Mott-Smith described its distribution as a combination of the upstream and downstream Maxwellians~\citep{MottSmith1951}, and measurements in argon and helium shocks confirmed the bimodal character of the distribution in strong shocks~\citep{MuntzHarnett1969,HoltzMuntz1983,PhamVanDiep1989}, which DSMC reproduces in detail~\citep{Bird1970,Erwin1991}. The two-Maxwellian picture nevertheless leaves part of the distribution undescribed. Salwen, Grosch, and Ziering gave a method for adding terms to the Mott-Smith distribution and computed shock thicknesses for Maxwell molecules with three terms~\citep{Salwen1964}, and Segal and Ferziger constructed a model equation with a trimodal gain function~\citep{SegalFerziger1972}. Monte Carlo simulations showed that the bimodal ansatz is reasonably accurate for the shock thickness but not for the distribution function~\citep{Bird1970,Cercignani1999}. Vilquin, Boudet, and Kellay found that velocity distributions measured in shocks in granular gases are not strictly bimodal and proposed a trimodal description in which the intermediate population consists of particles that have undergone one collision between the supersonic and subsonic streams; they showed that the same description fits distributions measured in argon and helium shocks and computed by molecular dynamics, at two positions for each of three Mach numbers~\citep{Vilquin2016}. The bimodal distribution has also been used to close the moment equations of strong shocks~\citep{SolovchukSheu2011}. The direct numerical solution of the Boltzmann equation for hard spheres by Ohwada, up to Mach 3, agrees with the Mott-Smith profiles of density, velocity, and temperature but shows a distinct difference in the distribution function~\citep{Ohwada1993},
and in the limit of infinite Mach number the distribution separates into the upstream beam and a remainder that is singular at the beam velocity~\citep{Grad1969,Cercignani1999,Takata2000,PomeauTran2019}. For hard spheres, this remainder first appears next to the beam velocity and then develops a central maximum, so that it has two maxima inside the shock~\citep{Takata2000}. For the BGK model, Liepmann, Narasimha, and Chahine obtained shock profiles from the integral form of the kinetic equation~\citep{Liepmann1962}. 

For a weak shock and a relaxation model with constant collision time, Lyubarskii found that ahead of the shock the density, velocity, and temperature approach their limits like $\exp(-C|x|^{2/3})$ and not exponentially, and attributed this to fast particles that penetrate far from the front almost without collisions, so that the structure far from the front depends on how the collision time varies with velocity at large velocities~\citep{Lyubarskii1961}. Narasimha obtained the distribution of these molecules by matched asymptotic expansions: on the cold side they form a precursor of the hot side, the downstream Maxwellian attenuated as in a molecular beam whose free path is proportional to the molecular speed. A heuristic extension to inverse-power potentials gave $\exp(-|x|^{l})$ with $l=\lambda/(\lambda+2)$ for a potential proportional to $r^{-\lambda}$, that is, $2/3$ for Maxwell molecules and unity for hard spheres~\citep{Narasimha1968}. These results concern weak shocks, in which the slow decay is reached only far upstream. For strong shocks, Liepmann, Narasimha, and Chahine noted a long upstream tail in their numerical solutions of the BGK equation and expected the model to exaggerate it except for Maxwell molecules~\citep{Liepmann1965}; Narasimha's later account of that work~\citep{Narasimha2014} recalls that the tail is the more pronounced the higher the order of the moment, and that such tails have since been found in solutions of the Boltzmann equation. Rigorous analyses of weak shock profiles have recently obtained sub-exponential remainders with the same exponent~\citep{Duan2026,Albritton2024}. To our knowledge, the decay law has not been measured in kinetic solutions of strong shocks. The estimates with which such solutions can be compared, the precursor distribution of Narasimha and its extension to other collision laws (Eqs.~\eqref{eq:halfspace} and \eqref{eq:phi} below), take a molecule to leave the upstream-moving group at its first collision. This is exact for a relaxation model. In a gas a collision redistributes both partners, as the collision operators of the Boltzmann and Landau equations treated in the rigorous analyses do, and the consequence of this redistribution for the decay ahead of a strong shock has not been examined. Likewise, a three-population description exists~\citep{Salwen1964,Vilquin2016}, so that the number three is not new; what has not been determined is where in the shock and from what Mach number each population is required, what each contributes to the transport, whether the populations of a fit are sets of molecules with a common collision history, and how the answers depend on the collision law. A tool for the question of history exists for confined rarefied flows: Stefanov and co-workers divided the DSMC solution into a ballistic part, the molecules that have not collided since they left a wall, and a collision part, and obtained the contribution of each to the flow rate, the temperature, and the heat flux~\citep{Vargas2014,Tatsios2015,AkhlaghiRoohiStefanov2018}. It is applied here to the shock, with the incoming stream in the place of the wall, and extended by a record of how recently a scattered molecule or its partners belonged to the stream.

A second line of work concerns compact representations of kinetic solutions. A deterministic discrete velocity method (DVM) stores a distribution on the tensor product of a physical grid and a velocity grid, so that even a one-dimensional shock with three velocity coordinates contains hundreds of millions of values. Gaussian kernels are a natural localised basis for such data. Alekseenko, Grandilli, and Wood developed an ultra-sparse isotropic-Gaussian approximation for spatially homogeneous distributions~\citep{AlekseenkoGrandilliWood2020}, Gaussian-mixture compression has been applied to plasma distribution functions~\citep{HuPlasmaGMM2025}, and Shenoy and Frankel fitted Gaussian primitives to snapshots of turbulent velocity fields and evaluated derivative-sensitive quantities~\citep{ShenoyFrankel2026}. Gaussian primitives are also prominent in continuous scene and signal representations~\citep{Kerbl2023,Sitzmann2020}, while reduced-order fluid modelling has long relied on proper orthogonal decomposition and projection methods~\citep{Sirovich1987,Berkooz1993,Benner2015}. Dynamical low-rank and tensor methods, reduced-order models of the BGK equation, and neural sparse representations reduce the cost of evolving phase-space solutions~\citep{EinkemmerLubich2018,ConservativeBernard,LiNeuralSparse2024}, learned collision operators target the collision step~\citep{MillerRobertsBondCyr2022,XiaoFrank2021,XiaoFrank2023}, and operator-learning and physics-informed approaches map problem inputs to solutions~\citep{Brunton2020,LuDeepONet2021,LiFNO2021,LouMengKarniadakis2021,DeFlorio2022,Tucny2024,Tucny2025,RoohiMahdaviDeepONet2026,RoohiShojaSaniPoF2026,RoohiShojaSaniAST2026}. An earlier preprint introduced positive Gaussian phase-space representations of shock distributions~\citep{Roohi2026a}, and a second preprint examined their accuracy in the transport moments and their range of validity for normal shocks and lid-driven cavities~\citep{RoohiTransport}. The 512-kernel representation summarised in Sec.~\ref{sec:baseline} and reproduced in the supplementary material is taken from the latter, restricted to the shocks: its fits at Mach 3 and 5, the comparison at equal storage, and the Mach-conditioned models with their range check are the results reported there, and the reference BGK solutions are the same. It is included here as the reference against which the few-population description and the predictions in scaled variables are judged. The reduced distributions, the DSMC, the population analysis, the collision histories, the similarity and the prediction, and the analysis of the precursor (Secs.~\ref{sec:two}--\ref{sec:prec}) are new.

These two lines of work lead to the three questions of this paper, which are taken in this order. First, what does the distribution inside a shock consist of: how many Gaussian populations does it contain, where in the shock does each one matter, what does each contribute to the transport, to what sets of molecules do they correspond, and which of these properties belong to the gas rather than to the kinetic model? Second, how does the distribution depend on shock strength, and can this dependence be used to predict a shock at a Mach number for which no solution exists, including Mach numbers beyond those available? Third, by what mechanism do upstream-moving molecules penetrate ahead of the shock, and how does their decay depend on the collision law? A compact Gaussian representation of the whole phase-space function, with several hundred kernels in physical variables, is carried through the paper as a reference for the second question: it shows what a general-purpose representation achieves for a given shock and where it stops being valid when the shock strength changes. The two approaches differ in more than the choice of variables, and they are compared as whole descriptions (Sec.~\ref{sec:baseline}).

\begin{figure}[!htbp]
\includegraphics[width=\textwidth]{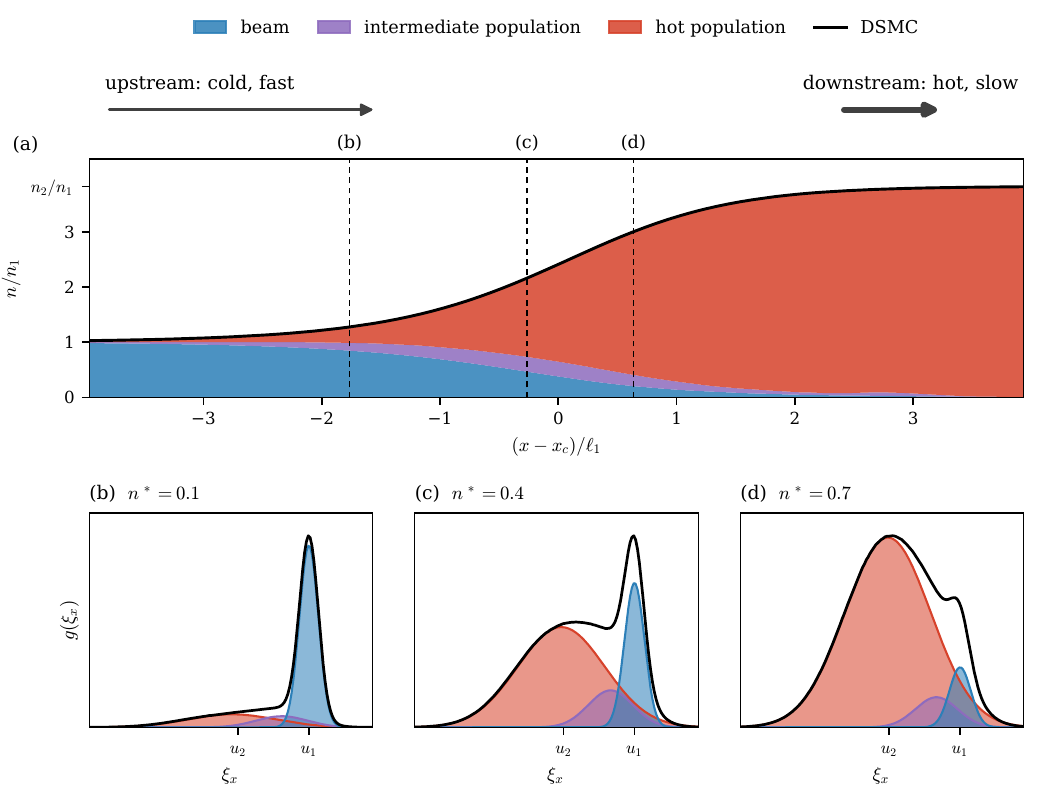}
\caption{The three populations of a hard-sphere shock at Mach 8. (a) Number density through the shock from DSMC
(line) and its division among the beam, the intermediate population, and the hot population of the
three-population fit (shaded); the flow is from left to right and $x_c$ is the position at which $n^*=0.5$.
(b)--(d) Streamwise reduced distribution $g$ at the three positions marked in (a), from DSMC (line), with the three
fitted Gaussians (shaded); $u_1$ and $u_2$ are the upstream and downstream velocities. The beam holds 67\%, 22\%,
and 7\% of the mass at the three positions and the intermediate population 11\%, 13\%, and 7\%; the sum of the
three reproduces $g$ to within 1.1\% throughout the shock.}
\label{fig:overview}
\end{figure}

Figure~\ref{fig:overview} shows the answer to the first question for a hard-sphere gas at Mach 8, and the three
terms used throughout the paper. They are suggested by the kinetic equation. Followed along a molecular path, the
distribution at any position is the sum of the molecules of the upstream flow that have arrived there without a
collision and of the molecules emitted by collisions on the way; for the relaxation models this is
Eq.~\eqref{eq:integral}. The first part is the upstream Maxwellian, attenuated but with its velocity $u_1$ and
temperature $T_1$ unchanged: this is the \emph{beam}. Far downstream the gas is in equilibrium, so that the
scattered molecules approach the Maxwellian of the downstream state: this is the \emph{hot population}. These are
the two terms of Mott-Smith. Inside the shock, however, the scattered molecules are not in equilibrium. Those that
left the stream recently retain part of its momentum, and the distribution of the scattered molecules is skewed
towards $u_1$, which a Maxwellian centred near $u_2$ does not reproduce. The third term, the \emph{intermediate
population}, is a Gaussian centred between $u_2$ and $u_1$ that represents this part; it stands where
\citet{Salwen1964} added a third term to the Mott-Smith distribution and where the gain function of
\citet{SegalFerziger1972} has its third maximum. Each population has its own streamwise and transverse
temperatures, because molecules scattered between two streams are spread differently along and across the flow.
The fit of figure~\ref{fig:overview} is consistent with this reading: on the upstream half of the shock the beam
is centred at $u_1$ to within $0.001\,(u_1-u_2)$ and has the width of the upstream Maxwellian to within 8\%,
and the hot population has the width of the downstream Maxwellian to within 6\% throughout, although none of these values is
imposed. That three terms are sufficient, and what molecules each of them contains, are not assumed: the first is
established by the number of populations required at each position and Mach number (Sec.~\ref{sec:pop}), and the
second by the collision records (Sec.~\ref{sec:meaning}).

The contributions follow the same order. (i) The reference solutions are verified through their conserved fluxes and through the integral form of the kinetic equation, and the DSMC through refinement and through two published direct solutions of the Boltzmann equation. (ii) The reduced distributions are decomposed into at most four anisotropic Gaussian populations between Mach 1.5 and 12; the number required at each position and Mach number and the contribution of each population to the heat flux and stress are determined. The stability of the decomposition is tested with independent data, random restarts, and a changed fit criterion, its physical content with collision histories recorded in DSMC, which also divide the viscous stress and the heat flux between the unscattered stream and the scattered molecules, and its accuracy beyond the fitted reduced distributions with the transverse fourth moment. (iii) The same analysis is applied to Shakhov and ellipsoidal-statistical (ES-BGK) solutions and to DSMC velocity distributions of hard-sphere and variable-hard-sphere gases between Mach 2 and 25, which separates properties of the gas from properties of the BGK model. (iv) The distribution is shown to be nearly independent of Mach number in variables scaled with the Rankine--Hugoniot states, and this similarity, closed with one moment equation, is used to predict the distribution and the profile of shocks at Mach numbers between and beyond those used. Three ways of using the similarity are compared (three populations, the scaled distributions themselves, and the scaled distributions with the beam treated separately), which shows what the Gaussian form contributes; the conserved fluxes of the predictions, their errors in physical space, and the range of source Mach numbers over which each method applies are reported. (v) The upstream precursor is analysed with the integral form of the kinetic equation for the relaxation models and compared with the asymptotic theory of Lyubarskii and Narasimha; for the gas, the composition of the precursor is obtained from collision histories, a linear cascade model without adjustable parameters is compared with the simulations, and the decay exponent is determined with its statistical uncertainty and its dependence on the fit window and the domain length. (vi) A positive mixture of 512 Gaussians is fitted to $\log f(x,\bm\xi)$ for Mach-3 and Mach-5 BGK shocks and conditioned on Mach number over nine shock strengths; tests at withheld positions and with withheld complete solutions, a comparison with regular grids having the same number of coefficients, and a check of the fitted Mach-number interval determine where this representation is valid.

Section~\ref{sec:reference} describes the kinetic solutions, the DSMC, and their verification. Sections~\ref{sec:two} and \ref{sec:pop} establish the departure from two Maxwellians and the population decomposition for the BGK model, and Secs.~\ref{sec:models} and \ref{sec:dsmc} show what changes with the kinetic model and in the gas, and what the populations are. Sections~\ref{sec:sim} and \ref{sec:predictor} give the similarity across Mach number and the prediction built on it. Section~\ref{sec:prec} treats the upstream precursor, first for the relaxation models and then for the gas. Section~\ref{sec:baseline} summarises, for comparison, the many-kernel phase-space representation of \citet{RoohiTransport}. The details of the population fits and of the collision record and the cascade model are given in the appendices, and the complete account of the many-kernel study in the supplementary material.

\section{Kinetic reference problems}
\label{sec:reference}

\subsection{BGK and Shakhov models and discrete velocity moments}

All reference solutions were computed by the author with graphics-processing-unit (GPU) discrete-velocity solvers. The errors reported for the Gaussian descriptions are measured against the DVM distributions and their quadrature moments. The normal shocks use the BGK relaxation model,
\begin{equation}
\frac{\partial f}{\partial t}+\bm\xi\cdot\nabla_{\bm x}f=\frac{f^S-f}{\tau},
\label{eq:bgk}
\end{equation}
with $f^S=f^M$ in the shock calculations. Here $f=f(\bm x,\bm\xi,t)$ is the molecular velocity distribution, $\bm x$ is physical position, $\bm\xi$ is molecular velocity, $t$ is time, and $\tau$ is the relaxation time. In the shock solver, time is measured in units of the upstream relaxation time, and the relaxation time is the same constant in every cell,
\begin{equation}
\begin{aligned}
\tau_{\mathrm{shock}}&=1,\\
f^{n+1}&=f^M+(f^*-f^M)\exp(-\Delta t/\tau_{\mathrm{shock}})\\
&=f^M+(f^*-f^M)\exp(-\Delta t),
\end{aligned}
\label{eq:exact_relaxation}
\end{equation}
where $f^*$ is the distribution after the transport step. The relaxation factor in each time step is thus $\exp(-\Delta t)$, and the relaxation time does not depend on temperature. The local Maxwellian is
\begin{equation}
 f^M=\frac{\rho}{(2\pi T)^{3/2}}\exp\left(-\frac{|\bm c|^2}{2T}\right).
\label{eq:maxwellian}
\end{equation}
The Shakhov solutions used for the model comparison of Sec.~\ref{sec:models} instead use the target
\begin{equation}
 f^S=f^M\left[1+(1-\mathrm{Pr})\frac{\bm c\cdot\bm q}{5pT}
 \left(\frac{|\bm c|^2}{T}-5\right)\right],
\label{eq:shakhov_target}
\end{equation}
with Prandtl number $\mathrm{Pr}=2/3$. In these expressions $\rho$, $\bm u$, and $T$ are the density, bulk velocity, and temperature, $\bm c=\bm\xi-\bm u$ is the peculiar velocity, $p=\rho T$ is the pressure, and $\bm q$ is the heat flux. The relaxation coefficient in Eq.~\eqref{eq:exact_relaxation} was verified directly in the shock solver. The shock solutions are therefore solutions of the BGK model with constant $\tau$ and $\mathrm{Pr}=1$, and the Shakhov solutions have $\mathrm{Pr}=2/3$~\citep{BGK1954,Shakhov1968,Mieussens2000,XuHuang2010}.

Let $\{(\bm\xi_n,w_n)\}_{n=1}^{N_v}$ denote the discrete velocity nodes and quadrature weights, and let $f_n=f(\bm\xi_n)$. The macroscopic fields are
\begin{align}
\rho &= \sum_{n=1}^{N_v} f_nw_n, &
\rho u_i &= \sum_{n=1}^{N_v}\xi_{i,n}f_nw_n, \label{eq:macro1}\\
3\rho T &= \sum_{n=1}^{N_v}|\bm c_n|^2f_nw_n, &
\bm c_n&=\bm\xi_n-\bm u. \label{eq:macro2}
\end{align}
The temperature and the peculiar velocity in Eq.~\eqref{eq:macro2} are defined with the bulk velocity of Eq.~\eqref{eq:macro1}, and all higher moments below are taken about the same velocity.
The transport and higher-order observables used throughout the paper are
\begin{align}
\sigma_{ij}&=\sum_{n=1}^{N_v}c_{i,n}c_{j,n}f_nw_n-\rho T\delta_{ij}, \label{eq:stress}\\
q_i&=\frac12\sum_{n=1}^{N_v}|\bm c_n|^2c_{i,n}f_nw_n, \label{eq:heatflux}\\
M_{300}^{\mathrm{neq}}&=\sum_{n=1}^{N_v}c_{x,n}^3f_nw_n, \label{eq:m300}\\
M_{400}^{\mathrm{neq}}&=\sum_{n=1}^{N_v}c_{x,n}^4f_nw_n-3\rho T^2. \label{eq:m400}
\end{align}
Here $i$ and $j$ denote Cartesian components and $\delta_{ij}$ is the Kronecker delta. Equation~\eqref{eq:stress} is the traceless part of the pressure tensor, Eq.~\eqref{eq:heatflux} the heat flux, and Eq.~\eqref{eq:m300} the streamwise skewness. In Eq.~\eqref{eq:m400} the equilibrium value $3\rho T^2$ is subtracted, so that errors of the fourth moment are measured against its nonequilibrium part.

\subsection{Normal shocks}

The reference problem is the stationary monatomic normal shock, computed at nine upstream Mach numbers between 1.5 and 12. The single-state tests of the 512-kernel representation in Sec.~\ref{sec:baseline} use Mach 3 (M3) and Mach 5 (M5). Shock variables are nondimensionalised by the upstream state, denoted by subscript 1; velocities are scaled by the upstream thermal speed $c_1=\sqrt{T_1}$, so heat flux, stress, and the third- and fourth-order moments are reported in units of $\rho_1c_1^3$, $\rho_1c_1^2$, $\rho_1c_1^3$, and $\rho_1c_1^4$, respectively. The plotted coordinate is $x/\lambda_1=x^*/\mathrm{Kn}_{\mathrm{eff}}$, where $x^*\in[-1/2,1/2]$ is the solver coordinate, $\lambda_1$ is the upstream mean free path, and $\mathrm{Kn}_{\mathrm{eff}}=\lambda_1/L$ is the effective Knudsen number based on the domain length $L$. Rankine--Hugoniot (RH) states are imposed at the domain boundaries. The M3 reference contains 1800 grid points in $x$ and 64,009 velocity nodes arranged as $121\times23\times23$; the M5 reference contains 2200 grid points in $x$ and 102,789 velocity nodes arranged as $141\times27\times27$. Their distributions contain approximately $1.15\times10^8$ and $2.26\times10^8$ scalar values.

The two shocks differ in scale: the M3 transition is narrower, and the M5 profiles have larger ranges of downstream temperature and of the fourth moment. The reconstructions of both are summarised in Sec.~\ref{sec:baseline} and given in the supplementary material.

The parametric study uses nine independently converged BGK-DVM shocks at $M=1.5$, 2, 2.5, 3, 4, 5, 6, 8, and 12. Their physical grids contain 1400--4200 points and their velocity grids contain 35,017--270,725 nodes. Together the nine distributions contain $2.59\times10^9$ values (10.36 GB in single precision). In each test one Mach number is withheld, and the model is fitted to the other eight solutions and evaluated on all nine. M3 and Mach 6 (M6) are interpolation tests, because the fitted Mach numbers lie on both sides of them; Mach 12 (M12) is an extrapolation test, because withholding it reduces the fitted interval from $[1.5,12]$ to $[1.5,8]$.

\subsection{Convergence and internal verification}
\label{sec:verification}

\begin{table}
\caption{Reference DVM configurations and numerical verification. GL3 denotes composite three-point Gauss--Legendre velocity quadrature. For M3 and M5, the verification entry is the largest discrepancy when moments are recomputed from the stored distributions; for M6 and M12, it is the largest cross-grid discrepancy among the low-order and nonequilibrium moments of table~\ref{tab:shock_grid_cert}.}
\label{tab:dvm_setup}
\centering
\small
\resizebox{\textwidth}{!}{%
\begin{adjustbox}{max width=\linewidth}
\begin{tabular}{lcccccc}
\toprule
Case & model & physical grid & velocity grid & domain & rarefaction & verification\\
\midrule
M3 shock & BGK & $N_x=1800$ & $121\times23\times23$ & $|\xi|\le16$ & $\mathrm{Kn}_{\rm eff}=1/120$ & stored-$f$ $<1.4\times10^{-6}$\\
M5 shock & BGK & $N_x=2200$ & $141\times27\times27$ & $|\xi|\le22$ & $\mathrm{Kn}_{\rm eff}=1/160$ & stored-$f$ $<9.4\times10^{-7}$\\
M6 shock & BGK & $N_x=2880$ & $75\times72\times72$ & composite GL3 & $x/\lambda_1\in[-95,95]$ & medium--fine $<0.12\%$\\
M12 shock & BGK & $N_x=4200$ & $63\times60\times60$ & composite GL3 & $x/\lambda_1\in[-240,240]$ & coarse--medium $<0.97\%$\\
\bottomrule
\end{tabular}
\end{adjustbox}%
}
\end{table}

As summarised in table~\ref{tab:dvm_setup}, direct recomputation of $\rho$, $u_x$, $T$, $q_x$, and $\sigma_{xx}$ from the stored M3 and M5 distributions reproduces the archived profiles with maximum relative $L_2$ discrepancies below $1.4\times10^{-6}$ and $9.4\times10^{-7}$, respectively.

\begin{table}
\caption{Independent BGK-DVM grid-refinement discrepancies for the high-Mach references. Entries are relative $L_2$ differences in percent. The maximum difference is 0.11448\% for M6 medium--fine and 0.96139\% for M12 coarse--medium.}
\label{tab:shock_grid_cert}
\centering
\small
\begin{adjustbox}{max width=\linewidth}
\begin{tabular}{lccccccc}
\toprule
Grid pair & $\rho$ & $u_x$ & $T$ & $q_x$ & $\sigma_{xx}$ & $M_{300}^{\rm neq}$ & $M_{400}^{\rm neq}$\\
\midrule
M6 medium--fine & 0.00569 & 0.00828 & 0.01238 & 0.10165 & 0.11448 & 0.09541 & 0.09708\\
M12 coarse--medium & 0.08202 & 0.05513 & 0.10359 & 0.27290 & 0.74370 & 0.65464 & 0.96139\\
\bottomrule
\end{tabular}
\end{adjustbox}
\end{table}

High-Mach sensitivity was checked independently by repeating M6 and M12 with composite three-point Gauss--Legendre (GL3) velocity grids and double-precision (float64) arithmetic. The M6 medium and fine grids use 226,800 and 388,800 velocity nodes; their relative discrepancies are 0.0057--0.0124\% for $\rho$, $u_x$, and $T$ and 0.095--0.114\% for $q_x$, $\sigma_{xx}$, $M_{300}^{\mathrm{neq}}$, and $M_{400}^{\mathrm{neq}}$. The M12 coarse and medium grids use 34,816 and 226,800 velocity nodes; the corresponding discrepancies are 0.055--0.104\% and 0.273--0.961\%. Table~\ref{tab:shock_grid_cert} reports every quantity. The converged M6 fine and M12 medium profiles are used for the independent re-evaluation of the moments.

\subsection{Additional solutions, reduced distributions, verification, and DSMC}
\label{sec:data}

The analysis of Secs.~\ref{sec:two}--\ref{sec:prec} uses all nine BGK shocks and, to assess the dependence on the collision model, steady solutions of the Shakhov model at Mach 2--5 ($\mathrm{Pr}=2/3$, viscosity $\mu\propto T$), one Shakhov solution at Mach 8 with $\mu\propto T^{0.81}$, and solutions of the ES-BGK model~\citep{Holway1966} at Mach 2 and 3 ($\mathrm{Pr}=2/3$, $\mu\propto T$). These were computed for the reduced distributions defined below with first-order upwind DVM schemes in an earlier study~\citep{RoohiJFM}. The DSMC solutions are described in Sec.~\ref{sec:dsmcsim}. All kinetic-model solutions are steady solutions of
\begin{equation}
\xi_x\frac{\partial f}{\partial x}=\nu\,(f_t-f),
\label{eq:kinetic}
\end{equation}
where $f_t$ is the target distribution of the model and the collision frequency $\nu$ does not depend on molecular velocity; $\nu=1/\tau_{\mathrm{shock}}$ for the BGK shocks. Lengths in these sections are measured in the upstream relaxation length $\ell_1=\sqrt{RT_1}\,\mu_1/p_1$, with $R=1$ the gas constant in the present units; for the BGK shocks $\ell_1$ coincides with $\lambda_1$. For Mach 6 and 12 the population analysis uses the GL3 solutions of table~\ref{tab:dvm_setup}, for the reason given below.

\subsubsection{Reduced distributions and scaled variables}

For a plane shock, all moments that involve only streamwise transport follow from the two reduced
distributions introduced by Chu~\citeyearpar{Chu1965},
\begin{equation}
g(x,\xi_x)=\int f\,d\xi_y\,d\xi_z,\qquad
h(x,\xi_x)=\int(\xi_y^2+\xi_z^2)\,f\,d\xi_y\,d\xi_z .
\end{equation}
With $c=\xi_x-u$, the density is $\rho=\int g\,d\xi_x$, the streamwise and transverse temperatures
are $\rho\Tpar=\int c^2g\,d\xi_x$ and $2\rho\Tperp=\int h\,d\xi_x$, $T=(\Tpar+2\Tperp)/3$, the normal
stress is $\sigma_{xx}=\rho(\Tpar-T)$, and the heat flux is $q_x=\tfrac12\int c\,(c^2g+h)\,d\xi_x$.
The fourth-order moment reported below is $M_{400}^{\rm neq}=\int c^4g\,d\xi_x-3\rho T^2$. For the
BGK solutions $g$ and $h$ were obtained by integrating the stored three-dimensional distributions
over the transverse velocities; the moments recomputed from $g$ and $h$ agree with those stored by
the solver to within $6\times10^{-6}$ on the uniform grids and $1.4\times10^{-4}$ on the
Gauss--Legendre grids.

Positions are labelled by the normalised density $n^*=(\rho-\rho_1)/(\rho_2-\rho_1)$, and molecular
velocities by
\begin{equation}
\xih=\frac{\xi_x-u_2}{u_1-u_2},
\end{equation}
so that the upstream and downstream bulk velocities are at $\xih=1$ and $\xih=0$ for every Mach
number. A molecule moves upstream, against the flow, when $\xi_x<0$, that is, when
$\xih<-u_2/(u_1-u_2)=-1/(\rho_2/\rho_1-1)$; this limit is $-0.78$ at Mach 2, $-0.5$ at Mach 3, and tends
to $-1/3$ for strong shocks. A negative $\xih$ alone means only that the molecule is slower than the
downstream gas. The mass fraction of upstream-moving molecules,
$\beta=\rho^{-1}\int_{\xi_x<0}g\,d\xi_x$, is always evaluated with the limit $\xi_x=0$. This is the total upstream-moving mass fraction. It includes the molecules
of the upstream equilibrium itself, $\beta_1=\tfrac12\,\mathrm{erfc}(M\sqrt{\gamma/2})$, which is $5.4\times10^{-5}$ at Mach 3,
$1.2\times10^{-7}$ at Mach 4, and $5.4\times10^{-11}$ at Mach 5. Where this constant is subtracted, the result is called the
excess $\beta-\beta_1$; this is done only for the BGK curves of figure~\ref{fig:prec}(b) and the slopes derived from
them, which include Mach 3 and 4. All DSMC results for the precursor are at Mach 5 and above and use the total $\beta$.

\subsubsection{Verification}
\label{sec:verify}

\begin{table*}
\centering\small
\caption{Kinetic solutions and verification. $N_x$ is the number of cells, $L$ the half-length of
the domain in units of $\ell_1$ (the Mach-8 Shakhov domain is $[-48,29]$), and $\delta$ the
maximum-slope density thickness. ``Flux'' is the largest relative deviation of the mass, momentum
or energy flux from its upstream value; ``integral'' is the largest relative $L_2$ difference
between the computed $g$ and its reconstruction from Eq.~\eqref{eq:integral} for $0.01<n^*<0.99$.
GL denotes composite three-point Gauss--Legendre velocity grids.}
\label{tab:data}
\begin{adjustbox}{max width=\linewidth}
\begin{tabular}{llccccccc}
\toprule
Model & $M$ & $N_x$ & velocity grid & $L$ & $\delta/\ell_1$ & flux (\%) & integral (\%) & $(\Tpar/\Tperp)_{\rm in}$\\
\midrule
BGK, constant $\tau$ & 1.5 & 1400 & $97\times19\times19$ & 35 & 10.0 & 0.05 & 0.03 & 1.000\\
 & 2 & 1600 & $97\times19\times19$ & 40 & 7.7 & 0.07 & 0.08 & 1.001\\
 & 2.5 & 1600 & $111\times21\times21$ & 50 & 7.7 & 0.15 & 0.17 & 1.002\\
 & 3 & 1800 & $121\times23\times23$ & 60 & 8.1 & 0.16 & 0.21 & 1.003\\
 & 4 & 2000 & $131\times25\times25$ & 70 & 9.6 & 0.17 & 0.23 & 1.005\\
 & 5 & 2200 & $141\times27\times27$ & 80 & 11.3 & 0.18 & 0.31 & 1.026\\
 & 6 & 2880 & $75\times72\times72$ (GL) & 95 & 13.2 & 0.13 & 0.18 & 1.000\\
 & 8 & 2800 & $171\times31\times31$ & 120 & 17.1 & 0.29 & 1.38 & 1.180\\
 & 12 & 4200 & $63\times60\times60$ (GL) & 240 & 25.0 & 0.12 & 0.18 & 1.000\\
Shakhov, $\mu\propto T$ & 2 & 1280 & 321 & 30 & 6.4 & 0.11 & 0.08 & 1.000\\
 & 3 & 1920 & 801 & 45 & 5.8 & 0.17 & 0.14 & 1.000\\
 & 4 & 2560 & 801 & 60 & 6.5 & 0.17 & 0.15 & 1.000\\
 & 5 & 3200 & 801 & 75 & 7.4 & 0.16 & 0.15 & 1.000\\
Shakhov, $\mu\propto T^{0.81}$ & 8 & 1200 & 401 & -- & 6.1 & 0.29 & 0.29 & 1.029\\
ES-BGK, $\mu\propto T$ & 2 & 640 & 801 & 30 & 5.8 & 0.24 & 0.21 & 1.000\\
 & 3 & 960 & 801 & 45 & 5.1 & 0.37 & 0.43 & 1.001\\
\bottomrule
\end{tabular}
\end{adjustbox}
\end{table*}

Three tests were applied to every solution; the results are in table~\ref{tab:data}.

\emph{Conserved fluxes.} In a steady shock the fluxes of mass, momentum and energy, computed from
$g$ and $h$, must be independent of $x$ and equal to their upstream values. The largest deviation
over the whole domain is below 0.3\% for the BGK and Shakhov solutions and below 0.4\% for the
ES-BGK solutions. The downstream density and temperature agree with the RH values to within
0.1\%.

\emph{Integral form.} Integrating Eq.~\eqref{eq:kinetic} along a characteristic gives, for
$\xi_x<0$,
\begin{equation}
g(x,\xi_x)=\int_x^{\infty} g_t(x',\xi_x)\,
\exp\!\left[-\frac{\theta(x,x')}{|\xi_x|}\right]\frac{\nu(x')\,dx'}{|\xi_x|},
\qquad \theta(x,x')=\int_x^{x'}\nu\,dx'' ,
\label{eq:integral}
\end{equation}
and the analogous expression with integration from upstream for $\xi_x>0$; the same holds for $h$.
The reduced target $g_t$ and the collision frequency depend on $x$ only through the low-order
moments, so the right-hand side can be evaluated from the moments of a computed solution and
compared with its distribution. The difference measures any residual time dependence together
with the discretisation error of the solver, without reference to a second computation. For all
solutions except one, the reconstructed and computed $g$ differ by at most 0.5\% (relative $L_2$
norm in velocity at the worst position, median 0.02--0.09\%), and the heat flux and stress of the
reconstruction agree with the computed ones to within 1.2\% of their peaks. The exception is the
Mach-8 BGK solution (1.4\%), discussed next.

\emph{Upstream equilibrium.} The BGK solutions at Mach 1.5--5 and 8 use velocity grids that are
uniform in all three directions. At high Mach number the transverse spacing of these grids
becomes comparable to the upstream thermal speed, and the discrete upstream Maxwellian acquires an
artificial anisotropy: $\Tpar/\Tperp$ at the inflow boundary is within 0.5\% of unity up to Mach 4
but 1.026 at Mach 5 and 1.18 at Mach 8. Solutions on composite Gauss--Legendre grids are available
at Mach 6 and 12 and are free of this error ($\Tpar/\Tperp-1<3\times10^{-5}$); they are used
throughout. The effect on the streamwise distribution $g$ is small: at Mach 6 the uniform-grid
and Gauss--Legendre-grid $g$ are indistinguishable, and at Mach 12, where the artificial anisotropy
of the uniform grid is far larger than at Mach 8, they differ by about 5\% of the peak value.
To determine what the error at Mach 8 does to the results, that shock was recomputed for the reduced
distributions themselves: Eq.~\eqref{eq:kinetic} for $g$ and $h$ with the reduced Maxwellians $g_M$ and
$h_M=2Tg_M$, on the same spatial grid ($\Delta x=0.086\,\ell_1$) and 321 streamwise velocities, uniformly spaced between
$\pm32\sqrt{RT_1}$ and integrated with the trapezoidal rule, with first-order upwind transport and exact
relaxation. The boundary values are the Rankine--Hugoniot Maxwellians, and the solution is marched from the
archived one for 300 relaxation times, at the end of which the density still changes at a rate of $4.3\times10^{-6}$ per relaxation time. This residual is a slow drift of the shock through the grid, at $2.6\times10^{-5}\,\ell_1$ per relaxation time, and not a change of its structure: continuing the march for a further 150 relaxation times moves the shock centre by 0.004\,$\ell_1$ and changes $g$ and $h$ at equal $n^*$ by less than 0.0002\% and 0.0005\%, and the profiles of the heat flux, the stress, and the temperature against $n^*$ by less than 0.0003\%; the shock thickness and the peak of $\Tpar/\Tperp$ change in the sixth significant figure.

Because the transverse velocities are integrated
analytically, the discrete equilibrium has no transverse error. In the recomputed solution $\Tpar/\Tperp$ at
the inflow is 1.0002, the fluxes are constant to within 0.13\%, and the downstream state agrees with the RH
values to $10^{-5}$. At equal $n^*$ it differs from the archived solution by at most 1.6\% in $g$ and 1.2\% in
$h$ (relative $L_2$ norm; the largest values occur at the foot of the shock, $n^*=0.02$, and at the shock centre
the differences are 0.3\% and 0.2\%), by 0.2\% in the profiles of the heat flux, the stress, and the
temperatures, by 0.2\% in the peak of $\Tpar/\Tperp$ (3.124 against 3.130), and by less than 0.0003 in the
upstream-moving mass fraction; the scaled parameters of the three populations of Sec.~\ref{sec:pop} change by
at most 0.03. The same solver applied at Mach 6 and 12 reproduces the Gauss--Legendre solutions to within
0.6\% and 0.8\% in $g$, 0.9\% and 1.9\% in $h$, and 0.02\% in the profiles of the moments, which bounds the difference due to the solver itself. Two tests bound its discretisation error. With twice the cell size the fluxes are constant to within 0.26\%, and the solution changes by at most 0.3\% in $g$, 0.7\% in $h$, and 0.73\% in the profiles of the heat flux, stress, and temperature; with half the number of velocities the changes are 0.5\%, 0.4\%, and less than 0.01\%. With twice the number of velocities they are 0.13\%, 0.11\%, and 0.001\%. The scheme is of first order in the cell size, so that the error of the solution used is about that of the first test; the agreement with the Gauss--Legendre solutions on the same cells shows that this error is common to the archived and the recomputed solutions, and therefore to the sources and the target of a prediction.  The artificial anisotropy is thus confined to the cold gas ahead of the shock, whose transverse
temperature is a few percent of $T_2$, and the reported comparisons quantify its influence on the observables tested below. The tables and figures use
the archived solution; the predictions that take Mach 8 as a source are repeated with the recomputed one in
Sec.~\ref{sec:predictor}.

\subsubsection{Direct simulation Monte Carlo}
\label{sec:dsmcsim}

To separate properties of the gas from properties of the relaxation models, normal shocks were also
computed by DSMC~\citep{Bird1994} for a variable-hard-sphere (VHS) gas with viscosity exponents
$\omega=0.5$ (hard spheres), 0.81 and 1 (Maxwell-type) at Mach 2, 3, 4, 5, 6, 8, 10, 12 and 25, and the reduced distributions $g$ and $h$ were sampled in every cell. The reference cross-section is chosen so that
the first-approximation VHS viscosity at $T_1$ equals $\mu_1$; lengths are then in the same unit
$\ell_1$ as for the kinetic models, and a DSMC gas with exponent $\omega$ has the same viscosity law
as a relaxation model with $\mu\propto T^\omega$. The domain is $-60\le x/\ell_1\le30$ ($-90\le x/\ell_1\le40$ at Mach 10 and 12; at Mach 25, $-60\le x/\ell_1\le30$, $-160\le x/\ell_1\le60$, and $-300\le x/\ell_1\le100$ for
$\omega=0.5$, 0.81, and 1) with cells of $0.1\,\ell_1$, 200 simulated molecules per upstream cell, and time steps of 0.001--0.01\,$\ell_1/\sqrt{RT_1}$.
Collisions use the no-time-counter scheme. Both ends are equilibrium reservoirs at the RH states.
The shock position is estimated from the total number of molecules, and the molecules are shifted
by one cell whenever it has moved by half a cell, the emptied end cell being refilled from the
reservoir. After 15\,000 steps (30\,000 at Mach 25), $g$ and $h$ are accumulated every second step for 200\,000 steps (215\,000 at Mach 25)
on a velocity grid of spacing $0.2\sqrt{RT_1}$; eight independent runs per case give
$1.6\times10^8$ samples per upstream cell.

The upstream and downstream densities, velocities and temperatures agree with the RH states to
within 0.25\% (0.5\% and 0.8\% for the upstream temperature at Mach 8 and 12 with $\omega=1$, where the
precursor reaches the inflow region), and the fluxes of mass, momentum and energy are constant to within 0.35\%. The
shock position in four consecutive quarters of the sampling time varies by at most
$0.06\,\ell_1$.

The cell is largest relative to the local mean free path for hard spheres downstream of the
strongest shocks, where $\Delta x/\lambda_2\approx0.3$; the time step is at most 0.08 of the local mean
collision time. The sensitivity to these parameters was tested at Mach 8, 12 and 25. At Mach 8, for hard spheres,
halving both the cell and the time step changes the width of the density profile by 0.4\%, the
heat-flux and stress profiles by 0.6\% and 0.8\%, the peak of $\Tpar/\Tperp$ by 0.5\%, and the
distributions at $n^*=0.1$, 0.5 and 0.9 by 0.2--0.4\%; halving the time step alone gives changes of
the same size, so that the cell size contributes little. With 50 instead of 200 molecules per cell
the width changes by 1.3\% and the distributions by 0.4--0.6\%. For $\omega=1$, halving the cell and
the time step changes the width by 0.25\%, the heat flux and stress by 0.2\%, and the
distributions by 0.2--0.3\%. At Mach 12 the same halving changes the width by 0.6\% for hard spheres and 0.3\% for
$\omega=1$, the heat-flux and stress profiles by at most 0.7\%, the peak of $\Tpar/\Tperp$ by at most 0.4\%, and the
distributions by 0.2--0.5\%; with 400 instead of 200 molecules per cell the changes are at most 0.5\% in the width and
0.3\% in the heat flux, the stress, and the distributions. The distance over which $\beta$ falls from $10^{-2}$ to
$10^{-4}$ ahead of the shock, which sets the decay rates of Sec.~\ref{sec:prec}, changes by at most 0.6\% in these four
tests. At Mach 25, for hard spheres, halving both the cell and the time step changes the width by 0.6\%, the
heat-flux and stress profiles by at most 0.8\%, the peak of $\Tpar/\Tperp$ by 0.4\%, and the distributions by
0.3--0.7\%; with 400 molecules per cell the changes are at most 0.2\% in the width, the heat flux, the stress, and the
distributions. For $\omega=1$ at the same Mach number the two tests change the width by 0.5\% and 0.4\%, the heat-flux
and stress profiles by at most 0.2\%, the peak of $\Tpar/\Tperp$ by 0.1\%, and the distributions by at most 0.5\%.
The distance over which $\beta$ falls from $10^{-2}$ to $10^{-4}$ changes by 1.6\% and 1.7\% in the two hard-sphere
tests and by 2.9\% and 0.3\% for $\omega=1$, well inside the uncertainty of the Mach-25 exponents of table~\ref{tab:exponents}. The statistical scatter of these comparisons is 0.05--0.1\%. The
excess density of 0.1--0.25\% at both ends is inversely proportional to the number of molecules
per cell.

Three independent solutions are available for comparison. For hard spheres, the density and
temperature profiles at Mach 2 and 3 were compared with the direct numerical solution of the
Boltzmann equation by Ohwada~\citeyearpar{Ohwada1993} (his Tables III and V), whose length unit
$(\sqrt{\pi}/2)\,l_0$, with $l_0$ the upstream mean free path, equals $1.131\,\ell_1$. With the two
solutions aligned at $n^*=0.5$, the density differs by at most 0.3\%, the temperature by at most
0.2\%, and the width of the density profile between $n^*=0.2$ and 0.8 by 0.1\% at Mach 2 and
0.3\% at Mach 3. At Mach 3 and 8 the profiles were also compared with the solution of the Boltzmann equation by
Dodulad \& Tcheremissine~\citeyearpar{Dodulad2013} (their table, with lengths in units of $l_0=1.277\,\ell_1$): the density
and the temperature differ by at most 0.2\% of their rise across the shock, the longitudinal temperature $\Tpar$ by at
most 0.9\%, and the maximum-slope width of the density profile by 1.0\%. For $\omega=1$ the results were compared with an independent DSMC
computation~\citep{RoohiJFM}: the peak values of $\Tpar/\Tperp$ are 1.369, 1.734 and 2.153 at Mach
2, 3 and 5, against 1.369, 1.733 and 2.147, and the heat-flux minima agree to within 0.8\%. The
shock thickness for $\omega=1$, $\delta/\ell_1=6.5$, 6.0 and 7.6 at Mach 2, 3 and 5, is within 4\% of that of the Shakhov solutions with $\mu\propto T$ (6.4, 5.8 and 7.4).

\FloatBarrier
\section{Departure from the two-Maxwellian picture}
\label{sec:two}

\begin{figure}[!htbp]
\includegraphics[width=\textwidth]{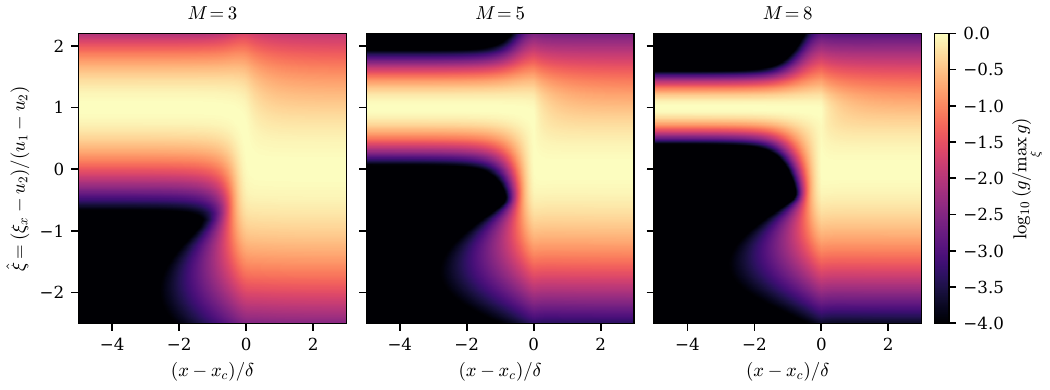}
\caption{Streamwise distribution $g$ of the BGK solutions at Mach 3, 5 and 8, normalised by its
maximum over velocity at each position. $x_c$ is the position where $n^*=0.5$ and $\delta$ the
maximum-slope density thickness. The distribution is interpolated in velocity for display.}
\label{fig:maps}
\end{figure}

\begin{figure}[!htbp]
\includegraphics[width=\textwidth]{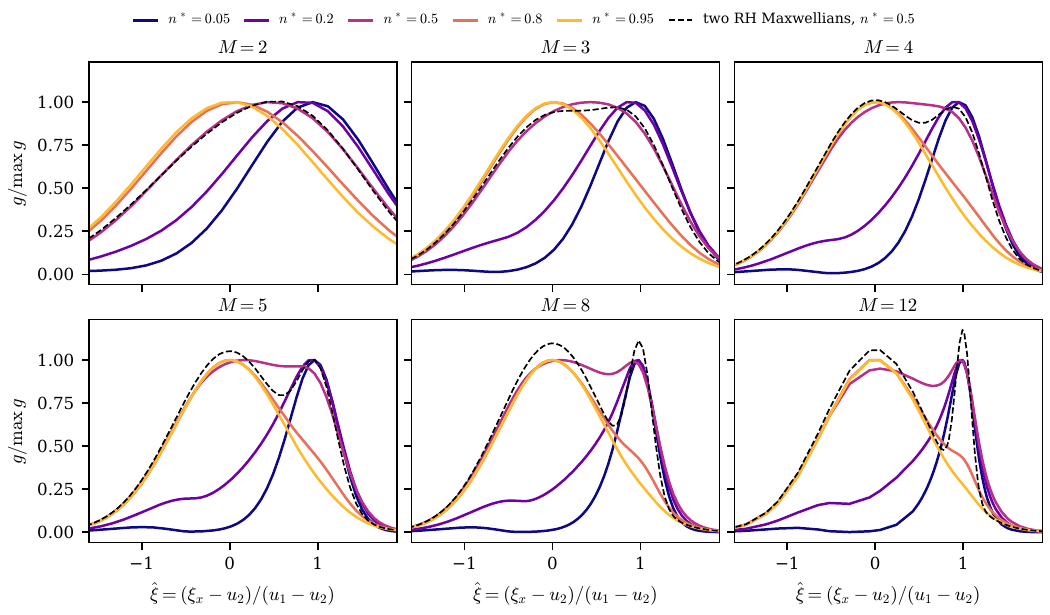}
\caption{Streamwise distribution of the BGK solutions at five positions in the shock. The dashed
line is the best non-negative combination of the two Rankine--Hugoniot Maxwellians at $n^*=0.5$.}
\label{fig:profiles}
\end{figure}

Figure~\ref{fig:maps} shows the streamwise distribution of the BGK solutions as a function of
position and velocity, and figure~\ref{fig:profiles} shows $g$ at five positions for six Mach
numbers. Three features are visible. First, upstream of the shock centre a group of molecules with
$\xi_x<0$, that is, moving against the flow, extends several shock thicknesses ahead of the density
rise; from Mach 2.5 upward it forms a separate maximum of $g$ near $\xih\approx-0.8$. Second, the
gap between the upstream beam at $\xih=1$ and the hot gas near $\xih=0$ is filled: at Mach 3 and 4
the distribution at the shock centre has a single broad maximum, and two maxima appear at the
centre only from Mach 5 upward. Third, the beam itself broadens and shifts slightly as it crosses
the shock.

\begin{table}
\centering\small
\caption{Ranges of $n^*$ (within $0.02<n^*<0.98$) over which the streamwise distribution of the BGK
solutions has two maxima.}
\label{tab:bimodal}
\begin{adjustbox}{max width=\linewidth}
\begin{tabular}{lcc}
\toprule
$M$ & upstream side & centre\\
\midrule
1.5, 2 & none & none\\
2.5 & 0.02--0.05 & none\\
3 & 0.02--0.10 & none\\
4 & 0.02--0.15 & none\\
5 & 0.02--0.18 & 0.47--0.50\\
6 & 0.02--0.18 & 0.46--0.54\\
8 & 0.02--0.21 & 0.45--0.60\\
12 & 0.02--0.21 & 0.43--0.69\\
\bottomrule
\end{tabular}
\end{adjustbox}
\end{table}

Table~\ref{tab:bimodal} gives the ranges of $n^*$ over which $g$ has two maxima. The upstream
bimodality, caused by the upstream-moving group, starts at Mach 2.5 and at Mach 8--12 extends to
$n^*\approx0.2$. The central bimodality, which is the one usually associated with the Mott-Smith
picture, is confined to $0.47<n^*<0.50$ at Mach 5 and widens to $0.43<n^*<0.69$ at Mach 12.

\begin{figure}[!htbp]
\includegraphics[width=\textwidth]{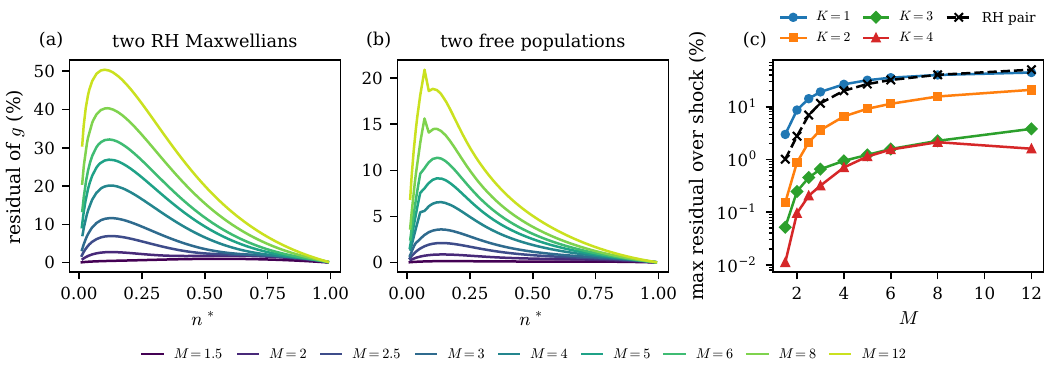}
\caption{Relative $L_2$ residual of $g$ for the BGK solutions. (a) Best non-negative combination
of the two Rankine--Hugoniot Maxwellians. (b) Two free bi-Maxwellian populations. (c) Largest
residual over $0.01<n^*<0.99$ for one to four populations and for the RH pair.}
\label{fig:count}
\end{figure}

To measure the distance from the two-Maxwellian picture we fit $g$ at every position by a
non-negative combination of the two RH Maxwellians, $g\simeq a\,g_{M1}+b\,g_{M2}$, with $a$ and $b$
chosen independently at each position by least squares. This is more favourable than the
Mott-Smith solution itself, in which $a$ and $b$ are tied to a single profile. The relative $L_2$
residual is shown in figure~\ref{fig:count}(a). Its maximum over the shock grows from 1.0\% at Mach
1.5 to 11.6\% at Mach 3, 27\% at Mach 5 and 50\% at Mach 12. The maximum is located at
$n^*\approx0.1$ for all Mach numbers above 1.5, on the upstream side, where the upstream-moving
group and the filling of the gap are both present; at the shock centre the residual is 3.5\% at
Mach 3 and 20\% at Mach 12.

\section{Population decomposition}
\label{sec:pop}

\subsection{Method}

At each position the distribution is approximated by $K$ populations, each an anisotropic Gaussian
(bi-Maxwellian) with number density $n_k$, streamwise mean velocity $u_k$, and streamwise and
transverse temperatures $T_{\parallel k}$ and $T_{\perp k}$. In terms of the reduced distributions,
\begin{equation}
g\simeq\sum_{k=1}^{K}g_k,\qquad g_k=\frac{n_k}{\sqrt{2\pi T_{\parallel k}}}
\exp\!\left[-\frac{(\xi_x-u_k)^2}{2T_{\parallel k}}\right],\qquad h\simeq\sum_{k=1}^{K}2T_{\perp k}\,g_k .
\label{eq:mixture}
\end{equation}
The $4K$ parameters are found by minimising the sum of the squared relative $L_2$ residuals of $g$
and $h$ on the velocity grid of the solution, with the Levenberg--Marquardt method and analytic
derivatives. About 100 positions are fitted in each solution. For each $K$ the fit is started from
several initial guesses (the RH states, the solution at the neighbouring position, and the
$K-1$ solution with one population added at the maximum of its residual), and the fit with the
smallest residual is kept. For $K=3$ we also follow a single solution branch continuously from
the upstream end of the shock, each position being started from its upstream neighbour only; this
branch has the same residual as the best fit at all but a few positions and is used wherever
parameters are compared between positions or Mach numbers. One population is a local bi-Maxwellian, the distribution of the Gaussian (ten-moment) closure~\citep{LevermoreMorokoff1998}, which carries no heat flux;
two free populations generalise the Mott-Smith ansatz by releasing all of its parameters.

Two points of interpretation apply to what follows. First, the fit is made to the reduced distributions
$g$ and $h$, which do not determine the distribution $f(\xi_x,\xi_y,\xi_z)$ uniquely. Statements about
accuracy refer to $g$, $h$, and the moments that follow from them; Sec.~\ref{sec:meaning} adds a test of
the transverse fourth moment, which $g$ and $h$ do not contain. Second, a term of the fit need not be
a set of molecules with a common history. We call every Gaussian term a population, in the sense of an
effective population with a density, a mean velocity, and two temperatures, and name the three terms of the
$K=3$ fit the beam, the intermediate population, and the hot population; how stable the
decomposition is, and to what the populations correspond physically, is examined in
Sec.~\ref{sec:meaning} with collision histories recorded in DSMC.

All moments of the mixture~\eqref{eq:mixture} are algebraic. With $d_k=u_k-u$, the heat flux and
stress are sums of contributions from each population,
\begin{equation}
q_x=\sum_k\tfrac12 n_kd_k\,(d_k^2+3T_{\parallel k}+2T_{\perp k}),\qquad
\sigma_{xx}=\sum_k\tfrac23 n_k\,(T_{\parallel k}+d_k^2-T_{\perp k}),
\label{eq:contrib}
\end{equation}
which allows the transport to be attributed to populations.

\subsection{Number of populations}

Figure~\ref{fig:count}(b) shows the residual of the free two-population fit and figure~\ref{fig:count}(c) the largest residual over the shock for $K=1$ to 4. Releasing the
parameters of the two Maxwellians reduces the error by a factor of two to three (to 3.6\% at Mach
3 and 21\% at Mach 12) but does not change its location or its growth with Mach number. A third
population reduces the largest residual to 0.05\% at Mach 1.5, 0.65\% at Mach 3, 1.2\% at Mach 5
and 3.8\% at Mach 12. A fourth population brings a further improvement only below Mach 4 and at
Mach 12.

\begin{figure}[!htbp]
\includegraphics[width=\textwidth]{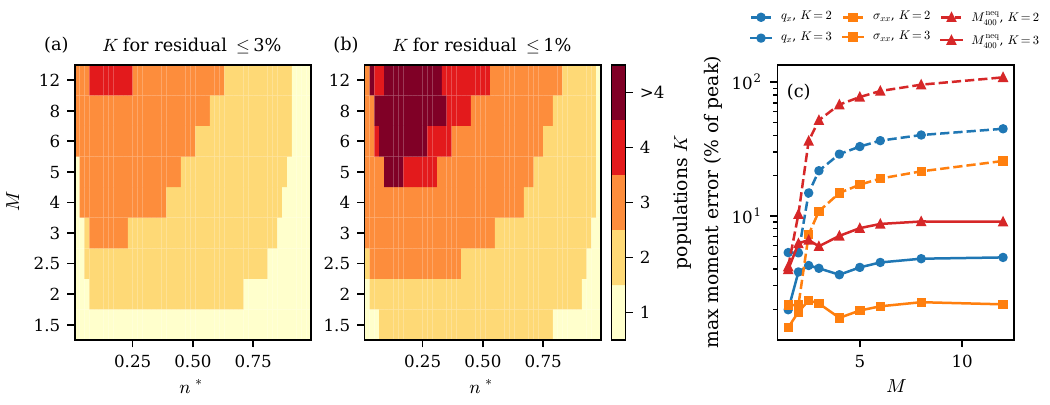}
\caption{(a), (b) Smallest number of populations that represents $g$ to within 3\% and 1\%.
(c) Largest errors of the heat flux, normal stress and fourth-order moment of the two- and
three-population fits, relative to the peak value of each moment.}
\label{fig:map}
\end{figure}

The consequences for the moments are more severe than the residuals of $g$ suggest
(figure~\ref{fig:map}c). With two free populations the largest errors of the heat flux, stress and
fourth-order moment, relative to their peak values, are 22\%, 11\% and 52\% at Mach 3 and 45\%,
26\% and 109\% at Mach 12. With three populations they are at most 4.9\%, 2.3\% and 9.1\% over the
whole range of Mach numbers. Figures~\ref{fig:map}(a) and \ref{fig:map}(b) show the smallest
number of populations that represents $g$ to within 3\% and 1\% as a function of position and Mach
number. One population suffices only at the edges of weak shocks; two suffice on the downstream
half; three are needed on the upstream half from Mach 2.5 (at the 1\% level) or Mach 3 (at the 3\%
level) upward. The region near $n^*\approx0.1$--0.3 at Mach 5 and above cannot be represented to
1\% even with four Gaussians, which we attribute to the structure of the distribution next to the
beam, known to be singular in the limit of infinite Mach number~\citep{Cercignani1999,Takata2000}.

\subsection{Identity of the populations and scaling with Mach number}

\begin{figure}[!htbp]
\includegraphics[width=\textwidth]{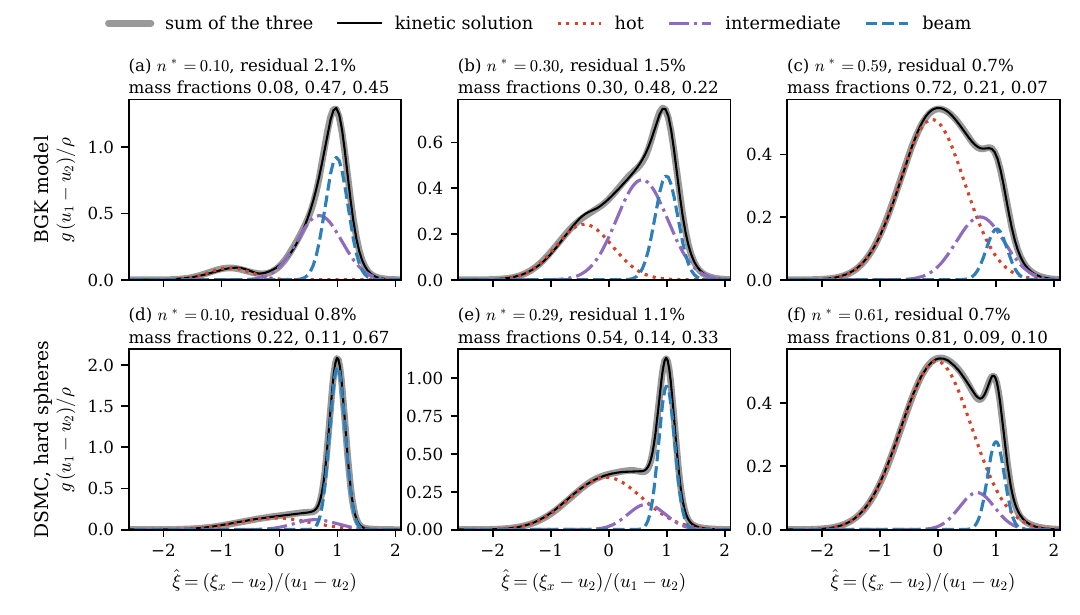}
\caption{The three fitted populations on the scaled streamwise distribution at three positions of the Mach-8 shock: BGK model (top) and DSMC of hard spheres (bottom). Thin solid line: kinetic solution; thick grey line: sum of the three populations; dotted, dash-dotted, and dashed lines: hot population, intermediate population, and beam. Above each panel: position, relative $L_2$ residual of $g$, and the mass fractions of the hot, intermediate, and beam populations. In the BGK solution the hot population at $n^*=0.1$ is a separate group of upstream-moving molecules and the intermediate population holds almost half of the mass; in DSMC the hot population is centred near the downstream velocity and the intermediate one is small (Secs.~\ref{sec:models} and \ref{sec:dsmc}).}
\label{fig:decomp}
\end{figure}

\begin{figure}[!htbp]
\includegraphics[width=0.92\textwidth]{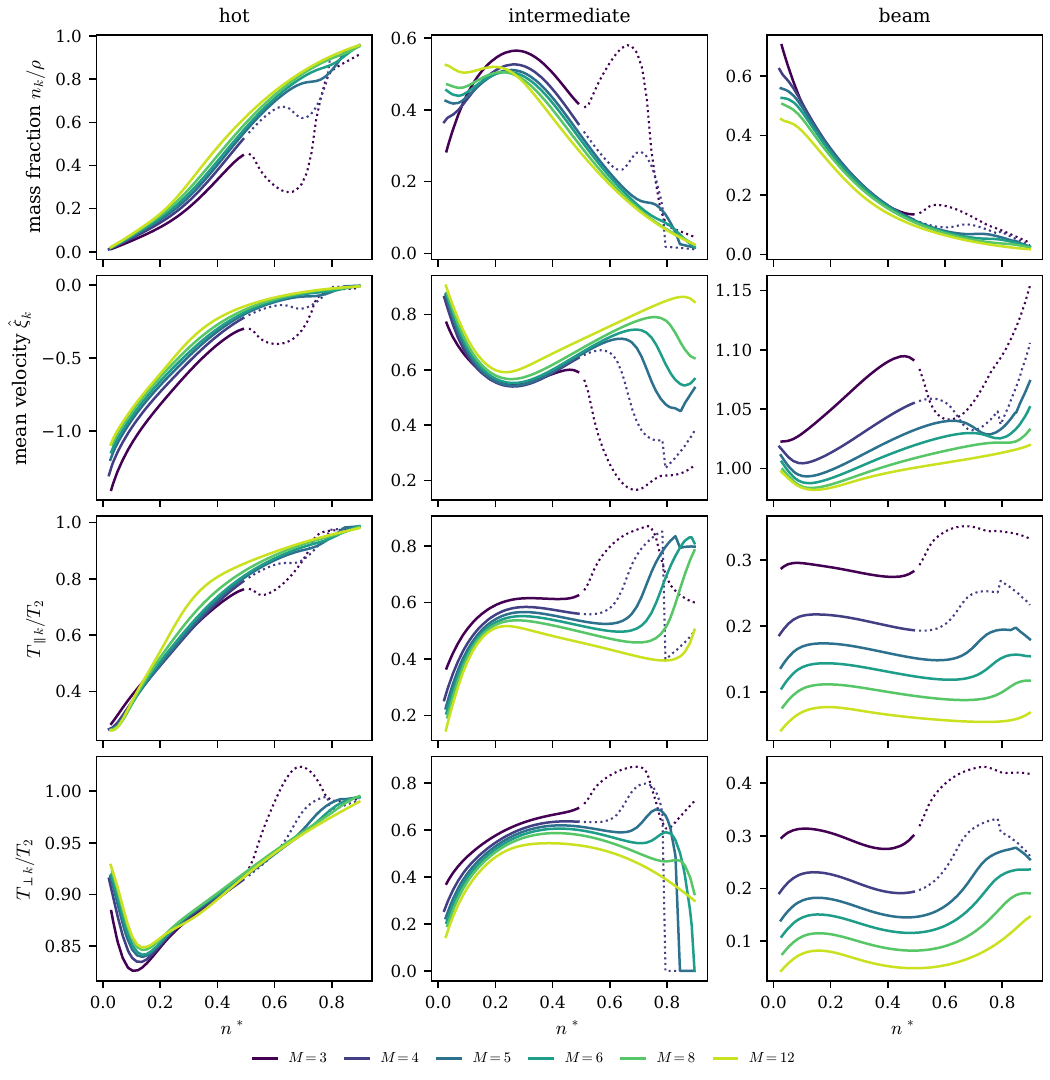}
\caption{Parameters of the three populations (continuous branch) for the BGK solutions. The curves for Mach 3 and 4 are dotted downstream of $n^*=0.5$, where the populations overlap and the decomposition is not unique.}
\label{fig:params}
\end{figure}

Figure~\ref{fig:decomp} shows the three populations of figure~\ref{fig:overview} on the distribution itself at three positions of the Mach-8 shock, for the BGK model and for DSMC, and figure~\ref{fig:params} their parameters through the shock for Mach 3 to 12. Ordered by mean velocity at the upstream end, they are as follows.

The \emph{hot population} has a negative mean velocity far upstream ($\xih_k$ between
$-1.4$ and $-1.1$ at $n^*=0.02$), which rises to the downstream bulk velocity as its mass fraction
grows from zero to one. Its transverse temperature stays between $0.83\,T_2$ and $T_2$ throughout, while
its streamwise temperature is as low as $0.27\,T_2$ at the upstream end. For a relaxation model the
integral form of the kinetic equation identifies it (Sec.~\ref{sec:prec}): it consists of molecules
emitted in the hot part of the shock, of which only those with large upstream-directed velocity reach a
given upstream position, which selects a narrow range of streamwise velocities without affecting
the transverse ones (Sec.~\ref{sec:prec}).

The \emph{intermediate population} has a mean velocity between the two bulk velocities,
$\xih_k\approx0.55$--0.9, and temperatures of 0.15--0.6\,$T_2$. It carries 40--55\% of the mass on
the upstream side and disappears downstream of $n^*\approx0.8$. For the BGK model we interpret it as
molecules emitted forward by the partly relaxed gas inside the shock. A third population between the two
streams was added to the Mott-Smith description by Vilquin, Boudet and Kellay~\citeyearpar{Vilquin2016}, who
attributed it to particles that have undergone a single collision between the supersonic and the subsonic
streams. In their description the two streams keep fixed shapes and the distribution of the third
population is computed from single binary collisions; here all three populations are fitted at every
position, which shows how their parameters change through the shock and with Mach number. Whether the
fitted intermediate population of a gas is a set of molecules that have collided once is a separate
question, which Sec.~\ref{sec:meaning} answers in the negative.

The \emph{beam} has a mean velocity within $0.04\,(u_1-u_2)$ of $u_1$ at Mach 5 and above
(within $0.10\,(u_1-u_2)$ at Mach 3) for $0.05<n^*<0.6$. Its streamwise temperature is
1.0--1.25\,$T_1$ at Mach 3, 1.3--1.5\,$T_1$ at Mach 5 and 2.5--3.6\,$T_1$ at Mach 12, so that at
high Mach number it is broader than the upstream Maxwellian; it represents the unscattered
molecules together with those scattered close to the beam.

For Mach 5 and above the mass fractions, the scaled mean velocities, and the temperatures of the
hot and intermediate populations in units of $T_2$ depend only weakly on Mach number.
At $n^*=0.1$, between Mach 5 and 12 the hot population has a mass fraction of 0.07--0.08,
$\xih_k=-0.93$ to $-0.85$, $T_{\parallel k}/T_2=0.35$--0.36 and $T_{\perp k}/T_2=0.85$--0.86; the
intermediate population has a mass fraction of 0.44--0.50 and $\xih_k=0.67$--0.72. We interpret this
as the approach to the strong-shock limit of Grad~\citeyearpar{Grad1969}, in which $T_1$ drops out of the
problem. At Mach 3 and 4 the same three populations are present on the upstream half, with
parameters that differ by 10--30\% from the strong-shock values, but downstream of $n^*\approx0.5$
they overlap and the decomposition is no longer unique.

\subsection{Transport carried by each population}

\begin{figure}[!htbp]
\includegraphics[width=\textwidth]{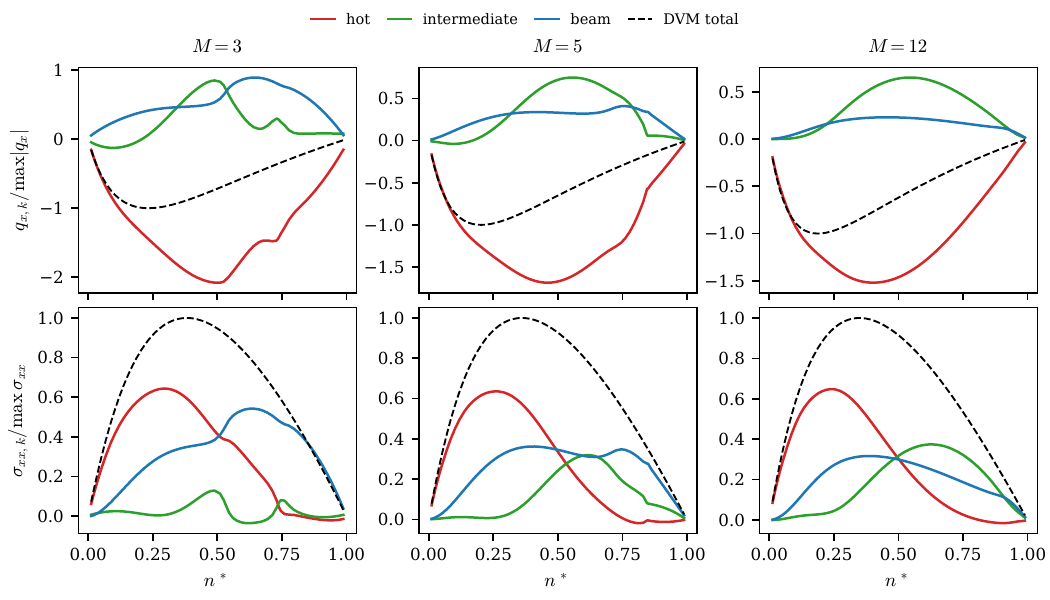}
\caption{Contributions of the three populations to the heat flux (top) and the normal stress
(bottom), Eq.~\eqref{eq:contrib}, normalised by the peak of the total. Dashed lines: totals from
the kinetic solutions.}
\label{fig:transport}
\end{figure}

Figure~\ref{fig:transport} shows the contributions of the three populations to the heat flux and
the normal stress, from Eq.~\eqref{eq:contrib}. The heat flux is the sum of a large negative
contribution of the hot population, 1.5--2 times the total at its peak, and positive contributions
of the intermediate population and the beam, which move downstream relative to the local mean
velocity. The upstream part of the heat-flux profile ($n^*<0.1$) is carried almost entirely by the
hot population even though its mass fraction there is below 8\%. The stress on the upstream side
is likewise dominated by the hot population (about 0.65 of the peak stress at $n^*\approx0.25$),
while the beam and, at high Mach number, the intermediate population dominate downstream. This is
why a description that omits the hot population as a separate entity on the upstream side loses a
large part of the heat flux and the fourth-order moment even where its error in $g$ is modest.

\section{Dependence on the collision model}
\label{sec:models}

\begin{figure}[!htbp]
\includegraphics[width=\textwidth]{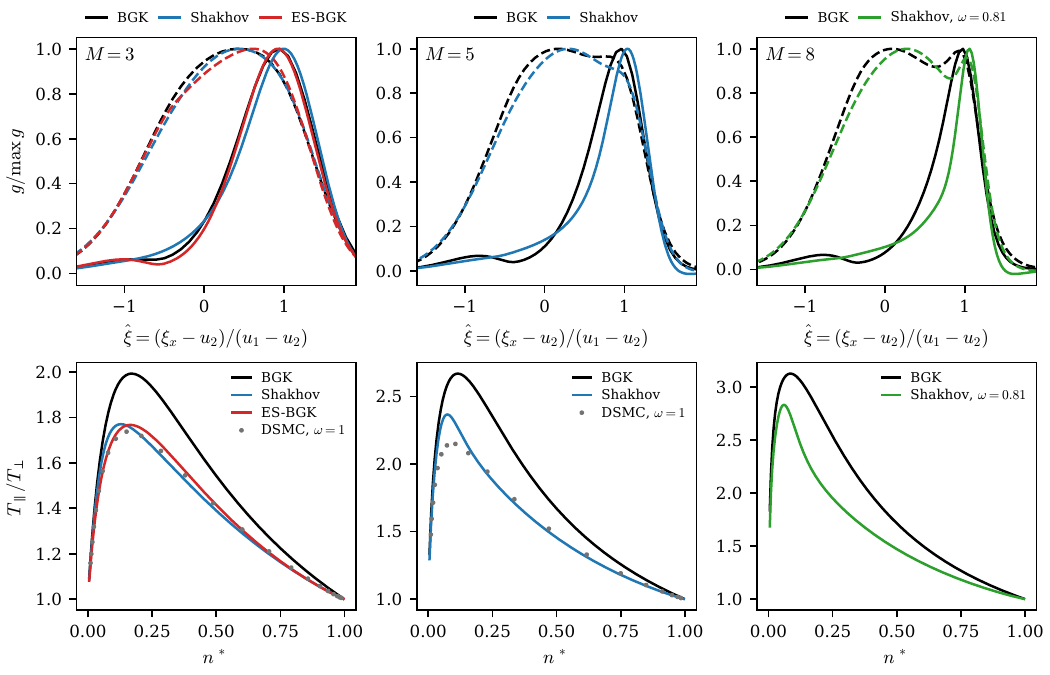}
\caption{Dependence on the collision model. Top: $g$ at $n^*=0.1$ (solid) and 0.5 (dashed).
Bottom: temperature anisotropy; symbols are DSMC results for a variable-hard-sphere gas with
$\omega=1$.}
\label{fig:models}
\end{figure}

\begin{table}
\centering\small
\caption{Comparison of collision models. Residuals are the largest relative $L_2$ residuals of $g$
over $0.01<n^*<0.99$, in percent, for the RH pair and for two and three populations. The last
four columns give the mass fraction and scaled mean velocity of the hot population at $n^*=0.1$
and the upstream-moving mass fraction $\beta$ at $n^*=0.05$ and 0.2.}
\label{tab:models}
\begin{adjustbox}{max width=\linewidth}
\begin{tabular}{llcccccccc}
\toprule
Model & $M$ & $(\Tpar/\Tperp)_{\max}$ & RH pair & $K=2$ & $K=3$ & $n_{\rm hot}/\rho$ & $\xih_{\rm hot}$ & $\beta_{0.05}$ & $\beta_{0.2}$\\
\midrule
BGK & 2 & 1.46 & 2.8 & 0.9 & 0.25 & 0.03 & $-1.40$ & 0.020 & 0.068\\
Shakhov & 2 & 1.37 & 1.6 & 1.3 & 0.21 & 0.07 & $-0.70$ & 0.024 & 0.075\\
ES-BGK & 2 & 1.37 & 3.4 & 1.2 & 0.26 & 0.03 & $-1.51$ & 0.022 & 0.075\\
BGK & 3 & 1.99 & 11.6 & 3.6 & 0.65 & 0.05 & $-1.09$ & 0.026 & 0.107\\
Shakhov & 3 & 1.77 & 7.8 & 4.7 & 1.26 & 0.11 & $-0.50$ & 0.026 & 0.098\\
ES-BGK & 3 & 1.77 & 13.2 & 4.4 & 0.78 & 0.05 & $-1.13$ & 0.026 & 0.099\\
BGK & 5 & 2.67 & 26.9 & 9.1 & 1.22 & 0.07 & $-0.93$ & 0.034 & 0.127\\
Shakhov & 5 & 2.37 & 20.4 & 10.4 & 4.20 & 0.16 & $-0.34$ & 0.032 & 0.118\\
BGK & 8 & 3.13 & 40.3 & 15.6 & 2.25 & 0.08 & $-0.87$ & 0.038 & 0.135\\
Shakhov, $\omega=0.81$ & 8 & 2.83 & 32.7 & 15.5 & 7.13 & 0.19 & $-0.25$ & 0.035 & 0.126\\
\bottomrule
\end{tabular}
\end{adjustbox}
\end{table}

The BGK model with constant relaxation time has a Prandtl number of unity and a viscosity
proportional to pressure, neither of which is realistic. To see which of the results above depend
on these choices, figure~\ref{fig:models} and table~\ref{tab:models} compare the BGK solutions with
the Shakhov and ES-BGK solutions ($\mathrm{Pr}=2/3$, $\mu\propto T$) at the same Mach numbers,
with the Mach-8 Shakhov solution for $\omega=0.81$, and with DSMC moment profiles for a
variable-hard-sphere gas with $\omega=1$ at Mach 3 and 5~\citep{RoohiJFM}. The DSMC data contain
moments only.

The following results are common to the three relaxation models. The two-Maxwellian residual, the
need for a third population on the upstream side, and the growth of both with Mach number are the
same within a factor of about two. The upstream-moving mass fraction at a given $n^*$ agrees between
models to within 10\% at Mach 3, 5 and 8 (0.026 at $n^*=0.05$ and 0.10--0.11 at $n^*=0.2$ for all
three at Mach 3) and to within 20\% at Mach 2; for the BGK and Shakhov solutions its decay follows Eq.~\eqref{eq:stretched} (Sec.~\ref{sec:prec}). The central bimodality appears at Mach 5 in
both the BGK and the Shakhov solutions.

The following results depend on the model. The ES-BGK solutions have populations almost identical
to those of the BGK solutions at the same Mach number, with a separate upstream maximum of $g$. In
the Shakhov solutions the upstream-moving molecules do not form a separate maximum at $n^*=0.1$:
the hot population obtained from the fit is two to three times more massive and slower, and the
three-population residual is larger (4.2\% against 1.2\% at Mach 5). The Shakhov distribution is
also slightly negative at the high-speed edge of the beam, a known property of its
target~\citep{Shakhov1968}; the largest negative mass fraction is 0.6\% at Mach 5 and 2.1\% at Mach
8. The temperature anisotropy differs most clearly: the peak of $\Tpar/\Tperp$ at Mach 3 is 1.99
for the BGK solution, 1.77 for both the Shakhov and ES-BGK solutions and 1.73 for DSMC; at
Mach 5 the values are 2.67 (BGK), 2.37 (Shakhov) and 2.15 (DSMC). The constant-relaxation-time BGK
model thus overestimates the peak anisotropy by 15\% and 24\%, while the Shakhov model with
$\mu\propto T$ is within 2\% of DSMC at Mach 3 and 10\% at Mach 5.

These comparisons involve relaxation models and moments of earlier DSMC solutions. Section~\ref{sec:dsmc} extends the analysis to the velocity distributions sampled in the DSMC of
Sec.~\ref{sec:dsmcsim}.

\section{Velocity distributions from DSMC: gas and model}
\label{sec:dsmc}

\subsection{What is common to DSMC and the relaxation models}

\begin{figure}[!htbp]
\includegraphics[width=\textwidth]{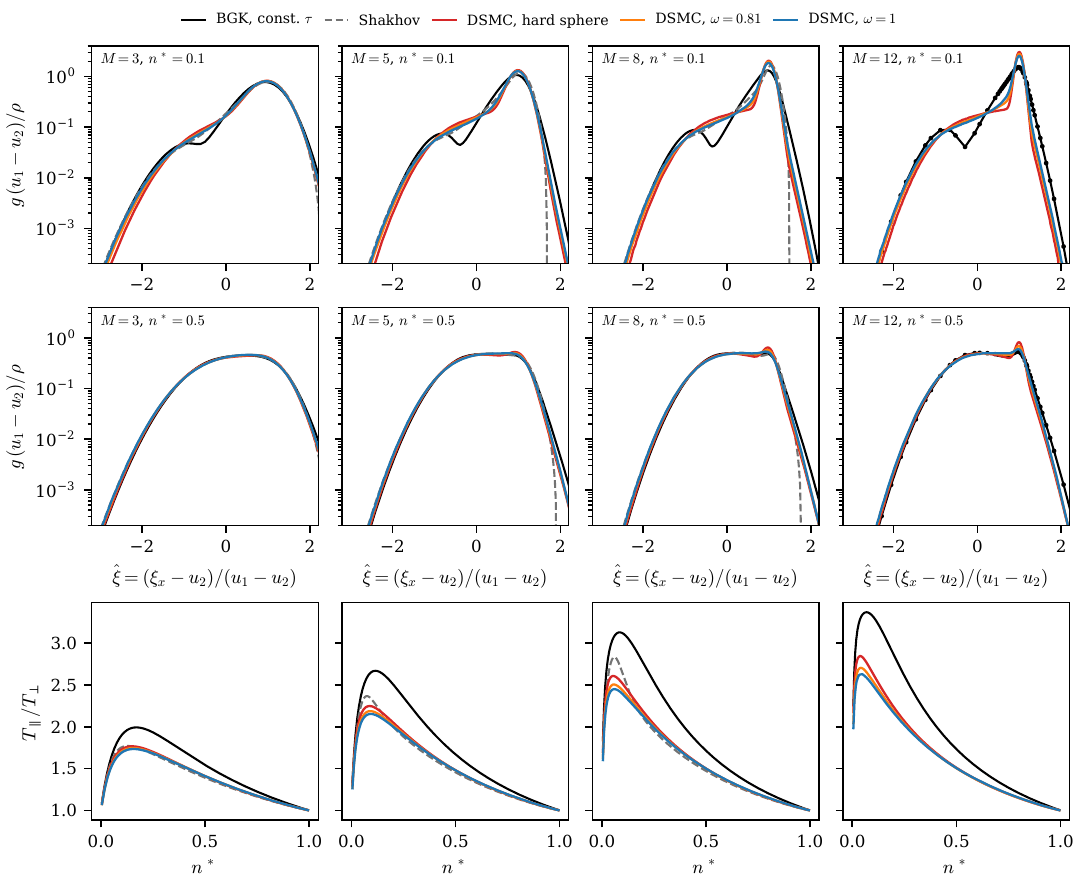}
\caption{DSMC distributions for three viscosity exponents compared with the BGK and Shakhov
solutions at Mach 3, 5, 8 and 12 (the Shakhov solution at Mach 8 has $\omega=0.81$; none is available at Mach 12). Top and middle: scaled streamwise
distribution at $n^*=0.1$ and 0.5. Bottom: temperature anisotropy. Symbols on the Mach-12 BGK
curves mark the velocity nodes of that solution.}
\label{fig:dsmc}
\end{figure}

\begin{table}
\centering\small
\caption{DSMC results for a VHS gas. Residuals are the largest relative $L_2$ residuals of $g$
over $0.02<n^*<0.98$, in percent; the statistical noise contributes 0.03--0.05\%. $q_x$ and
$M_{400}^{\rm neq}$ are the largest errors of the three-population fit relative to the peak of
each moment, in percent.}
\label{tab:dsmc}
\begin{adjustbox}{max width=\linewidth}
\begin{tabular}{llccccccccc}
\toprule
$\omega$ & $M$ & $\delta/\ell_1$ & $(\Tpar/\Tperp)_{\max}$ & RH pair & $K=2$ & $K=3$ & $q_x$ & $M_{400}^{\rm neq}$ & $\beta_{0.05}$ & $\beta_{0.2}$\\
\midrule
0.5 & 2 & 5.1 & 1.38 & 1.0 & 0.4 & 0.14 & 1.1 & 1.3 & 0.027 & 0.081\\
0.5 & 3 & 3.7 & 1.77 & 2.6 & 1.3 & 0.17 & 1.0 & 2.9 & 0.032 & 0.103\\
0.5 & 4 & 3.3 & 2.04 & 6.0 & 2.4 & 0.41 & 1.3 & 3.0 & 0.037 & 0.116\\
0.5 & 5 & 3.2 & 2.25 & 8.6 & 3.5 & 0.62 & 1.8 & 4.2 & 0.041 & 0.127\\
0.5 & 6 & 3.1 & 2.40 & 10.4 & 4.4 & 0.80 & 2.3 & 5.2 & 0.043 & 0.131\\
0.5 & 8 & 3.0 & 2.61 & 12.6 & 5.8 & 1.12 & 3.1 & 6.8 & 0.044 & 0.133\\
0.5 & 10 & 3.0 & 2.75 & 13.8 & 6.6 & 1.40 & 3.6 & 7.9 & 0.045 & 0.136\\
0.5 & 12 & 3.0 & 2.85 & 14.5 & 7.2 & 1.64 & 4.1 & 8.8 & 0.046 & 0.136\\
0.5 & 25 & 2.9 & 3.11 & 15.7 & 8.3 & 2.36 & 5.4 & 11.3 & 0.047 & 0.140\\
0.81 & 2 & 5.9 & 1.37 & 0.8 & 0.6 & 0.09 & 0.6 & 1.9 & 0.026 & 0.079\\
0.81 & 3 & 5.0 & 1.75 & 4.9 & 2.0 & 0.17 & 0.7 & 1.4 & 0.031 & 0.101\\
0.81 & 4 & 5.1 & 2.00 & 9.6 & 3.6 & 0.40 & 1.5 & 3.2 & 0.036 & 0.114\\
0.81 & 5 & 5.4 & 2.19 & 13.0 & 5.0 & 0.65 & 2.2 & 4.6 & 0.040 & 0.125\\
0.81 & 6 & 5.8 & 2.32 & 15.6 & 6.3 & 0.95 & 2.7 & 5.9 & 0.042 & 0.129\\
0.81 & 8 & 6.7 & 2.51 & 19.0 & 8.4 & 1.46 & 3.7 & 7.8 & 0.043 & 0.130\\
0.81 & 10 & 7.5 & 2.62 & 21.1 & 9.8 & 1.96 & 4.6 & 9.6 & 0.044 & 0.134\\
0.81 & 12 & 8.3 & 2.71 & 22.5 & 10.9 & 2.39 & 5.3 & 10.8 & 0.044 & 0.135\\
0.81 & 25 & 12.8 & 2.91 & 25.7 & 14.0 & 3.92 & 8.2 & 16.4 & 0.046 & 0.138\\
1 & 2 & 6.5 & 1.37 & 1.3 & 0.7 & 0.11 & 1.0 & 2.2 & 0.025 & 0.079\\
1 & 3 & 6.0 & 1.73 & 6.5 & 2.4 & 0.19 & 0.9 & 1.8 & 0.030 & 0.100\\
1 & 4 & 6.7 & 1.98 & 11.9 & 4.1 & 0.44 & 1.6 & 3.2 & 0.035 & 0.113\\
1 & 5 & 7.6 & 2.15 & 16.2 & 5.6 & 0.75 & 2.3 & 4.6 & 0.039 & 0.123\\
1 & 6 & 8.7 & 2.28 & 19.5 & 7.0 & 1.09 & 2.9 & 5.9 & 0.041 & 0.128\\
1 & 8 & 11.0 & 2.45 & 24.1 & 9.3 & 1.76 & 4.0 & 8.0 & 0.042 & 0.129\\
1 & 10 & 13.5 & 2.56 & 27.3 & 11.1 & 2.42 & 5.1 & 10.0 & 0.043 & 0.132\\
1 & 12 & 15.9 & 2.63 & 29.6 & 12.6 & 3.08 & 6.0 & 11.9 & 0.043 & 0.133\\
1 & 25 & 32.3 & 2.81 & 37.4 & 18.2 & 5.75 & 10.3 & 20.3 & 0.045 & 0.136\\
\bottomrule
\end{tabular}
\end{adjustbox}
\end{table}

Figure~\ref{fig:dsmc} compares the DSMC distributions with the BGK and Shakhov solutions, and table~\ref{tab:dsmc} lists the corresponding quantities for DSMC.

At the shock centre ($n^*=0.5$) the distributions of all models and of DSMC for the three values
of $\omega$ coincide within the line width over three decades, apart from the high-speed edge of
the beam. The amount of upstream-moving mass is also common: for Mach 3--12, $\beta=0.030$--0.046 at
$n^*=0.05$ and 0.10--0.14 at $n^*=0.2$ in DSMC, against 0.026--0.038 and 0.10--0.14 for the
relaxation models.

The three-population description holds for DSMC as well. Two free populations leave residuals of
1.3--12.6\% for Mach 3--12, with errors of 6--14\% in the heat flux and 9--27\% in the fourth-order
moment; three populations reduce the residual to 0.2--3.1\% and the moment errors to 0.7--6\% and
1.4--12\%. The parameters of the three populations change little above Mach 5 (mass
fractions and scaled mean velocities by at most 0.06 between Mach 5 and 8 and by at most 0.05
between Mach 8 and 12), as for the BGK solutions.

\subsection{What depends on the collision law}

\emph{Size of the departure from two Maxwellians.} The residual of the best combination of the RH
Maxwellians is 2.6\%, 8.6\% and 12.6\% for hard spheres at Mach 3, 5 and 8; 4.9\%, 13.0\% and
19.0\% for $\omega=0.81$; and 6.5\%, 16.2\% and 24.1\% for $\omega=1$. At Mach 12 it reaches 14.5\%,
22.5\% and 29.6\%, against 50.4\% for the BGK model. At Mach 25 it is 15.7\%, 25.7\% and 37.4\%, and the residual of three populations is 2.4\%, 3.9\% and 5.8\% (table~\ref{tab:dsmc}). It grows with the softness of
the potential, and its maximum lies at $n^*\approx0.26$--0.37 rather than at 0.1. The Shakhov
solutions with $\mu\propto T$ (7.8\% and 20.4\% at Mach 3 and 5) are within 20--26\% of DSMC for
$\omega=1$, whereas the BGK model with constant relaxation time (11.6\%, 26.9\% and 40.3\%)
overestimates the departure by a factor of 1.7--1.8 relative to $\omega=1$ and of 3--4.4 relative to
hard spheres. The residual of 2.6\% for hard spheres at Mach 3 is consistent with the solution of
Ohwada~\citeyearpar{Ohwada1993}, in which the Mott-Smith profiles of density, velocity and temperature
agree well with the Boltzmann solution at Mach 2 and 3 while the distribution function at Mach 3
shows a distinct difference; Mach 3 was the largest Mach number of that study.

\emph{Identity of the populations on the upstream side.} In DSMC the distribution at $n^*=0.1$
has no separate maximum at $\xi_x<0$; the upstream-moving molecules form a smooth shoulder
(figure~\ref{fig:dsmc}, top row). At Mach 5--12 the three populations are: the beam, with 57--69\% of
the mass, $\xih_k=1.00$ and the upstream temperature; a hot population with 19--23\% of the mass,
a mean velocity close to the downstream one ($\xih_k\approx-0.1$) and temperatures of
1.05--1.35\,$T_2$; and an intermediate population with 10--22\% of the mass at
$\xih_k\approx0.6$--0.7, $T_{\parallel k}\approx0.3$--0.4\,$T_2$ and $T_{\perp k}\approx0.4$--0.8\,$T_2$. As a
fit, the distribution of the Boltzmann gas is therefore the two Mott-Smith Maxwellians, the hot one being
somewhat hotter than the downstream gas, plus one intermediate population; Sec.~\ref{sec:meaning} examines
what these populations are. In the BGK solutions the hot population at the same position is a narrow group with $\xih_k\approx-0.9$ and
$T_{\parallel k}\approx0.35\,T_2$ holding 7--8\% of the mass, and the intermediate population holds
almost half of the mass (Sec.~\ref{sec:pop}); the ES-BGK solutions, available at Mach 2 and 3, have the same structure (table~\ref{tab:models}). The separate upstream maximum of these models is
thus associated with relaxation, at a rate independent of the molecular velocity, towards a Gaussian target (a Maxwellian for the BGK model, an anisotropic Gaussian for the ES-BGK model), and is not a property of the gas. A related difference was reported by Yuan and
Wu~\citeyearpar{YuanWu2022}, who compared the Shakhov model with the Boltzmann equation for a Mach-5 shock
in argon with the Lennard-Jones potential. The number of molecules moving against the flow is
nearly the same in the two solutions (less than 3.8\% of the density at their positions, as for
$\beta$ here), but far upstream these molecules carry a much larger share of the thermal energy in
the Shakhov solution, with an additional peak in the energy-weighted distribution; they attributed
this to a collision frequency that does not depend on the molecular velocity. For the Boltzmann gas, the part of the distribution outside the beam has two maxima
inside a hard-sphere shock of infinite Mach number, one next to the beam velocity and a central
one~\citep{Takata2000}; the intermediate and the hot population are their counterparts in the
present description.

\emph{Temperature anisotropy.} The peak of $\Tpar/\Tperp$ in DSMC is 1.77, 2.25 and 2.61 for hard
spheres at Mach 3, 5 and 8, and 1.73, 2.15 and 2.45 for $\omega=1$; at Mach 12 it is 2.85 and 2.63.
It depends on $\omega$ by at most 8\%. The BGK model with constant relaxation time (3.37 at Mach 12)
overestimates it by 13--28\%, the Shakhov model
by 2\% and 10\% at Mach 3 and 5 ($\omega=1$) and by 13\% at Mach 8 ($\omega=0.81$). Because
$T_\parallel$ is the same function of the density in all models [Eq.~\eqref{eq:exact}], these
differences are differences in the transverse temperature.

\emph{Precursor.} The decay of the upstream-moving mass ahead of the shock also depends on the collision law, and
in a way that the relaxation models cannot represent; it is treated in Sec.~\ref{sec:prec}.

\subsection{Stability and physical content of the three populations}
\label{sec:meaning}

\emph{Stability of the fit.} Three tests were made at five positions ($n^*=0.1$, 0.2, 0.35, 0.5, 0.7) of the
Mach-8 and Mach-12 simulations for the three values of $\omega$. First, the eight runs were divided into two
independent halves and each half was fitted separately. The mass fractions of the populations differ between
the halves by at most 0.002, the scaled mean velocities by 0.006, the temperatures by
0.008\,$T_2$, and the contributions of the populations to the heat flux and the stress by
0.005 and 0.007 of the peak of each. Second, the fit was restarted thirty times from
strongly perturbed parameters (factors of about 1.6 in the densities and temperatures, $\pm0.25\,(u_1-u_2)$
in the velocities). Between 20 and 29 of the thirty starts return the same solution, no start
finds a lower cost, and the remaining starts end in other local minima, all with larger residuals than
the 0.4--3.0\% of the solution reported.
Third, the weight of the residual of $h$ relative to that of $g$ was changed by factors of 1/4 and 4; the mass
fractions change by at most 0.03, the scaled velocities by 0.05, the temperatures by
0.05\,$T_2$, and the shares of the heat flux and the stress by 0.06 and
0.05 of the peak. The three-population description is therefore a well-defined and reproducible
property of the distributions, and so is the attribution of the transport: at these positions the hot
population contributes between $-1.3$ and $-0.8$ times the peak heat flux, and the other two
contribute with the opposite sign.

\emph{Collision history.} A stable term of a fit is not thereby a set of molecules with a common history. In a
second series of DSMC runs every simulated molecule carries a record of its collisions
(Appendix~\ref{app:labels}). With it, the molecules of the hard-sphere gas can be
divided into those of the incoming stream that have not yet collided with a scattered molecule, and the
rest, classified by how recently they or their collision partners belonged to the stream. The beam of figure~\ref{fig:overview}
is the unscattered stream: its mass fraction exceeds that of the unscattered molecules by at most
0.054 at Mach 8 and 0.04 at Mach 12, the excess being scattered
molecules that lie under the peak. The other two populations are not collision classes. The stream molecules
that have collided exactly once hold 0.14, 0.17, 0.16, and 0.12 of the mass at $n^*=0.1$, 0.2,
0.35, and 0.5 (Mach 8); their mean velocity lies between the two streams ($\xih=0.38$--0.49), but
their streamwise temperature is 0.73--0.94\,$T_2$, far larger than that of the intermediate
population (0.25--0.37\,$T_2$), and the fit assigns most of them to the hot population. At $n^*=0.2$ the hot
population holds 41\% of the mass, whereas only 6\% of the molecules have had no recent
contact with the stream and 54\% are one or two collisions away from it. The intermediate
population, with 7--14\% of the mass (figure~\ref{fig:overview}a), is narrower than any collision class; it is the correction that turns the
sum of two Gaussians into the skewed distribution of the scattered molecules. For softer molecules the
correspondence weakens: the beam exceeds the unscattered stream by up to
0.15 of the mass for $\omega=0.81$ at Mach 8, and for Maxwell-type molecules, whose
collisions at low relative speed are as frequent as any other, the stream is eroded gradually and no
population can be identified with it. The hot and the intermediate population are therefore effective
populations only. In particular, the intermediate population of a fit is not the set of molecules that
have collided once, and the weights of a three-population fit of a measured distribution do not measure how
many molecules have collided how often.

\begin{figure}[!htbp]
\includegraphics[width=\textwidth]{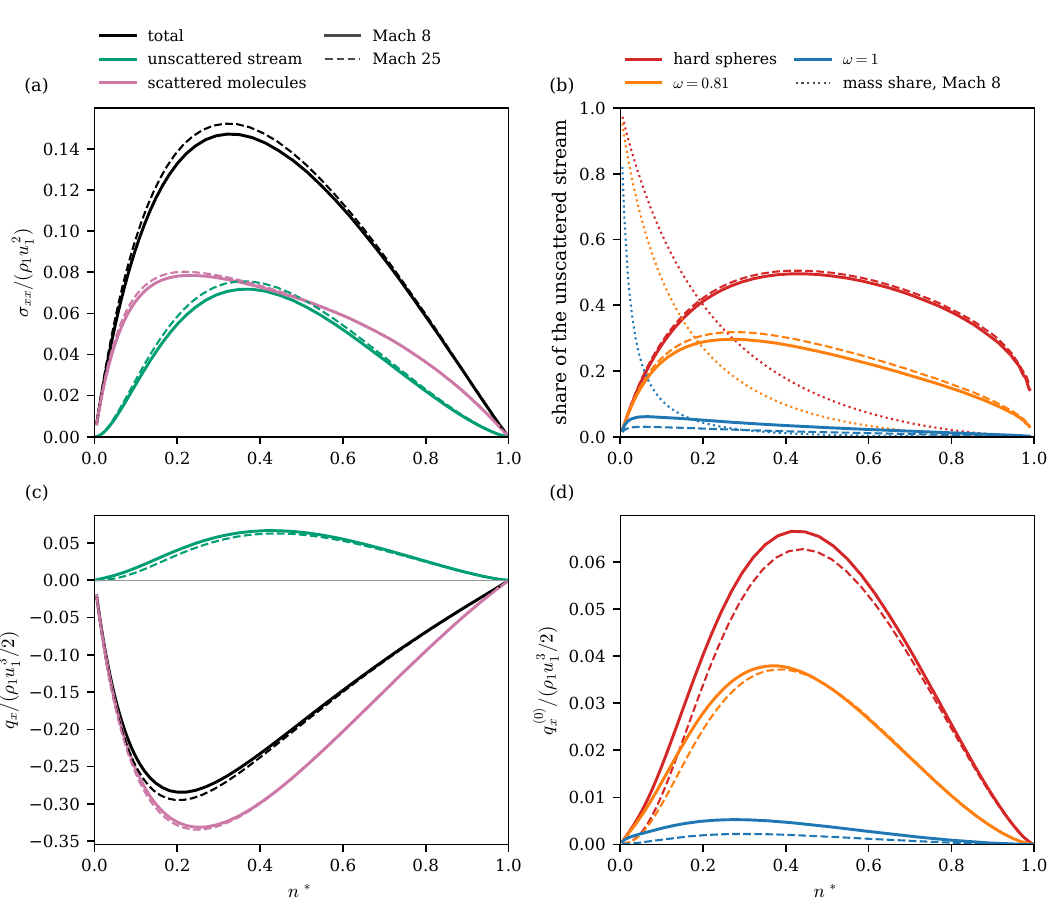}
\caption{Viscous stress and heat flux of the DSMC shocks divided by collision history. (a) Hard spheres at Mach 8 (solid) and Mach 25 (dashed): stress carried by the unscattered stream, stress carried by the scattered molecules, and their sum. (b) Share of the stress carried by the unscattered stream for the three collision laws at Mach 8 (solid) and Mach 25 (dashed), and the share of the mass held by the unscattered stream at Mach 8 (dotted). (c) Heat flux of the hard-sphere gas divided in the same way, with the lines of (a). (d) Heat flux carried by the unscattered stream for the three collision laws at Mach 8 (solid) and Mach 25 (dashed), with the colours of (b).}
\label{fig:split}
\end{figure}

\emph{Stress carried by the unscattered stream.} The collision record divides the distribution into the part
carried by the unscattered stream and the part carried by the scattered molecules, $f=f^{(0)}+f^{(s)}$. This is
the division into a ballistic and a collision part that Stefanov and co-workers introduced for DSMC of confined
flows, where the ballistic part consists of the molecules that have not collided since they left a
wall~\citep{Vargas2014,Tatsios2015,AkhlaghiRoohiStefanov2018}; here the incoming stream takes the place of the
wall, and collisions among its molecules, which leave it in equilibrium, are not counted. Every moment taken about
the mean velocity $u$ of the whole gas is the sum of the contributions of the two parts. For the viscous stress,
\begin{equation}
\sigma_{xx}=\sigma_{xx}^{(0)}+\sigma_{xx}^{(s)},\qquad
\sigma_{xx}^{(k)}=\frac23\int\Bigl[(\xi_x-u)^2-\tfrac12\,(\xi_y^2+\xi_z^2)\Bigr]f^{(k)}\,d\bm\xi .
\label{eq:split}
\end{equation}
Figure~\ref{fig:split} shows the two parts. The largest stress is nearly the same for the three collision laws,
0.146--0.147\,$\rho_1u_1^2$ at Mach 8, but it is carried by different molecules. For hard spheres the unscattered
stream carries 48\% of it with 24\% of the mass, and close to half of the stress over the central part of the
shock; for $\omega=0.81$ it carries 29\% with 14\% of the mass, and for $\omega=1$, 4\% with 2\%. The
molecules of generation 1, whose most recent collision involved a stream molecule, hold a third of the mass of the
hard-sphere gas at this position and carry 3\% of the stress: their velocities are distributed almost
isotropically about a mean close to that of the gas. For hard spheres the division is nearly independent of Mach number: at the position of the largest stress
the unscattered stream carries 47\%, 48\%, 48\%, and 49\% of it at Mach 5, 8, 12, and 25, and 44\% at Mach 3. For
$\omega=0.81$ its share rises slowly, 26\%, 29\%, 30\%, and 32\%; for $\omega=1$ it is a few percent throughout
and depends on the length of the upstream domain (Appendix~\ref{app:labels}). Around its maximum the stress of a hard-sphere shock is thus carried in nearly equal parts by molecules that
have not been scattered and by the scattered gas (over the whole shock the unscattered stream carries 35--37\% of
the integrated stress), whereas for Maxwell-type molecules it is carried by scattered molecules throughout: the
same largest stress arises from different sets of molecules. This is the counterpart, in the transport, of the result above that the beam is a set of molecules
with a common history only for hard spheres.

The heat flux divides differently (figure~\ref{fig:split}(c) and (d)). Its two parts,
$q_x^{(k)}=\tfrac12\int(\xi_x-u)\,|\bm\xi-u\bm e_x|^2f^{(k)}\,d\bm\xi$, were obtained from a repetition of the
collision-record runs in which the third-order sums are accumulated for every class. The part carried by the
unscattered stream is positive: the stream moves faster than the gas as a whole and carries its energy
downstream, against the net heat flux. The heat flux towards the upstream side is therefore carried entirely by
the scattered molecules and exceeds the net one. For hard spheres at Mach 8, where the net heat flux is largest
($n^*=0.20$) its value of $-0.284$ in units of $\rho_1u_1^3/2$ is the sum of $+0.041$ from the stream and
$-0.325$ from the scattered molecules, and the contribution of the stream reaches $+0.067$ at $n^*=0.41$, nearly a
quarter of the largest net heat flux. Its maximum is $+0.038$ for $\omega=0.81$ and $+0.005$ for $\omega=1$, and for
hard spheres it changes little with Mach number: $+0.073$, $+0.067$, $+0.064$, and $+0.063$ at Mach 5, 8, 12, and
25. For $\omega=0.81$ the corresponding values are $+0.040$, $+0.038$, $+0.037$, and $+0.037$; for $\omega=1$ the
contribution is $+0.005$ at Mach 5 to 12 and $+0.002$ at Mach 25, following the small share of the stream in this
gas. Two parts of opposite sign whose sum is the net heat flux were also found for the rarefied flow in a
lid-driven cavity by \citet{AkhlaghiRoohiStefanov2018}.

\emph{Beyond the reduced distributions.} The fit uses $g$ and $h$ only. The bi-Maxwellian form of the
populations was tested beyond them with the reduced distribution of the transverse fourth moment,
$h_2(\xi_x)=\int(\xi_y^2+\xi_z^2)^2f\,d\xi_y\,d\xi_z$, which contains information on the transverse velocities that
the fit does not use. It was sampled in additional runs at Mach 5, 8, and 12 (four runs for each $\omega$,
statistically independent of those used for the fits), together with the mixed moment
$\int(\xi_x-u)^2(\xi_y^2+\xi_z^2)f\,d\bm\xi=\int(\xi_x-u)^2h\,d\xi_x$. The mixed moment follows from $h$, so that it
tests the fit on independent data, and $h_2$ tests the transverse shape. A mixture of bi-Maxwellians predicts
$h_2=\sum_k8T_{\perp k}^2g_k$ and, with $d_k=u_k-u$, the mixed moment $\sum_k2n_kT_{\perp k}(T_{\parallel k}+d_k^2)$
without further parameters; in equilibrium the two moments are $8\rho T^2$ and $2\rho T^2$, which are subtracted to
obtain their nonequilibrium parts. With the three populations
fitted to $g$ and $h$, the largest relative $L_2$ residual of $h_2$ over the shock is 6.2--19.4\%
(median 2.8--5.2\%), against 0.6--3.1\% and 0.7--3.6\% for $g$ and $h$ on the same
independent data; with two populations it is 13.1--20.7\%. The largest residuals occur at the foot of
the shock ($n^*<0.05$). The transverse fourth moment $\int h_2\,d\xi_x$ is reproduced to within
1.7--2.9\% of its peak and the mixed moment to within 1.0--2.3\%; relative to
the peaks of their nonequilibrium parts, which are 10--11\% of the totals for the transverse
moment, the errors are 16--29\% and 7--14\%. The
three-population description is thus accurate for the reduced distributions to which it is fitted and for
the moments that follow from them, and less accurate for the transverse structure of the distribution, of
which it gives the fourth moment to a few percent of its peak value but its departure from equilibrium only to
within a fifth to a third. Statements about the accuracy of the distribution in this paper refer to $g$ and
$h$.

\section{Similarity across Mach number}
\label{sec:sim}

Part of the similarity in the scaled variables is exact. Conservation of mass and momentum,
$\rho u=J$ and $\rho u^2+\rho RT_\parallel=P$, fixes the mean velocity and the streamwise temperature
as functions of the density alone, whatever the collision model (an argument attributed to Hicks and Yen by \citet{Cercignani1999}):
\begin{equation}
\frac{u-u_2}{u_1-u_2}=\frac{1-n^*}{1+n^*(r-1)},\qquad RT_\parallel=\frac{P}{\rho}-\frac{J^2}{\rho^2},
\qquad r=\frac{\rho_2}{\rho_1}.
\label{eq:exact}
\end{equation}
The zeroth, first and second velocity moments of $g$ at a given $n^*$ are therefore the same for
all models at the same Mach number. For strong shocks $r\to4$ and $P\to\rho_1u_1^2$, so that they also
become independent of Mach number: $T_\parallel/T_2\to(16/3)(\tilde\rho-1)/\tilde\rho^2$ with
$\tilde\rho=1+3n^*$, which has the maximum $4/3$ at $n^*=1/3$. At Mach 12 the maximum is 1.31--1.32 at
$n^*=0.34$ for the BGK solution and for DSMC alike, and Eq.~\eqref{eq:exact} is satisfied by all
solutions to within 1.5\%. Conservation does not fix the shape of
$g$ beyond its second moment, nor the transverse temperature, the heat flux and the higher
moments. The similarity of the populations concerns these quantities, and it is what is tested
below. For the macroscopic profiles, Erofeev and Fridlender~\citeyearpar{Erofeev2002} found from DSMC of
strong shocks that the stress, the heat flux and the velocity gradient are simple functions of the
reduced velocity $(u_1-u)/(u_1-u_2)$, with coefficients that depend on the Mach number and on the
interaction law.

\emph{Origin of the dependence on $1/M^2$.} The form of the remaining dependence on Mach number follows from
the kinetic equation. Let the velocities be measured from $u_2$ in units of $u_1-u_2$,
$\bm\xi=u_2\bm e_x+(u_1-u_2)\hat{\bm\xi}$, and let $f=\rho_1(u_1-u_2)^{-3}\hat f$. For a collision rate
proportional to a power of the relative speed, $c_r^{2-2\omega}$, as in the variable-hard-sphere gas, the
Boltzmann equation becomes
\begin{equation}
\Bigl(\hat\xi_x+\frac{1}{r-1}\Bigr)\frac{\partial\hat f}{\partial\hat x}=\hat Q(\hat f,\hat f),
\label{eq:scaledB}
\end{equation}
where $\hat Q$ is the collision operator in the scaled velocities and $\hat x$ is the position in a unit
proportional to $(u_1-u_2)^{2\omega-1}/\rho_1$; the relaxation models with a constant collision time or with a
collision frequency proportional to $\rho T^{1-\omega}$ scale analogously. The Mach number enters
Eq.~\eqref{eq:scaledB} only through the density ratio $r$ and through the two end states, which are Maxwellians of
density 1, velocity 1, and temperature $RT_1/(u_1-u_2)^2$ upstream and of density $r$, velocity 0, and temperature
$RT_2/(u_1-u_2)^2$ downstream. With $\varepsilon=1/M^2$ and $\gamma=5/3$ the Rankine--Hugoniot relations give
\begin{equation}
\frac1r=\frac{1+3\varepsilon}{4},\qquad
\frac{RT_2}{(u_1-u_2)^2}=\frac{(5-\varepsilon)(1+3\varepsilon)}{15\,(1-\varepsilon)^2},\qquad
\frac{RT_1}{(u_1-u_2)^2}=\frac{16\,\varepsilon}{15\,(1-\varepsilon)^2}.
\label{eq:rheps}
\end{equation}
All coefficients of the scaled problem are therefore smooth functions of $\varepsilon$, and the problem at
$\varepsilon=0$ is the shock of infinite Mach number, whose upstream state is a beam of zero
width~\citep{Grad1969,Takata2000}. For the scattered molecules, whose scaled velocities are spread over a width of
order one, a first correction linear in $\varepsilon$ is expected, and this is the form used in the test below
and in the predictions of Sec.~\ref{sec:predictor}. The beam behaves differently. Its thermal width in the scaled
velocity is $[RT_1]^{1/2}/(u_1-u_2)\simeq1.03\,\varepsilon^{1/2}$, so that near $\hat\xi_x=1$ the scaled
distribution is a Gaussian whose width vanishes like $\varepsilon^{1/2}$. Its moments are smooth functions of
$\varepsilon$, but its values at a fixed scaled velocity are not, and a linear combination of two such Gaussians
of different widths is not a Gaussian; extrapolated to a smaller $\varepsilon$, it is negative on the flanks of
the beam. This is why the scaled distributions combined directly give the moments accurately but not the shape
of the beam, and why the beam is treated separately, as a Maxwellian of the known upstream state, in the third
method of Sec.~\ref{sec:methods}.

The weak dependence of the scaled population parameters on Mach number suggests that the
distribution at one Mach number can be obtained from those at neighbouring Mach numbers. We test
this by withholding one Mach number at a time. Two predictions are built from the two neighbouring
solutions only, at fixed $n^*$ and by linear interpolation in $1/M^2$: (i) the three-population
parameters (mass fractions, $\xih_k$, and the logarithms of $T_{\parallel k}/T_2$ and
$T_{\perp k}/T_2$) are interpolated and the mixture is evaluated with the RH states of the withheld
Mach number; (ii) the scaled distributions $g\,(u_1-u_2)/\rho$ and $h\,(u_1-u_2)/(\rho T_2)$ are
interpolated directly on a common grid of $\xih$. Nothing from the withheld solution enters either
prediction. The comparison is made on $0.03\le n^*\le0.93$.

\begin{table}
\centering\small
\caption{Leave-one-Mach-out test for the BGK solutions. Largest relative $L_2$ residual of $g$ and
relative $L_2$ errors of the moment profiles over $0.03\le n^*\le0.93$, in percent, for
interpolated populations (P), interpolated scaled distributions (D), and a direct
three-population fit of the withheld solution (F).}
\label{tab:lomo}
\begin{adjustbox}{max width=\linewidth}
\begin{tabular}{llccccccccc}
\toprule
 & & \multicolumn{3}{c}{$g$ (max)} & \multicolumn{2}{c}{$q_x$} & \multicolumn{2}{c}{$\sigma_{xx}$} & \multicolumn{2}{c}{$M_{400}^{\rm neq}$}\\
$M$ & from & P & D & F & P & D & P & D & P & D\\
\midrule
3 & 2.5, 4 & 3.0 & 5.2 & 0.9 & 9.3 & 2.2 & 7.1 & 6.5 & 7.7 & 18.5\\
4 & 3, 5 & 4.3 & 4.2 & 1.2 & 7.4 & 1.1 & 3.4 & 2.8 & 5.3 & 8.5\\
5 & 4, 6 & 2.1 & 2.2 & 1.5 & 4.8 & 1.1 & 2.0 & 2.5 & 5.5 & 4.5\\
6 & 5, 8 & 1.8 & 1.7 & 1.6 & 3.1 & 0.4 & 0.8 & 0.5 & 5.4 & 1.4\\
8 & 6, 12 & 4.1 & 4.4 & 2.3 & 2.7 & 1.7 & 1.0 & 3.2 & 4.9 & 5.2\\
\bottomrule
\end{tabular}
\end{adjustbox}
\end{table}

\begin{figure}[!htbp]
\includegraphics[width=\textwidth]{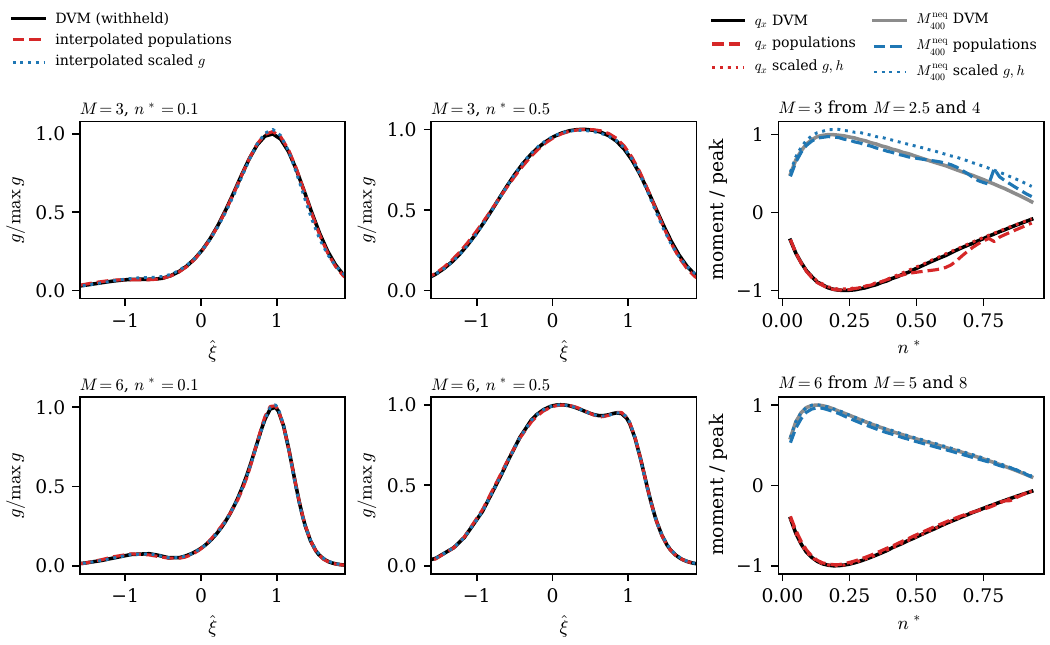}
\caption{Leave-one-Mach-out predictions at Mach 3 (top, from Mach 2.5 and 4) and Mach 6 (bottom,
from Mach 5 and 8): distribution at $n^*=0.1$ and 0.5, and profiles of the heat flux and the
fourth-order moment.}
\label{fig:lomo}
\end{figure}

Table~\ref{tab:lomo} and figure~\ref{fig:lomo} give the results. The largest residual of $g$ is
3--5\% at Mach 3 and 4, about 2\% at Mach 5 and 6, and 4\% at Mach 8, whose neighbours (Mach 6 and 12) are farthest apart (2.4\% when the temperatures are combined linearly, table~\ref{tab:predictor}); at Mach 5 and 6 it is close to the residual of a direct
three-population fit of the withheld solution. The heat flux is predicted to
within 3--9\% by the interpolated populations and to within 0.4--2.2\% by the interpolated scaled
distributions; the fourth-order moment is predicted more accurately by the populations at Mach 3
and 4 and by the scaled distributions at Mach 6. The two procedures therefore perform comparably,
which shows that the similarity lies in the scaled variables $(n^*,\xih)$ and the RH temperature
scale rather than in the Gaussian form. For comparison, the Mach-conditioned 512-kernel model of Sec.~\ref{sec:baseline}, tested on the same data, gave
errors of 11--43\% in the distribution and 23--83\% in the transport moments at withheld Mach 3
and 6 (table~\ref{tab:mach_holdout}).

The test gives the distribution as a function of $n^*$. The density profile $n^*(x)$ of the withheld Mach number is obtained in Sec.~\ref{sec:predictor} from one moment equation, together with its range of validity.

\section{Prediction of a shock from two others}
\label{sec:predictor}

The similarity of Sec.~\ref{sec:sim} gives the distribution at a new Mach number as a function of the normalised density. Two further steps turn it into a prediction of the whole shock structure: extrapolation beyond the Mach numbers used, and a relation between $n^*$ and position. The prediction is first made with the three populations (Secs.~\ref{sec:predpos}--\ref{sec:preddsmc}); Sec.~\ref{sec:methods} then repeats it without them, to separate what the scaled variables contribute from what the Gaussian form contributes.

\subsection{Position and prediction with three populations}
\label{sec:predpos}

\emph{Position.} The moment of Eq.~\eqref{eq:kinetic} with $\xi_x^2$ is $d(\int\xi_x^3g\,d\xi_x)/dx=\nu\,(\int\xi_x^2g_t\,d\xi_x-\int\xi_x^2g\,d\xi_x)$, so that it involves the second moment of the target and not only its density, velocity, and temperature. For the BGK model the target is the local Maxwellian, and for the Shakhov model the correction to the Maxwellian is odd in the peculiar velocity and does not contribute; in both
\begin{equation}
\frac{d}{dx}\int\xi_x^3\,g\,d\xi_x=-\nu\,\sigma_{xx},\qquad \nu=\frac{p}{\mu}.
\label{eq:moment}
\end{equation}
For the ES-BGK model the target is an anisotropic Gaussian whose stress is $(1-1/\mathrm{Pr})$ times that of $f$, the right-hand side is $-(\nu/\mathrm{Pr})\,\sigma_{xx}$, and since the collision frequency of that model is $\nu=\mathrm{Pr}\,p/\mu$ the equation takes the same form $-(p/\mu)\,\sigma_{xx}$ in terms of the viscosity; the predictor is not applied to the ES-BGK solutions here. For the three-population mixture both sides are algebraic functions of the population parameters: $\Phi_2\equiv\int\xi_x^3g\,d\xi_x=\sum_kn_k(u_k^3+3u_kT_{\parallel k})$, and $\sigma_{xx}$ is given by Eq.~\eqref{eq:contrib}. Since the parameters are known as functions of $n^*$,
\begin{equation}
\frac{dx}{dn^*}=\frac{d\Phi_2/dn^*}{-\nu\,\sigma_{xx}},
\label{eq:xofn}
\end{equation}
which is integrated from the shock centre. This is the step taken by Mott-Smith~\citeyearpar{MottSmith1951} for two Maxwellians; here it is applied to three populations whose parameters are taken from kinetic solutions. For the BGK shocks $\nu=1/\tau_{\mathrm{shock}}$. Applied to the DVM solutions themselves, with $\Phi_2$ and $\sigma_{xx}$ computed from the DVM distribution, Eq.~\eqref{eq:xofn} reproduces the positions to within 0.003 of the shock thickness $\delta$, which confirms the equation and its numerical evaluation. With the three-population fit of the same solution the positions are reproduced to within 0.01--0.02\,$\delta$.

\begin{table}
\centering\small
\caption{Predictions of the three-population model for BGK shocks. The population parameters of the source Mach numbers are combined linearly in $1/M^2$, and positions follow from Eq.~\eqref{eq:xofn}. Listed are the largest relative $L_2$ residual of $g$ (in parentheses: direct three-population fit of the target), the relative $L_2$ errors of the heat-flux and stress profiles, the largest temperature error, all in percent over $0.03\le n^*\le0.93$, the ratio of predicted to DVM width $x(n^*{=}0.8)-x(n^*{=}0.2)$, and the root-mean-square deviation of the predicted positions in units of the shock thickness. Temperatures are combined linearly here and logarithmically in table~\ref{tab:lomo}, which accounts for the differences between the two tables. Dashes mark tests without a monotone physical-space profile; only their fixed-$n^*$ distribution and moment errors are meaningful.}
\label{tab:predictor}
\begin{adjustbox}{max width=\linewidth}
\begin{tabular}{lllcccccc}
\toprule
 & $M$ & sources & $g$ (max) & $q_x$ & $\sigma_{xx}$ & $T$ & width ratio & position\\
\midrule
interpolation & 3 & 2.5, 4 & 2.3 (0.9) & 6.9 & 3.5 & 2.1 & 1.04 & 0.071\\
 & 4 & 3, 5 & 4.0 (1.1) & 6.2 & 3.3 & 2.4 & -- & --\\
 & 5 & 4, 6 & 1.9 (1.4) & 3.6 & 1.5 & 2.4 & 1.18 & 0.051\\
 & 6 & 5, 8 & 1.6 (1.6) & 3.3 & 0.7 & 2.2 & 0.98 & 0.014\\
 & 8 & 6, 12 & 2.4 (2.3) & 3.0 & 1.0 & 3.7 & 0.97 & 0.012\\
extrapolation & 8 & 5, 6 & 2.3 (2.3) & 5.6 & 1.8 & 3.2 & -- & --\\
 & 12 & 6, 8 & 4.7 (3.8) & 4.3 & 0.9 & 1.9 & 1.01 & 0.021\\
 & 12 & 5, 8 & 4.7 (3.8) & 5.0 & 1.3 & 2.0 & -- & --\\
 & 12 & 5, 6 & 5.2 (3.8) & 9.5 & 4.5 & 5.2 & -- & --\\
\bottomrule
\end{tabular}
\end{adjustbox}
\end{table}

\begin{figure}[!htbp]
\includegraphics[width=\textwidth]{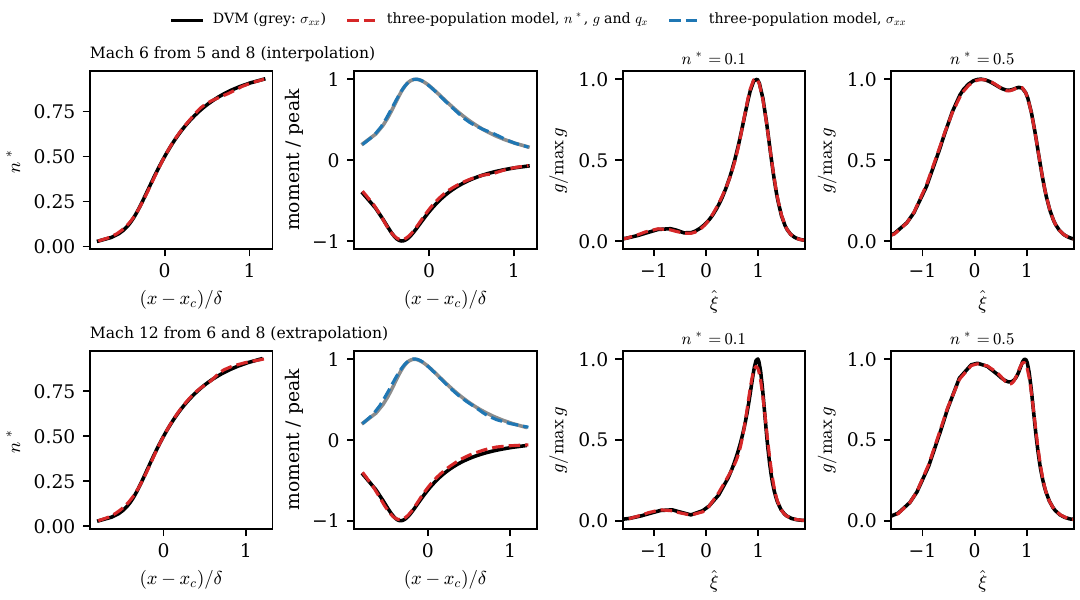}
\caption{Predictions with three populations (dashed) compared with the DVM solutions (solid): density profile, heat flux and stress, and the distribution at $n^*=0.1$ and 0.5. Top: Mach 6 from Mach 5 and 8. Bottom: Mach 12 from Mach 6 and 8.}
\label{fig:predictor}
\end{figure}

\emph{Prediction.} The population parameters of two source Mach numbers, in scaled form (mass fractions, $\hat\xi_k$, $T_{\parallel k}/T_2$, $T_{\perp k}/T_2$), are combined linearly in $1/M^2$ at fixed $n^*$ and evaluated with the RH states of the target Mach number; Eq.~\eqref{eq:xofn} then gives the profile. No information from the target solution is used. Table~\ref{tab:predictor} lists nine tests, in five of which the target lies between the sources and in four beyond them, and figure~\ref{fig:predictor} shows two of them.

\begin{table}
\centering\small
\caption{Conserved fluxes and physical-space errors of the predictions of tables~\ref{tab:predictor} and \ref{tab:predictor2} (Sec.~\ref{sec:preddsmc}), including tests not listed there. The first three columns give the largest deviation of the mass, momentum, and energy fluxes of the predicted mixture from their upstream values over $0.03\le n^*\le0.93$, in percent; these fluxes are not imposed. The last four give root-mean-square errors of the profiles as functions of the predicted position $x$, with the predicted and reference profiles aligned at $n^*=0.5$ and nothing else taken from the reference: the normalised density (percent of the density jump), the heat flux and the stress (percent of their peaks), and the temperature (percent of $T_2-T_1$).}
\label{tab:predictor3}
\begin{adjustbox}{max width=\linewidth}
\begin{tabular}{lllccccccc}
\toprule
 & $M$ & sources & mass & momentum & energy & $n^*(x)$ & $q_x(x)$ & $\sigma_{xx}(x)$ & $T(x)$\\
\midrule
BGK & 3 & 2.5, 4 & 1.5 & 1.0 & 2.4 & 3.3 & 4.5 & 6.1 & 3.0\\
 & 4 & 3, 5 & 4.3 & 3.4 & 3.7 & \multicolumn{4}{c}{no monotone profile}\\
 & 5 & 4, 6 & 2.2 & 1.9 & 3.1 & 2.0 & 3.5 & 2.9 & 0.9\\
 & 6 & 5, 8 & 0.9 & 0.4 & 0.8 & 0.5 & 1.7 & 1.1 & 0.6\\
 & 8 & 6, 12 & 0.4 & 0.4 & 0.4 & 0.6 & 1.5 & 0.8 & 0.7\\
 & 8 & 5, 6 & 2.0 & 2.2 & 5.7 & \multicolumn{4}{c}{no monotone profile}\\
 & 12 & 6, 8 & 0.6 & 0.7 & 1.8 & 0.7 & 2.9 & 2.1 & 1.4\\
 & 12 & 5, 8 & 1.0 & 1.4 & 3.5 & \multicolumn{4}{c}{no monotone profile}\\
 & 12 & 5, 6 & 5.3 & 5.9 & 13.9 & \multicolumn{4}{c}{no monotone profile}\\
DSMC, $\omega=1$ & 5 & 4, 6 & 0.5 & 0.3 & 0.5 & 0.3 & 0.9 & 0.4 & 0.7\\
 & 8 & 6, 12 & 0.9 & 0.4 & 0.8 & 0.5 & 0.6 & 1.1 & 1.2\\
 & 10 & 8, 12 & 1.1 & 0.6 & 1.4 & 0.7 & 0.9 & 1.8 & 0.8\\
 & 12 & 6, 8 & 1.2 & 0.9 & 1.8 & 0.9 & 1.3 & 2.1 & 0.7\\
 & 12 & 8, 10 & 1.2 & 0.8 & 1.8 & 0.8 & 1.4 & 2.4 & 0.6\\
 & 12 & 5, 6 & 1.4 & 1.2 & 3.3 & \multicolumn{4}{c}{no monotone profile}\\
 & 25 & 8, 12 & 1.5 & 1.3 & 2.9 & 1.3 & 3.5 & 4.6 & 1.0\\
DSMC, $\omega=0.81$ & 5 & 4, 6 & 0.5 & 0.2 & 0.6 & 0.7 & 2.7 & 1.9 & 1.9\\
 & 8 & 6, 12 & 0.9 & 0.4 & 1.0 & 0.6 & 2.2 & 1.8 & 2.1\\
 & 10 & 8, 12 & 1.1 & 0.6 & 1.5 & 0.6 & 1.8 & 2.0 & 1.7\\
 & 12 & 6, 8 & 1.2 & 0.9 & 2.0 & 0.6 & 1.5 & 2.1 & 1.4\\
 & 12 & 8, 10 & 1.2 & 0.8 & 1.9 & 0.6 & 1.6 & 2.2 & 1.5\\
 & 12 & 5, 6 & 1.2 & 1.5 & 3.0 & 0.9 & 1.6 & 2.5 & 1.1\\
 & 25 & 8, 12 & 1.5 & 1.2 & 2.9 & 0.7 & 2.4 & 3.7 & 1.2\\
DSMC, hard sphere & 5 & 4, 6 & 0.5 & 0.2 & 0.6 & 1.5 & 4.8 & 4.3 & 3.0\\
 & 8 & 6, 12 & 0.9 & 0.3 & 1.0 & 1.2 & 4.2 & 4.0 & 3.1\\
 & 10 & 8, 12 & 1.1 & 0.4 & 1.4 & 1.2 & 3.9 & 4.0 & 2.9\\
 & 12 & 6, 8 & 1.2 & 0.8 & 2.2 & 1.3 & 3.7 & 3.9 & 2.7\\
 & 12 & 8, 10 & 1.2 & 0.6 & 1.7 & 1.3 & 3.8 & 4.0 & 2.8\\
 & 12 & 5, 6 & 1.8 & 1.8 & 3.6 & 1.2 & 3.3 & 4.4 & 2.8\\
 & 25 & 8, 12 & 1.5 & 0.8 & 2.4 & 1.2 & 3.8 & 4.3 & 2.7\\
\bottomrule
\end{tabular}
\end{adjustbox}
\end{table}

When the sources are at Mach 6 or above, and for Mach 6 obtained from Mach 5 and 8, the prediction is nearly as accurate as a direct three-population fit of the target. The Mach-12 shock is obtained from Mach 6 and 8 with a largest residual of 4.7\% in the distribution (3.8\% for a direct fit of the Mach-12 solution), 4.3\% in the heat-flux profile, 0.9\% in the stress, 1.9\% in the temperature, and 1\% in the width of the density profile; the positions deviate by 0.02\,$\delta$. This Mach number lies outside the range in which the Mach-conditioned 512-kernel model of Sec.~\ref{sec:baseline} is valid. Interpolation to Mach 6 (from Mach 5 and 8) and to Mach 8 (from Mach 6 and 12) gives distribution residuals of 1.6\% and 2.4\% and widths within 3\%. In these three cases the fluxes of mass, momentum, and energy of the predicted mixtures, which are not imposed, deviate from their upstream values by at most 0.9\%, 0.7\%, and 1.8\% (table~\ref{tab:predictor3}).

\begin{table*}
\centering\small
\caption{Prediction of the Mach-12 BGK shock from Mach 6 and 8 with the archived and the recomputed solutions (Sec.~\ref{sec:verify}), by the three methods of Sec.~\ref{sec:methods}: three populations (P), direct combination of the scaled distributions (D), and the same with the beam treated separately (H). Columns as in table~\ref{tab:methods}, with the largest deviation of the energy flux added, in percent.}
\label{tab:m8}
\begin{adjustbox}{max width=\linewidth}
\begin{tabular}{llccccccccc}
\toprule
solutions & & $g$ (max) & $q_x$ & $\sigma_{xx}$ & $T$ & width & energy flux & $n^*(x)$ & $q_x(x)$ & $\sigma_{xx}(x)$\\
\midrule
archived Mach 6, 8, and 12 & P & 4.7 & 4.3 & 0.9 & 1.9 & 1.01 & 1.8 & 0.7 & 2.9 & 2.1\\
 & D & 7.9 & 1.4 & 1.9 & 1.9 & 1.02 & 0.7 & 0.5 & 1.0 & 1.1\\
 & H & 4.6 & 2.0 & 1.8 & 2.5 & 1.01 & 1.1 & 0.5 & 1.4 & 1.3\\
Mach 8 recomputed & P & 4.1 & 4.4 & 1.0 & 2.9 & 1.00 & 1.7 & 0.9 & 3.1 & 2.1\\
 & D & 6.0 & 1.3 & 2.1 & 0.4 & 1.02 & 0.3 & 0.4 & 0.9 & 1.2\\
 & H & 1.9 & 1.9 & 2.0 & 0.9 & 1.01 & 0.6 & 0.4 & 1.4 & 1.3\\
all three recomputed & P & 4.1 & 4.4 & 1.0 & 2.9 & 1.00 & 1.7 & 0.9 & 3.1 & 2.1\\
 & D & 6.2 & 0.1 & 0.3 & 0.1 & 1.00 & 0.1 & 0.1 & 0.6 & 0.4\\
 & H & 2.2 & 0.6 & 0.1 & 0.9 & 1.00 & 0.4 & 0.2 & 0.9 & 0.8\\
\bottomrule
\end{tabular}
\end{adjustbox}
\end{table*}

Two of these tests take the archived Mach-8 solution as a source, whose upstream equilibrium has an artificial anisotropy (Sec.~\ref{sec:verify}). They were repeated with the recomputed Mach-8 solution, which is free of it. The errors of the Mach-12 prediction from Mach 6 and 8 then change by at most one percentage point (4.1\% against 4.7\% in the distribution, 2.9\% against 1.9\% in the temperature), and those of Mach 6 from Mach 5 and 8 by at most 0.1. The three-population extrapolation to Mach 12 therefore does not depend on which Mach-8 solution is used. Table~\ref{tab:m8} lists the archived and recomputed results for the three methods of Sec.~\ref{sec:methods}; all other tables use the archived solutions.

These accuracies are obtained with both sources above Mach 5 and for the interpolation to Mach 6 from Mach 5 and 8. In the other tests with a source at Mach 5 or below the distribution residual is 1.9--5\%, the heat-flux error 4--10\%, and the width error up to 18\%. In four of the nine tests (Mach 4 from Mach 3 and 5; Mach 8 from Mach 5 and 6; Mach 12 from Mach 5 and 8; Mach 12 from Mach 5 and 6), all with a source at Mach 5 or below, $dx/dn^*$ from Eq.~\eqref{eq:xofn} is negative for $n^*$ between 0.63 and 0.83, so that the three populations give the distribution as a function of $n^*$ but not the profile in $x$. This comes from the representation and not from the numerical differentiation of the parameters: smoothing them over five positions in $n^*$ does not change it, the stress there is positive (0.36--0.78 of its peak), and it is the predicted third moment $\int\xi_x^3g\,d\xi_x$ that is not monotone in $n^*$; the direct three-population fit of the Mach-4 DVM solution behaves in the same way at one position. At these Mach numbers the three populations overlap downstream of $n^*\approx0.5$ and the decomposition is not unique (Sec.~\ref{sec:pop}), so that parameters from different Mach numbers do not correspond. In these tests the energy flux of the predicted mixture deviates by up to 14\%, and the errors in table~\ref{tab:predictor} for these rows describe the distribution as a function of $n^*$ only. The methods of Sec.~\ref{sec:methods} give a monotone profile in all of them. Extrapolation from Mach 5 and 6 to Mach 12, over a factor of two in Mach number, gives 5.2\% in the distribution and 10\% in the heat flux.

\subsection{Boltzmann gas and Shakhov model}
\label{sec:preddsmc}

\begin{table}
\centering\small
\caption{Predictions of the three-population model for DSMC and for the Shakhov solutions, in the format of table~\ref{tab:predictor}. The last column is the width ratio obtained when Eq.~\eqref{eq:xofn} is applied to the target solution itself, which measures the error of the closure alone. For $\omega=0.81$ and for hard spheres six of the thirteen tests are listed; the others follow those for $\omega=1$. Dashes mark the non-monotone $\omega=1$ prediction from Mach 5 and 6 to Mach 12.}
\label{tab:predictor2}
\begin{adjustbox}{max width=\linewidth}
\begin{tabular}{lllccccccc}
\toprule
Gas & $M$ & sources & $g$ (max) & $q_x$ & $\sigma_{xx}$ & $T$ & width ratio & position & closure\\
\midrule
DSMC, $\omega=1$ & 3 & 2, 4 & 3.2 (0.2) & 4.9 & 0.8 & 4.5 & 0.97 & 0.023 & 1.00\\
 & 4 & 3, 5 & 1.3 (0.5) & 1.4 & 0.9 & 2.1 & 0.99 & 0.009 & 1.00\\
 & 5 & 4, 6 & 0.9 (0.8) & 1.2 & 1.3 & 2.3 & 1.01 & 0.012 & 1.00\\
 & 6 & 5, 8 & 1.1 (1.1) & 1.4 & 1.9 & 2.8 & 1.02 & 0.017 & 1.00\\
 & 8 & 6, 12 & 1.8 (1.8) & 2.1 & 2.8 & 4.5 & 1.02 & 0.021 & 1.00\\
 & 10 & 8, 12 & 2.4 (2.5) & 2.5 & 3.1 & 5.2 & 1.04 & 0.020 & 1.00\\
 & 8 & 5, 6 & 2.5 (1.8) & 2.1 & 2.3 & 3.0 & 1.03 & 0.020 & 1.00\\
 & 12 & 6, 8 & 4.2 (3.1) & 2.6 & 3.0 & 4.5 & 1.05 & 0.022 & 1.00\\
 & 12 & 5, 8 & 5.2 (3.1) & 2.6 & 2.9 & 4.2 & 1.05 & 0.021 & 1.00\\
 & 12 & 8, 10 & 3.3 (3.1) & 2.8 & 3.3 & 5.8 & 1.05 & 0.021 & 1.00\\
 & 12 & 5, 6 & 9.5 (3.1) & 3.0 & 2.9 & 3.0 & -- & -- & 1.00\\
 & 25 & 8, 12 & 9.9 (5.8) & 3.9 & 3.9 & 8.2 & 1.07 & 0.028 & 1.00\\
 & 25 & 10, 12 & 8.3 (5.8) & 4.2 & 4.2 & 8.8 & 1.06 & 0.037 & 1.00\\
DSMC, $\omega=0.81$ & 5 & 4, 6 & 0.9 (0.7) & 1.1 & 1.5 & 2.1 & 0.98 & 0.027 & 0.98\\
 & 8 & 6, 12 & 1.6 (1.5) & 1.9 & 3.0 & 3.8 & 1.01 & 0.032 & 0.98\\
 & 8 & 5, 6 & 2.3 (1.5) & 1.9 & 2.3 & 2.4 & 1.00 & 0.023 & 0.98\\
 & 12 & 6, 8 & 3.7 (2.4) & 2.7 & 3.3 & 4.2 & 1.02 & 0.026 & 0.98\\
 & 12 & 8, 10 & 2.5 (2.4) & 2.7 & 3.5 & 5.1 & 1.01 & 0.027 & 0.98\\
 & 25 & 8, 12 & 7.3 (4.0) & 3.9 & 4.2 & 6.4 & 1.00 & 0.028 & 0.98\\
DSMC, hard sphere & 5 & 4, 6 & 0.9 (0.6) & 1.1 & 1.8 & 1.9 & 0.93 & 0.050 & 0.93\\
 & 8 & 6, 12 & 1.4 (1.2) & 1.6 & 3.1 & 3.1 & 0.96 & 0.052 & 0.92\\
 & 8 & 5, 6 & 1.9 (1.2) & 1.7 & 2.3 & 2.2 & 0.93 & 0.043 & 0.92\\
 & 12 & 6, 8 & 3.1 (1.7) & 2.2 & 3.4 & 2.6 & 0.94 & 0.046 & 0.92\\
 & 12 & 8, 10 & 1.7 (1.7) & 2.1 & 3.6 & 3.5 & 0.94 & 0.049 & 0.92\\
 & 25 & 8, 12 & 4.9 (2.4) & 2.9 & 4.1 & 4.4 & 0.93 & 0.048 & 0.92\\
Shakhov, $\mu\propto T$ & 3 & 2, 4 & 3.4 (1.3) & 5.6 & 5.0 & 4.2 & 1.01 & 0.041 & 1.00\\
 & 4 & 3, 5 & 11.2 (2.8) & 6.9 & 6.1 & 11.5 & 1.03 & 0.155 & 1.00\\
 & 5 & 3, 4 & 4.4 (4.1) & 8.5 & 5.3 & 12.0 & 1.09 & 0.062 & 1.00\\
 & 2 & 3, 4 & 5.1 (0.2) & 31.9 & 13.9 & 10.5 & 1.14 & 0.117 & 1.00\\
\bottomrule
\end{tabular}
\end{adjustbox}
\end{table}

The same procedure was applied to the DSMC distributions of Sec.~\ref{sec:dsmc}, which are available at eight Mach numbers between 2 and 12 and at Mach 25 for each of three viscosity exponents, with $\nu=\rho T^{1-\omega}$ (table~\ref{tab:predictor2}). Equation~\eqref{eq:moment} is first checked on the simulations themselves. For $\omega=1$ it reproduces the positions to 0.005\,$\delta$ and the width to 0.5\%, as expected, since for Maxwell-type molecules the stress relaxes at the rate $p/\mu$. For $\omega=0.81$ and for hard spheres the collision term is not of this form, and the same equation is an approximation that gives the shock width 1--2\% and 5--8\% too small.

For the Boltzmann gas, interpolated predictions are as accurate as a direct three-population fit of the target once both sources are at Mach 4 or above. For $\omega=1$, Mach 5 is predicted from Mach 4 and 6 with a largest distribution residual of 0.9\% (0.8\% for the direct fit), 1.2\% in the heat flux, 1.3\% in the stress, and 0.6\% in the width; Mach 6 from Mach 5 and 8 and Mach 8 from Mach 6 and 12 give 1.1\% and 1.8\% in the distribution and 1.4\% and 2.1\% in the heat flux. The Mach-12 shock is extrapolated from Mach 6 and 8 with 4.2\% in the distribution (3.1\% for the direct fit), 2.6\% in the heat flux, 3.0\% in the stress, and 5\% in the width, and from Mach 8 and 10 with 3.3\% in the distribution. The heat-flux and stress errors of the direct fits of the Mach-12 simulation are 3.0\% and 3.5\%, so that the accuracy of the prediction is that of the three-population representation and not of the combination of Mach numbers. The predicted $x(n^*)$ is monotone in all cases but one, the extrapolation from Mach 5 and 6 to Mach 12 for $\omega=1$, where it is not monotone at the downstream edge ($n^*>0.9$). The fluxes of mass, momentum, and energy of the predicted mixtures deviate from their upstream values by at most 1.1\%, 0.6\%, and 1.5\% for interpolation between Mach 4 and 12, and by 1.2\%, 0.9\%, and 2.2\% for the extrapolations to Mach 12 from Mach 6 and 8 or 8 and 10. The results for $\omega=0.81$ and for hard spheres are the same in the distribution and the heat flux (0.9--4.7\% and 1.1--2.7\% for sources at Mach 4 or above, the extrapolation from Mach 5 and 6 aside); the hard-sphere widths are 4--8\% smaller than in the simulations, which is the contribution of the closure. As for the BGK shocks, a source at Mach 2 or 3 reduces the accuracy (3.2\% in the distribution and 2.5--6\% in the heat flux for Mach 3 from Mach 2 and 4), and extrapolation from Mach 5 and 6 to Mach 12 gives the moments to within 3\% but a distribution residual of 9\%.

\emph{Mach 25.} The extrapolation was carried a factor of two farther, to Mach 25 from Mach 8 and 12, for which the weights are $-0.62$ and $1.62$. With three populations the heat flux and the stress are predicted to within 3.9\% and 4.2\% for the three gases and the width to within 7\%; the largest distribution residual is 4.9\% for hard spheres, 7.3\% for $\omega=0.81$, and 9.9\% for $\omega=1$, against 2.4\%, 4.0\%, and 5.8\% for the direct fits of the Mach-25 simulations (table~\ref{tab:predictor2}). The profile is monotone for all three gases.

\begin{figure}[!htbp]
\includegraphics[width=\textwidth]{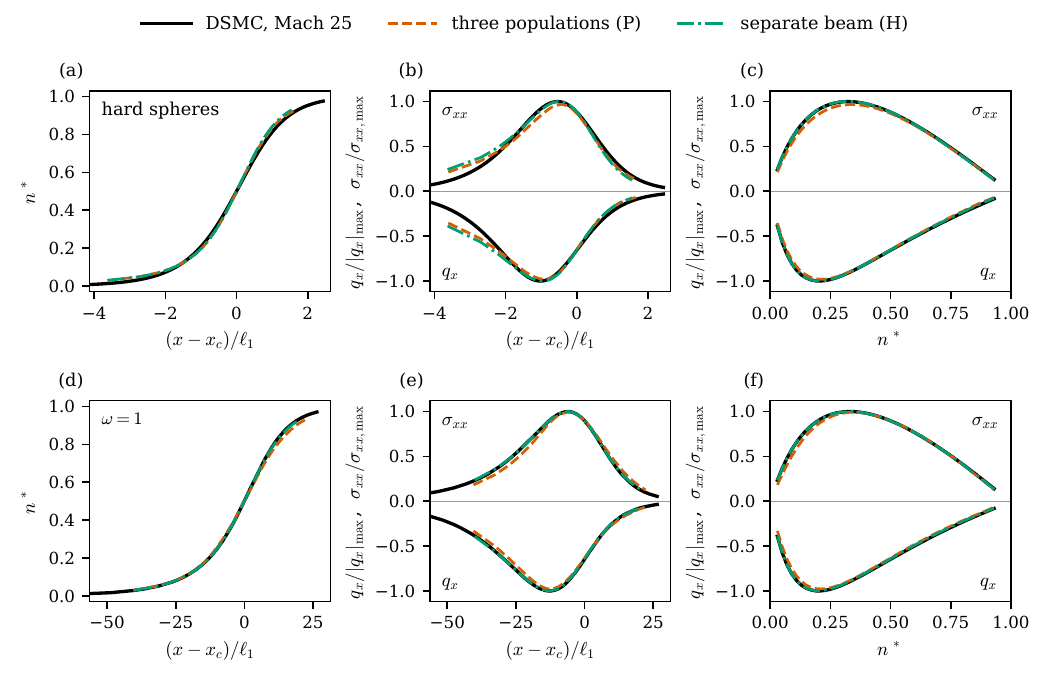}
\caption{Prediction of the Mach-25 shock from the Mach-8 and Mach-12 shocks, compared with DSMC at Mach 25 (solid lines): three populations (P, dashed) and direct combination with the beam treated separately (H, dash-dotted). Top row: hard spheres; bottom row: $\omega=1$. (a), (d) Normalised density against position, with $x(n^*)$ of the predictions obtained from Eq.~\eqref{eq:xofn} and all profiles aligned at $n^*=0.5$, where $x=x_c$. (b), (e) Heat flux and normal stress, each divided by the peak of the DSMC profile, against position. (c), (f) The same against $n^*$. The predictions cover $0.03\le n^*\le0.93$. For hard spheres the difference between (b) and (c) is the reconstruction of position; for $\omega=1$, where Eq.~\eqref{eq:moment} is closed exactly, the H prediction and the simulation coincide in both.}
\label{fig:m25x}
\end{figure}

Figure~\ref{fig:m25x} shows the Mach-25 prediction as profiles in physical space. Its error has two parts. The first is that of the similarity, measured at equal $n^*$: for hard spheres the heat flux and the stress differ from the simulation by 0.7\% and 0.5\% with method H and by 0.3\% and 0.2\% with method D of Sec.~\ref{sec:methods}, and by 2.9\% and 4.1\% with three populations. The second is the reconstruction of position from Eq.~\eqref{eq:xofn}. As functions of $x$ the root-mean-square differences of method H are 5.0\% and 4.6\% of the peaks for hard spheres, against 0.5\% and 0.3\% in the same measure at equal $n^*$, and the predicted width is 0.92 of the simulated one, which is the ratio that Eq.~\eqref{eq:xofn} gives when it is applied to the hard-sphere simulation itself; for $\omega=1$, where the equation is exact, the differences in $x$ are 0.4\% and 0.3\%.

These differences are to be read beside the accuracy of the reference. At equal $n^*$ the statistical scatter of the Mach-25 hard-sphere simulation, from its four time blocks, is 0.02\% in the heat flux and 0.06\% in the stress; doubling the number of molecules per cell changes the two profiles by 0.1\%, and halving the cell and the time step changes them by 0.7\% and 1.0\% (0.6\% and 0.8\% in the measure of Sec.~\ref{sec:dsmcsim}, which compares the profiles as functions of $x$). The differences of 0.2--0.7\% between the D and H predictions and the simulation are therefore no larger than the change of the reference under refinement: prediction and reference agree to within the numerical accuracy of the reference. The change under refinement is common to the three Mach numbers: when the two sources and the target are all computed with the halved cell and time step, method D gives 0.1\% in both moments. The differences of the three-population prediction and the position error for hard spheres are several times larger than the change under refinement and are properties of the methods.

\emph{Shakhov model.} For the Shakhov solutions with $\mu\propto T$, which cover Mach 2--5, Eq.~\eqref{eq:moment} reproduces the positions to 0.003\,$\delta$, and the predictions have errors of 3--11\% in the distribution and 6--9\% in the heat flux between Mach 3 and 5; the extrapolation from Mach 3 and 4 down to Mach 2 gives the heat flux to 32\%. The Shakhov distribution is partly negative, and three positive Gaussians represent it with larger residuals than the BGK and DSMC distributions (table~\ref{tab:models}).

The errors quoted so far compare profiles at equal $n^*$. Table~\ref{tab:predictor3} gives the errors of the profiles in physical space, where the predicted positions enter and only the location of $n^*=0.5$ is taken from the reference. For the five BGK tests with a monotone profile the root-mean-square errors are 0.5--3.3\% of the density jump for $n^*(x)$, 1.5--4.5\% and 0.8--6.1\% of the peaks for the heat flux and the stress, and 0.6--3.0\% of the temperature rise. For DSMC with sources at Mach 4 or above they are 0.3--0.9\%, 0.6--1.4\%, 0.4--2.4\%, and 0.6--1.2\% for $\omega=1$, and 1.1--1.5\%, 3.3--4.8\%, 3.6--4.4\%, and 2.6--3.2\% for hard spheres (all eleven tests), where the error of the closure in the shock width enters.

\subsection{Is the Gaussian form needed?}
\label{sec:methods}

The predictor combines the scaled variables with the three-population form, and the tests above do not separate them. The scaled variables can be used without any fit: the scaled reduced distributions $g\,(u_1-u_2)/\rho$ and $h\,(u_1-u_2)/(\rho T_2)$ of the two sources, on a common grid of $\hat\xi$ and at fixed $n^*$, are combined linearly in $1/M^2$ and evaluated with the RH states of the target, as in the test of Sec.~\ref{sec:sim}; the moments of the combined distribution then give $\sigma_{xx}$ and $\int\xi_x^3g\,d\xi_x$, and Eq.~\eqref{eq:xofn} gives the positions exactly as for the populations. We call this the direct method (D). One property of this method is known in advance: the beam has the width of the upstream thermal speed, which in the variable $\hat\xi$ decreases as $1/M$, so that a linear combination of two beams of different widths is not a beam of the right width. A third method (H) therefore treats the beam separately: the beam population of the three-population fit is subtracted from each source, the remainders are combined directly, and a beam with combined parameters is added back.

\begin{table*}
\centering\small
\caption{Three ways of predicting a shock from two others in the scaled variables: three populations (P), direct combination of the scaled reduced distributions (D), and the same with the beam treated separately (H). Errors at equal $n^*$ as in table~\ref{tab:predictor}; width ratio; and root-mean-square errors of $n^*(x)$, $q_x(x)$, and $\sigma_{xx}(x)$ in physical space as in table~\ref{tab:predictor3}. All in percent except the width ratio.}
\label{tab:methods}
\begin{adjustbox}{max width=\linewidth}
\begin{tabular}{llllcccccccc}
\toprule
 & $M$ & sources & & $g$ (max) & $q_x$ & $\sigma_{xx}$ & $T$ & width & $n^*(x)$ & $q_x(x)$ & $\sigma_{xx}(x)$\\
\midrule
BGK & 4 & 3, 5 & P & 4.0 & 6.2 & 3.3 & 2.4 & -- & \multicolumn{3}{c}{no monotone profile}\\
 &  &  & D & 4.3 & 1.0 & 2.8 & 0.6 & 0.97 & 0.6 & 1.4 & 1.3\\
 &  &  & H & 2.4 & 2.6 & 2.0 & 2.1 & 0.98 & 0.8 & 2.8 & 1.9\\
 & 6 & 5, 8 & P & 1.6 & 3.3 & 0.7 & 2.2 & 0.98 & 0.5 & 1.7 & 1.1\\
 &  &  & D & 1.7 & 0.3 & 0.5 & 0.6 & 1.00 & 0.1 & 0.4 & 0.5\\
 &  &  & H & 0.4 & 0.3 & 0.4 & 0.2 & 1.00 & 0.1 & 0.4 & 0.4\\
 & 8 & 6, 12 & P & 2.4 & 3.0 & 1.0 & 3.7 & 0.97 & 0.6 & 1.5 & 0.8\\
 &  &  & D & 4.4 & 1.7 & 3.2 & 0.7 & 0.97 & 0.6 & 0.7 & 1.6\\
 &  &  & H & 1.7 & 2.1 & 3.1 & 1.3 & 0.97 & 0.6 & 1.1 & 1.6\\
 & 12 & 6, 8 & P & 4.7 & 4.3 & 0.9 & 1.9 & 1.01 & 0.7 & 2.9 & 2.1\\
 &  &  & D & 7.9 & 1.4 & 1.9 & 1.9 & 1.02 & 0.5 & 1.0 & 1.1\\
 &  &  & H & 4.6 & 2.0 & 1.8 & 2.5 & 1.01 & 0.5 & 1.4 & 1.3\\
 & 12 & 5, 6 & P & 5.2 & 9.5 & 4.5 & 5.2 & -- & \multicolumn{3}{c}{no monotone profile}\\
 &  &  & D & 13.4 & 6.2 & 5.6 & 0.8 & 0.97 & 0.8 & 3.4 & 3.1\\
 &  &  & H & 5.5 & 4.4 & 6.1 & 2.6 & 0.95 & 1.1 & 2.9 & 3.5\\
DSMC, $\omega=1$ & 3 & 2, 4 & P & 3.2 & 4.9 & 0.8 & 4.5 & 0.97 & 1.0 & 3.5 & 2.2\\
 &  &  & D & 14.0 & 5.3 & 21.9 & 3.1 & 0.81 & 3.7 & 11.1 & 9.2\\
 &  &  & H & 3.1 & 5.6 & 16.2 & 3.8 & 0.74 & 4.7 & 9.7 & 10.6\\
 & 5 & 4, 6 & P & 0.9 & 1.2 & 1.3 & 2.3 & 1.01 & 0.3 & 0.9 & 0.4\\
 &  &  & D & 4.4 & 0.2 & 0.9 & 0.2 & 1.00 & 0.1 & 0.3 & 0.6\\
 &  &  & H & 0.5 & 0.6 & 0.6 & 0.5 & 0.99 & 0.3 & 0.9 & 0.6\\
 & 8 & 6, 12 & P & 1.8 & 2.1 & 2.8 & 4.5 & 1.02 & 0.5 & 0.6 & 1.1\\
 &  &  & D & 12.6 & 0.1 & 0.4 & 0.1 & 1.00 & 0.1 & 0.4 & 0.5\\
 &  &  & H & 1.3 & 0.7 & 0.3 & 0.4 & 1.00 & 0.2 & 0.9 & 0.7\\
 & 12 & 6, 8 & P & 4.2 & 2.6 & 3.0 & 4.5 & 1.05 & 0.9 & 1.3 & 2.1\\
 &  &  & D & 18.7 & 0.1 & 0.4 & 0.2 & 1.01 & 0.2 & 0.8 & 0.5\\
 &  &  & H & 3.3 & 0.9 & 0.3 & 0.7 & 1.01 & 0.3 & 0.4 & 0.5\\
 & 12 & 8, 10 & P & 3.3 & 2.8 & 3.3 & 5.8 & 1.05 & 0.8 & 1.4 & 2.4\\
 &  &  & D & 6.5 & 0.1 & 0.1 & 0.1 & 1.00 & 0.1 & 0.5 & 0.4\\
 &  &  & H & 0.9 & 0.2 & 0.1 & 0.2 & 1.01 & 0.1 & 0.4 & 0.4\\
 & 25 & 8, 12 & P & 9.9 & 3.9 & 3.9 & 8.2 & 1.07 & 1.3 & 3.5 & 4.6\\
 &  &  & D & 37.6 & 0.1 & 0.1 & 0.1 & 1.00 & 0.1 & 0.6 & 0.4\\
 &  &  & H & 8.7 & 0.8 & 0.2 & 0.3 & 1.01 & 0.2 & 0.4 & 0.3\\
DSMC, hard sphere & 8 & 6, 12 & P & 1.4 & 1.6 & 3.1 & 3.1 & 0.96 & 1.2 & 4.2 & 4.0\\
 &  &  & D & 14.3 & 0.2 & 0.6 & 0.1 & 0.92 & 1.6 & 4.9 & 4.5\\
 &  &  & H & 0.7 & 0.9 & 0.6 & 0.3 & 0.92 & 1.7 & 5.1 & 4.6\\
 & 12 & 6, 8 & P & 3.1 & 2.2 & 3.4 & 2.6 & 0.94 & 1.3 & 3.7 & 3.9\\
 &  &  & D & 20.7 & 0.1 & 0.5 & 0.2 & 0.92 & 1.6 & 5.2 & 4.8\\
 &  &  & H & 3.1 & 1.2 & 0.7 & 0.7 & 0.93 & 1.5 & 5.0 & 4.6\\
 & 25 & 8, 12 & P & 4.9 & 2.9 & 4.1 & 4.4 & 0.93 & 1.2 & 3.8 & 4.3\\
 &  &  & D & 42.0 & 0.3 & 0.2 & 0.1 & 0.91 & 1.7 & 5.0 & 4.6\\
 &  &  & H & 4.9 & 0.7 & 0.5 & 0.2 & 0.92 & 1.6 & 5.0 & 4.6\\
\bottomrule
\end{tabular}
\end{adjustbox}
\end{table*}

\begin{figure}[!htbp]
\includegraphics[width=\textwidth]{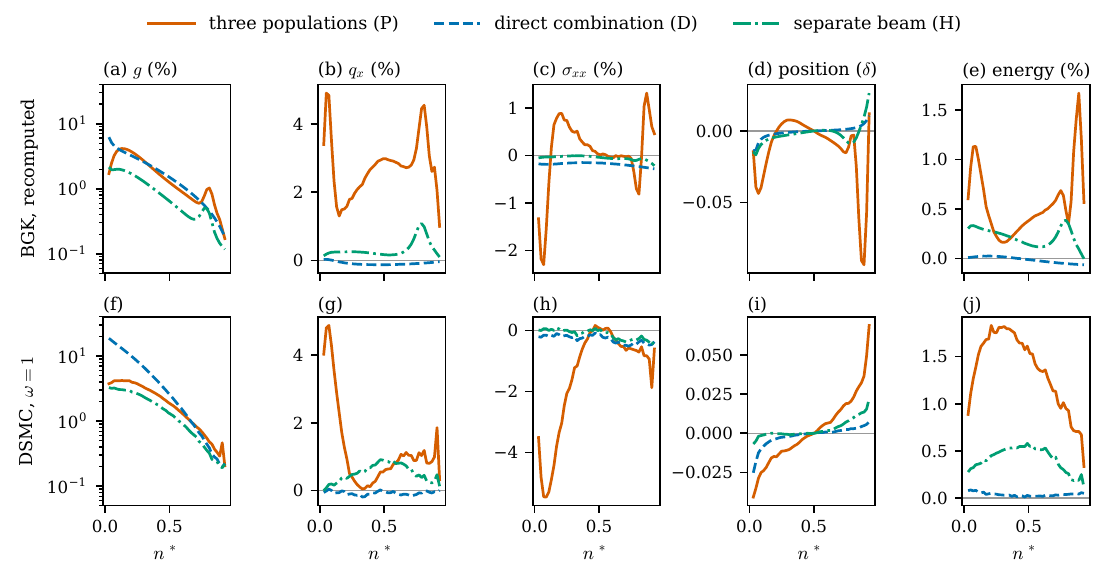}
\caption{Errors of the three predictions of the Mach-12 shock from Mach 6 and 8 as functions of the normalised density $n^*$: three populations (P, solid), direct combination of the scaled distributions (D, dashed), and the same with the beam separate (H, dash-dotted). Top row: BGK model, with source and target solutions all recomputed by the reduced solver. Bottom row: DSMC, $\omega=1$. (a), (f) Relative $L_2$ residual of $g$, in percent. (b), (g) and (c), (h) Error of the heat flux and of the normal stress, in percent of the peak of each. (d), (i) Error of the position obtained from Eq.~\eqref{eq:xofn}, in units of the shock thickness $\delta$, with predicted and reference profiles aligned at $n^*=0.5$. (e), (j) Deviation of the energy flux from its upstream value, in percent; this flux is not imposed.}
\label{fig:methods}
\end{figure}

Table~\ref{tab:methods} compares the three methods, and figure~\ref{fig:methods} shows their errors as functions of $n^*$ for the extrapolation to Mach 12. For DSMC with $\omega=1$ and sources at Mach 4 or above, the direct method gives the heat flux, the stress, and the temperature at equal $n^*$ to within 0.2\%, 0.9\%, and 0.2\%, against 1.2--2.8\%, 1.3--3.3\%, and 2.3--5.8\% for the three populations; for the extrapolation to Mach 12 from Mach 6 and 8 the three errors are 0.1\%, 0.4\%, and 0.2\% (D) against 2.6\%, 3.0\%, and 4.5\% (P), the width ratio is 1.01 against 1.05, and the conserved fluxes deviate by 0.2\% against 1.8\%. The direct method gives a monotone profile in every test, including the four BGK tests in which the three populations do not. Its residual is concentrated at the beam, as expected: the largest residual of $g$ is 19\% for that extrapolation, against 4.2\% for the populations. The method with a separate beam combines the two: 3.3\% in the distribution, 0.9\% and 0.3\% in the heat flux and stress, and a monotone profile in every test. For the BGK shocks the ranking is the same with smaller differences (tables~\ref{tab:methods} and \ref{tab:m8}). When source and target solutions all come from the same solver (the recomputed solutions at Mach 6, 8, and 12 of Sec.~\ref{sec:verify}), the BGK result is the same as for DSMC: the direct method gives the heat flux, stress, and temperature of the Mach-12 shock to within 0.1\%, 0.3\%, and 0.1\%, and the method with a separate beam gives 2.2\% in the distribution, 0.6\% in the heat flux, and 0.1\% in the stress, against 4.1\%, 4.4\%, and 1.0\% for the three populations; the larger differences in table~\ref{tab:methods} for the BGK model thus include differences of about 1\% between solutions computed on different velocity grids. For hard spheres the three methods give the same errors in physical space, about 5\% in the heat-flux and stress profiles, because these are set by the closure for the shock width, which is common to the three representations. The extrapolation to Mach 25 from Mach 8 and 12 shows the same ranking more strongly (table~\ref{tab:methods}): the direct method gives the heat flux, the stress, and the temperature at equal $n^*$ to within 0.3\% for the three gases and conserves the fluxes to within 0.2\%, with a distribution residual of 38--42\% at the beam; with the beam separated the distribution residual is 4.9--8.7\% and the heat flux and the stress are within 1.0\% and 0.5\%.

For weak shocks the ranking is reversed. When the sources are at Mach 2 and 4, the scaled distribution changes its shape between them, and the direct combination gives the stress at Mach 3 with an error of 22\% and a width 19\% smaller than the target, against 0.8\% and 3\% for the populations, whose parameters vary more smoothly with Mach number than the distribution itself.

\begin{table*}
\centering\small
\caption{Positivity of the extrapolated predictions before any lower bound is applied. D and H: negative part of the combined $g$ and $h$ at the worst position, in percent of the mass (of $\int g\,d\xi_x$ and $\int h\,d\xi_x$); for H, whether the lower bound on the extrapolated beam parameters is active. P: smallest extrapolated mass fraction and temperatures (in units of $T_2$) of the three populations over all positions, and whether the lower bounds of the implementation (mass fraction $10^{-9}$, temperatures $10^{-4}\,T_2$) are active. The errors of tables~\ref{tab:predictor}--\ref{tab:methods} are those of the predictions with these lower bounds applied.}
\label{tab:positivity}
\begin{adjustbox}{max width=\linewidth}
\begin{tabular}{lllccccccccc}
\toprule
 & & & \multicolumn{2}{c}{D} & \multicolumn{3}{c}{H} & \multicolumn{4}{c}{P}\\
 & $M$ & sources & $g^-$ & $h^-$ & $g^-$ & $h^-$ & lower bound & mass & $T_\parallel$ & $T_\perp$ & lower bound\\
\midrule
BGK & 12 & 6, 8 & 0.007 & 0.017 & 0.004 & 0.012 & no & 0.012 & 0.054 & 0.049 & no\\
BGK, recomputed & 12 & 6, 8 & 0.009 & 0.022 & 0.004 & 0.013 & no & 0.013 & 0.048 & 0.049 & no\\
BGK & 12 & 5, 8 & 0.080 & 0.175 & 0.015 & 0.053 & no & 0.012 & 0.053 & 0.050 & no\\
 & 8 & 5, 6 & 0.068 & 0.132 & 0.006 & 0.026 & no & 0.011 & 0.074 & 0.000 & yes\\
 & 12 & 5, 6 & 0.632 & 1.560 & 0.105 & 0.382 & no & 0.011 & 0.042 & 0.000 & yes\\
DSMC, $\omega=1$ & 12 & 6, 8 & 0.574 & $<0.001$ & $<0.001$ & $<0.001$ & no & 0.022 & 0.027 & 0.028 & no\\
 & 12 & 8, 10 & $<0.001$ & $<0.001$ & $<0.001$ & $<0.001$ & no & 0.024 & 0.025 & 0.026 & no\\
 & 12 & 5, 6 & 1.632 & 1.535 & 0.002 & 0.003 & no & -0.020 & 0.032 & 0.032 & yes\\
 & 25 & 8, 12 & 0.745 & $<0.001$ & $<0.001$ & $<0.001$ & no & 0.015 & 0.008 & 0.001 & no\\
DSMC, $\omega=0.81$ & 12 & 6, 8 & 0.748 & $<0.001$ & $<0.001$ & $<0.001$ & no & 0.016 & 0.025 & 0.026 & no\\
 & 12 & 8, 10 & $<0.001$ & $<0.001$ & $<0.001$ & $<0.001$ & no & 0.014 & 0.023 & 0.024 & no\\
 & 12 & 5, 6 & 1.785 & 1.979 & 0.002 & 0.003 & no & 0.004 & 0.029 & 0.030 & no\\
 & 25 & 8, 12 & 1.160 & $<0.001$ & $<0.001$ & $<0.001$ & yes & 0.011 & 0.006 & -0.048 & yes\\
DSMC, hard sphere & 12 & 6, 8 & 0.944 & 0.195 & $<0.001$ & $<0.001$ & no & 0.014 & 0.024 & 0.007 & no\\
 & 12 & 8, 10 & 0.091 & $<0.001$ & $<0.001$ & $<0.001$ & yes & 0.011 & 0.023 & -0.003 & yes\\
 & 12 & 5, 6 & 2.030 & 2.694 & 0.003 & 0.004 & no & 0.004 & 0.002 & 0.017 & no\\
 & 25 & 8, 12 & 1.998 & $<0.001$ & $<0.001$ & $<0.001$ & yes & 0.005 & 0.005 & -0.096 & yes\\
\bottomrule
\end{tabular}
\end{adjustbox}
\end{table*}

\emph{Positivity and lower bounds.} Outside the interval of the sources the combination has a negative weight ($-0.71$ and $1.71$ for Mach 12 from Mach 6 and 8, $-1.7$ and $2.7$ from Mach 5 and 6), so that neither a non-negative distribution for D and H nor positive densities and temperatures for P are guaranteed. The scaled distributions are combined on a common grid of 3841 points in $-7\le\hat\xi\le5$, by linear interpolation in $\hat\xi$ and in $n^*$ and with zero outside the tabulated velocities of a source. The combined distributions of D and H are not clipped or renormalised. The implementation has two lower bounds: the extrapolated parameters of the beam in H, and the extrapolated mass fractions and temperatures in P, are bounded below by small positive values. Table~\ref{tab:positivity} gives, for every extrapolation, the negative part of each prediction before any lower bound is applied and states whether a lower bound is active; the errors quoted elsewhere are those of the predictions with the lower bounds applied.

For sources at Mach 6 and above, the directly combined $g$ is negative next to the beam: at the worst position its negative part holds 0.6\%, 0.7\%, and 0.9\% of the mass for $\omega=1$, 0.81, and hard spheres (0.007\% for the BGK model, 0.009\% with the recomputed solutions). With the beam separated the negative parts are below 0.001\% of the mass in DSMC and 0.004\% in $g$ and 0.013\% in $h$ for the BGK model. No lower bound is active in H or in P, with one exception: for hard spheres from Mach 8 and 10 the extrapolated transverse temperature of the beam, which is about $0.02\,T_2$ in the sources, falls to $-0.003\,T_2$ at its lowest, and the lower bound is then used in both methods. For these sources the separate beam removes the negative part of the direct method. In the extrapolation to Mach 25 from Mach 8 and 12 the negative part of D holds 0.7\%, 1.2\%, and 2.0\% of the mass for $\omega=1$, 0.81, and hard spheres and is again removed by the separate beam; the lower bound on the transverse temperature of the beam is active for $\omega=0.81$ and for hard spheres.

With a source at Mach 5 the separate beam removes the negative part in DSMC but not for the BGK model. From Mach 5 and 6 the negative part of D grows to 1.6--2.0\% of the mass in DSMC and 0.63\% in $g$ and 1.56\% in $h$ for the BGK model; with the separate beam it is below 0.004\% in DSMC but still 0.105\% in $g$ and 0.382\% in $h$ for the BGK model. In P, the extrapolated transverse temperature of one population reaches zero in the two BGK extrapolations from Mach 5 and 6, and one extrapolated mass fraction is negative in the DSMC extrapolation from Mach 5 and 6 for $\omega=1$; the lower bounds are then active, and the smallest mass fraction over the BGK tests is 0.0109. These are the tests with a source at Mach 5 described in Sec.~\ref{sec:predpos}.

\emph{Sensitivity to the lower bound.} In the extrapolation to Mach 25 from Mach 8 and 12 the lower bound acts on one parameter only, the transverse temperature of the beam, at 27 of the 46 positions ($0.35\le n^*\le0.87$ for $\omega=0.81$ and $0.41\le n^*\le0.93$ for hard spheres). This parameter does not enter $g$, so that the distribution residuals do not depend on the bound. When the bound is varied by a factor of a hundred, from $10^{-5}\,T_2$ to $10^{-3}\,T_2$, the heat-flux and stress errors change by at most 0.007 and 0.005 percentage points; when it is removed and the extrapolated value is used as it stands, they change by at most 0.13 and 0.03 points (2.72\% and 4.03\% in place of 2.85\% and 4.06\% for hard spheres). The reason is that at these positions the beam holds 0.5--19\% of the mass but at most 2.1\% of the transverse thermal energy, so that its transverse temperature has little weight in the moments; the direct fits of the Mach-25 simulations themselves return values below $10^{-4}\,T_2$ over most of this range ($n^*\ge0.49$ for hard spheres and $n^*\ge0.37$ for $\omega=0.81$). The scaling of Sec.~\ref{sec:sim} also gives a form that is positive by construction: the beam temperature belongs to the upstream state, and when $\ln(T_\perp/T_1)$ of the beam is combined linearly in $1/M^2$, with all other parameters treated as before, no bound is needed. The smallest transverse temperature of the beam is then $1.23\,T_1$, $0.74\,T_1$, and $0.26\,T_1$ for $\omega=1$, 0.81, and hard spheres, the distribution residuals are unchanged, and the heat-flux and stress errors are 3.8\% and 3.8\%, 3.9\% and 4.1\%, and 2.9\% and 4.1\%, against 3.9\% and 3.9\%, 3.9\% and 4.2\%, and 2.9\% and 4.1\% with the bound. The Mach-25 results of the three-population method thus do not depend on the bound.

For strong shocks the Gaussian form is therefore not needed, and it is not the source of the accuracy of the prediction. The accuracy comes from the scaled variables together with a separate treatment of the beam, the one part of the distribution that does not scale. The three populations remain the most compact form of the prediction (twelve numbers per position), the only one of the three that is interpretable term by term, and the most accurate one when a source is a weak shock; for the moments of strong shocks the direct combination is more accurate, and it gives a profile in every test.

\subsection{Range of validity}

The prediction of a shock from two others therefore carries over from the BGK model to the Boltzmann gas. Three conditions set its range of validity. It uses solutions at two Mach numbers of the same gas. It applies where the scaled distribution has reached its strong-shock form, that is, for sources at Mach 4 or above in DSMC and at Mach 6 or above for the BGK model. And Eq.~\eqref{eq:moment} is closed exactly for a collision term whose second moment is $-\nu\sigma_{xx}$, which holds for the relaxation models and for Maxwell-type molecules; for hard spheres it gives the shock width to within 5--8\%.

\FloatBarrier
\section{The upstream precursor}
\label{sec:prec}

\subsection{Relaxation models: survival}

The upstream-moving molecules ahead of the shock can be analysed directly with
Eq.~\eqref{eq:integral}. Far upstream of the hot region, the source of molecules with $\xi_x<0$ is
concentrated where the gas is hot and dense. If that region is modelled as a half-space of
downstream equilibrium gas beginning at $x_c$, Eq.~\eqref{eq:integral} reduces to
\begin{equation}
g(x,\xi_x<0)\simeq g_{M2}(\xi_x)\,\exp\!\left[-\frac{\theta}{|\xi_x|}\right],\qquad
\theta=\int_x^{x_c}\nu\,dx',
\label{eq:halfspace}
\end{equation}
with $g_{M2}$ the reduced downstream Maxwellian. This is the precursor distribution that
Narasimha~\citeyearpar{Narasimha1968} obtained for a weak shock by the same replacement of the source by a
discontinuity [his Eq.~(7.14)], written here with the optical distance of a shock of arbitrary
strength. The exponential factor removes slow molecules:
because the collision frequency does not depend on speed, the free path of a molecule is
proportional to its speed, and the fraction that survives an optical distance $\theta$ is
$\exp(-\theta/|\xi_x|)$. The product of this factor with the Maxwellian has a maximum at
$|\xi_x|\simeq(\theta T_2)^{1/3}$, which moves to larger speeds with distance. This is the origin of
the separate maximum of $g$ at $\xi_x<0$ and of the low streamwise temperature of the hot
population; the transverse velocities are not filtered, which is why its transverse temperature
remains close to $T_2$.

Integrating Eq.~\eqref{eq:halfspace} over $\xi_x<0$ by the saddle-point method gives the
upstream-moving mass fraction
\begin{equation}
\beta(x)=\frac1\rho\int_{\xi_x<0}g\,d\xi_x\;\propto\;
\exp\!\left[-\tfrac32\,s-\frac{u_2}{\sqrt{T_2}}\,s^{1/2}\right],\qquad
s=\left(\frac{\theta}{\sqrt{T_2}}\right)^{2/3},
\label{eq:stretched}
\end{equation}
up to algebraic factors. The decay is a stretched exponential in distance, slower than any
exponential. For the weak shock it was found by Lyubarskii~\citeyearpar{Lyubarskii1961} from the integral
equations for the moments and derived from the precursor distribution by
Narasimha~\citeyearpar{Narasimha1968}, whose Eq.~(7.17) is the leading term $-\tfrac32s$ of
Eq.~\eqref{eq:stretched} in his variables. In a weak shock this decay is reached only far upstream,
after a region of exponential decay~\citep{Narasimha1968}.
Equation~\eqref{eq:halfspace} suggests that the same law holds for shocks of any strength, with the
downstream temperature as the scale, and the solutions below test this. Since $u_2/\sqrt{T_2}=\sqrt{\gamma}\,M_2$ tends to a constant for strong shocks,
Eq.~\eqref{eq:stretched} predicts that $\beta$ is a universal function of $s$ at high Mach number.

\begin{figure}[!htbp]
\includegraphics[width=\textwidth]{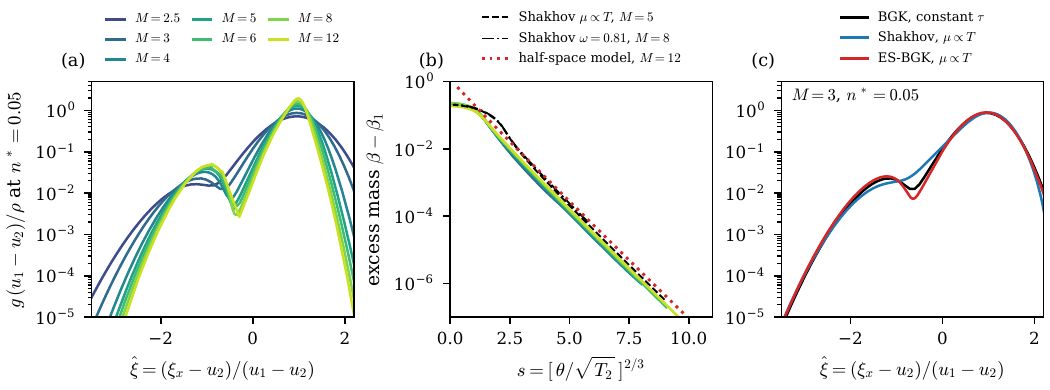}
\caption{Upstream precursor. (a) Scaled distribution at $n^*=0.05$ for the BGK solutions.
(b) Excess upstream-moving mass fraction $\beta-\beta_1$ against $s=(\theta/\sqrt{T_2})^{2/3}$, with $\theta$ the optical
distance upstream of the shock centre: BGK solutions at Mach 3--12 (colours as in (a)), two Shakhov
solutions, and the half-space model~\eqref{eq:halfspace}. (c) Distribution at Mach 3 and
$n^*=0.05$ for three collision models.}
\label{fig:prec}
\end{figure}

Figure~\ref{fig:prec}(b) shows the excess $\beta-\beta_1$ against $s$ for the BGK solutions at Mach 3--12. The curves for Mach 4--12 collapse, and between
$\beta=10^{-2}$ and $10^{-6}$ they are straight lines with slopes between $-1.638$ and $-1.649$.
Numerical evaluation of the half-space model~\eqref{eq:halfspace} for the Mach-12 downstream state
gives a slope of $-1.62$ over the same interval, without adjustable parameters; the model lies
above the data by a constant factor, as expected since the real source is not a sharp half-space.
The Shakhov solutions, with $\theta$ computed from their position-dependent collision frequency,
follow the same law (slopes $-1.66$ at Mach 5 and $-1.69$ at Mach 8). Figure~\ref{fig:prec}(a)
shows that the scaled distribution at $n^*=0.05$ is nearly the same for Mach 5--12.

Two remarks are in order. First, Eq.~\eqref{eq:stretched} relies on a collision frequency that is
independent of molecular speed. For hard spheres the collision frequency grows with speed, the
free path of fast molecules is bounded, and a faster decay, closer to an exponential, is expected. DSMC
shows this dependence (Sec.~\ref{sec:precgas}): fitted as $\ln\beta=a-bN^{p}$ in the column density $N$, the
decay has $p=0.68$ for Maxwell-type molecules and $p=0.87$ for hard spheres. Equation~\eqref{eq:halfspace}
also assumes that a collision removes a molecule from the upstream-moving group. This is exact for a relaxation model, which replaces the molecule by one drawn from its target distribution; in a gas both partners of a
collision continue, and Sec.~\ref{sec:precgas} shows that most upstream-moving molecules ahead of the shock
have in fact collided with the incoming stream.
Second, the upstream-moving mass is not small inside the shock: for the BGK model at Mach 12, $\beta=0.04$ at $n^*=0.05$ and 0.12--0.13 at $n^*=0.2$.

\subsection{Precursor of the gas}
\label{sec:precgas}

In a VHS gas a molecule that moves upstream with streamwise speed $w$ meets the incoming stream at the
relative speed $w+u_1$ and collides with it at the rate $n\,A\,(w+u_1)^{2-2\omega}$, where $A\,c_r^{2-2\omega}$ is the product of the VHS cross-section and the relative speed $c_r$. If, as in Eq.~\eqref{eq:halfspace}, a
molecule is taken to leave the upstream-moving group at its first collision, the saddle-point estimate gives
$\beta\propto\exp(-\Lambda)$ with
\begin{equation}
\Lambda(N)=\min_{w}\left[\frac{(w+u_2)^2}{2T_2}+A\,N\,\frac{(w+u_1)^{2-2\omega}}{w}\right],\qquad N=\int_x^{x_c}n\,dx'.
\label{eq:phi}
\end{equation}
For $\omega=1$, and for the relaxation models with $AN$ replaced by the optical distance $\theta$, the relative
speed drops out and $\Lambda$ is the exponent of Eq.~\eqref{eq:stretched}. For other exponents, if the
minimising speed were much larger than $u_1$ the estimate would give $\Lambda\propto N^{2/(2\omega+1)}$, which is
the exponent $\lambda/(\lambda+2)$ obtained heuristically by Narasimha~\citeyearpar{Narasimha1968} for a potential
proportional to $r^{-\lambda}$ ($\omega=1/2+2/\lambda$) and the exponent of the sub-exponential remainder in a recent
existence proof for cutoff potentials~\citep{Duan2026}; if it were much smaller, the exponent $2/3$ would be
recovered for every potential. In strong shocks the minimising speed is 0.9--1.1\,$u_1$, so that neither limit
applies and Eq.~\eqref{eq:phi} is evaluated numerically. A self-similar analysis for infinite Mach number and
forces with an infinite total cross-section~\citep{PomeauTran2019} leads to a different behaviour, which the VHS
model, with its finite cross-section, does not represent.

\begin{figure}[!htbp]
\includegraphics[width=\textwidth]{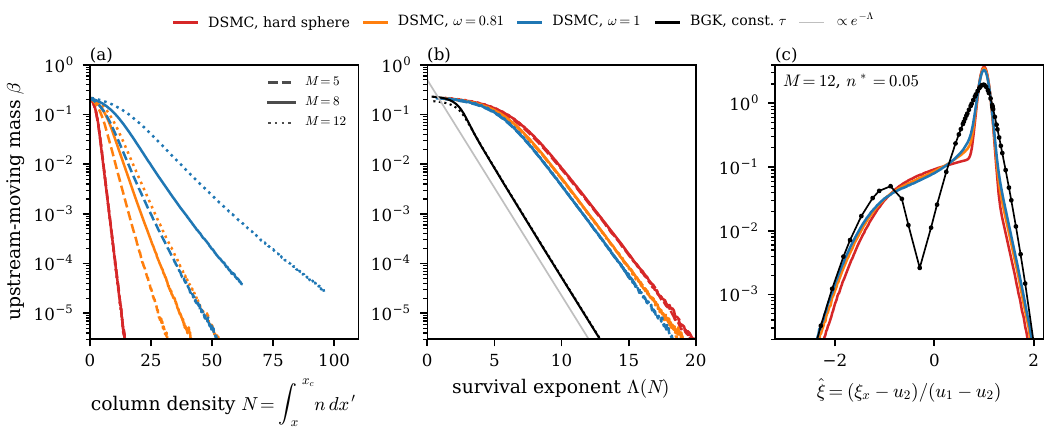}
\caption{Precursor in DSMC. (a) Upstream-moving mass fraction against the column density upstream
of the shock centre at Mach 5, 8 and 12. (b) The same against the survival exponent $\Lambda$ of Eq.~\eqref{eq:phi};
the black lines are the BGK solutions at Mach 8 and 12, for which the optical distance takes the place of $AN$. (c) Distribution at Mach 12 and $n^*=0.05$;
symbols mark the velocity nodes of the BGK solution.}
\label{fig:dsmcprec}
\end{figure}

Figure~\ref{fig:dsmcprec} tests the estimate at Mach 5, 8, and 12; Mach 6 and 10 behave in the same way. Plotted against $\Lambda$, the curves for the different Mach numbers coincide for each $\omega$ and are straight between $\beta=10^{-2}$ and
$2\times10^{-5}$ (root-mean-square deviation of $\ln\beta$ from a straight line 0.02--0.03). The slope is $-1.01$
for the BGK solutions at Mach 8 and 12, as it must be for a model in which survival is exact, and $-0.80$ to
$-0.84$ in DSMC for the three values of $\omega$ and the five Mach numbers. The precursor of the gas therefore
decays about 20\% more slowly in the exponent than survival predicts, by nearly the same factor whatever the
exponent $\omega$.

\begin{figure}[!htbp]
\includegraphics[width=\textwidth]{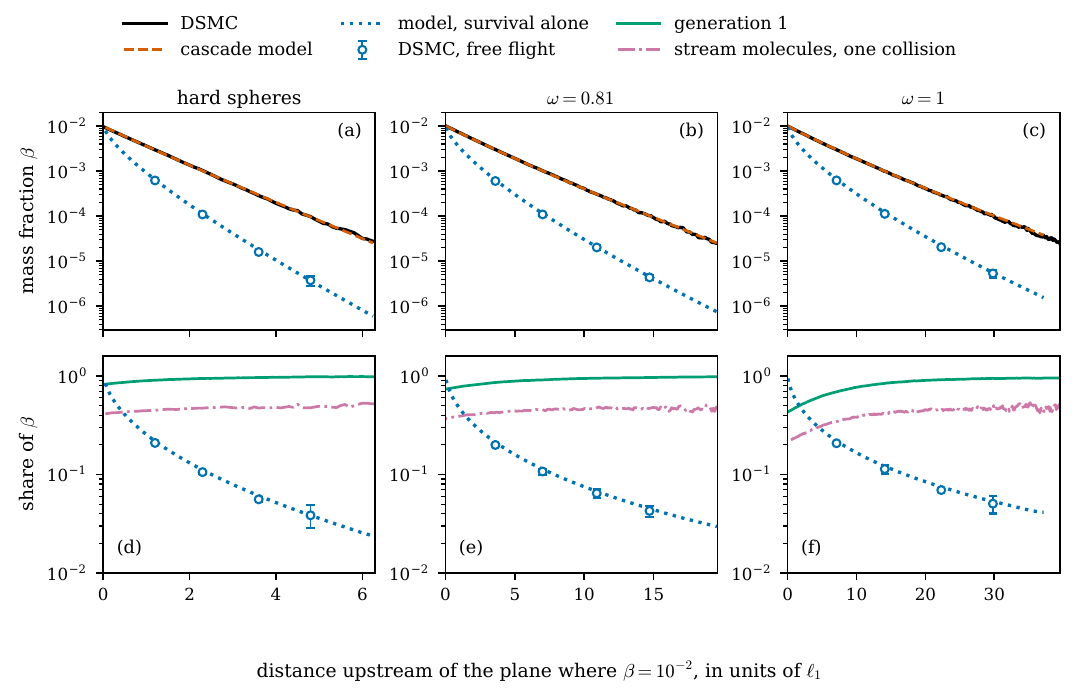}
\caption{Survival and recoil in the precursor of the gas at Mach 8, ahead of the plane at which $\beta=10^{-2}$: hard spheres, $\omega=0.81$, and $\omega=1$. (a)--(c) Upstream-moving mass fraction $\beta$ in DSMC (solid) and in the linear cascade model driven by the DSMC flux measured at that plane (dashed; the source is the measured flux, so that the comparison tests the transport ahead of the plane); mass fraction of the molecules that have come from the plane without any collision, in the model (dotted, survival alone) and in DSMC (circles: molecules carrying the free-flight flag, with the 95\% interval of four runs). (d)--(f) Shares of the upstream-moving molecules: free flight in the model (dotted) and in DSMC (circles); generation 1, whose most recent collision involved an unscattered stream molecule; and stream molecules after exactly one collision. Survival alone falls below the precursor by a factor of ten within one decade of its decay; the difference is made up by the products of collisions with the stream.}
\label{fig:mechanism}
\end{figure}

The reason is that a collision in a gas does not remove a molecule from the precursor. In a second series
of DSMC runs every simulated molecule carries a record of its collisions (Appendix~\ref{app:labels}). Where
$\beta=10^{-3}$, the most recent collision of 94\% of the upstream-moving molecules of the hard-sphere gas
(91--92\% for $\omega=0.81$ and 84\% for $\omega=1$) involved an unscattered molecule of the incoming
stream, and where $\beta=10^{-4}$ this holds for 94--98\%. A flag that marks the molecules crossing the plane where $\beta=10^{-2}$ and is removed at their next collision
shows that only 10.6--11.5\% of the upstream-moving molecules at the position where $\beta=10^{-3}$, and
3--6\% where $\beta=10^{-4}$, have come from that plane without a collision (figure~\ref{fig:mechanism}). With isotropic
scattering 41--47\% of the upstream-moving molecules are molecules of the stream itself, turned back by a single
collision. The composition of the precursor is thus that of a cascade of recoils fed by the few molecules that survive, and not that of a group of survivors (table~\ref{tab:composition}, Appendix~\ref{app:labels}).

\begin{figure}[!htbp]
\includegraphics[width=\textwidth]{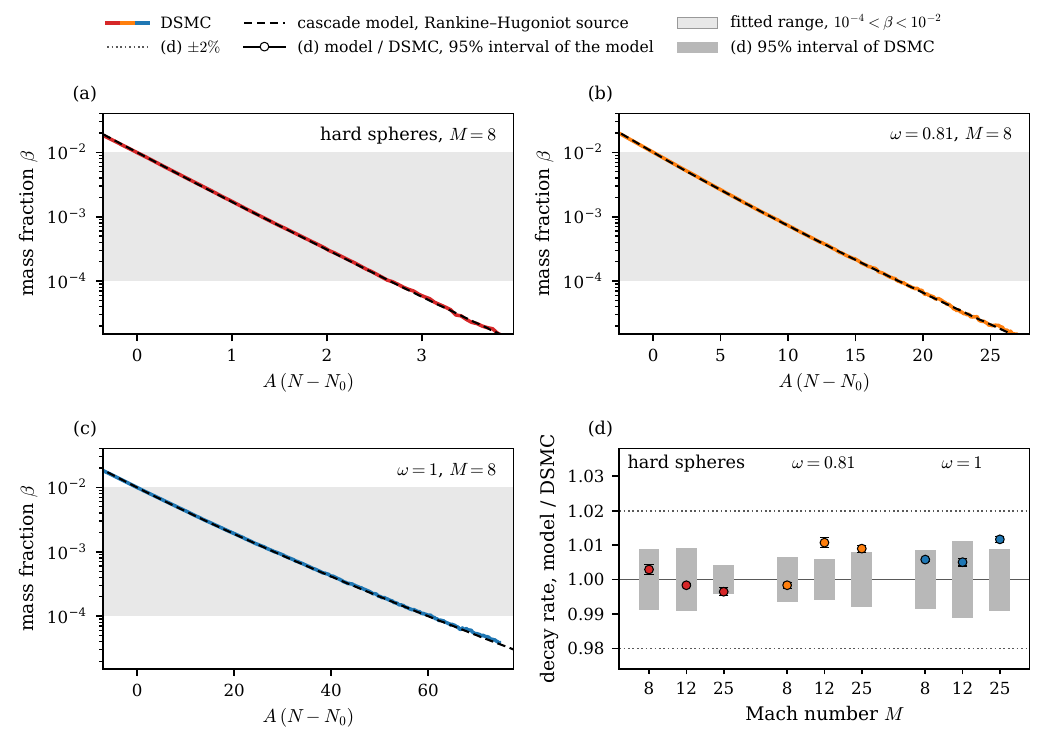}
\caption{Decay of the precursor in DSMC and in the linear cascade model whose source is a half-range Maxwellian of the downstream Rankine--Hugoniot state. (a)--(c) Upstream-moving mass fraction $\beta$ at Mach 8 for hard spheres, $\omega=0.81$, and $\omega=1$: DSMC (solid, eight runs pooled) and the model (dashed, mean of four runs) against $A(N-N_0)$. Here $N$ is the column density upstream of the shock centre in DSMC and the depth below the source plane in the model, and $N_0$ is its value where $\beta=10^{-2}$ on each curve; the model does not fix the position of its source plane relative to the shock, and the decay rate does not depend on this origin. The shaded band is the range $10^{-4}<\beta<10^{-2}$ over which the decay rate is fitted. (d) Ratio of the decay rate of the model to that of DSMC at Mach 8, 12, and 25, from the values of table~\ref{tab:cascade}: symbols, ratio with the 95\% interval of the model mean; grey bars, 95\% interval of the DSMC mean; dotted lines, $\pm2\%$.}
\label{fig:prediction}
\end{figure}

A linear model of this cascade, in which molecules enter a uniform stream and both partners of every collision are followed, was used in two ways. In the first, the source is a half-range Maxwellian of the downstream Rankine--Hugoniot state, so that nothing from the interior of the DSMC shock enters: this is an independent prediction, and figure~\ref{fig:prediction} shows it. At Mach 8 the model follows the profile of $\beta$ in DSMC over the two decades in which the decay rate is measured, for each of the three collision laws; the two curves share only the origin of the column density, placed where $\beta=10^{-2}$. For the nine cases at Mach 8, 12, and 25 the ratio of the decay rate of the model to that of DSMC lies between 0.996 and 1.012 (figure~\ref{fig:prediction}(d); Appendix~\ref{app:labels}, table~\ref{tab:cascade}): in six cases, among them all three of the hard-sphere gas, the model lies within the 95\% interval of the simulations, and in the other three it is higher by 0.9--1.2\%. Over the measured range the model thus reproduces the decay rates of the simulations, without adjustable parameters, from the downstream state and the collision law alone. In the second, the source is the flux of upstream-moving molecules measured in DSMC at the plane where $\beta=10^{-2}$ (figure~\ref{fig:mechanism}): the flagged molecules of the simulation then follow the model's curve of survival alone, and the precursor itself, which decays far more slowly, is reproduced when the products of collisions are followed. This second comparison is conditioned on the measured source and tests the transport ahead of the plane, that is, the collision kinematics and the linearity of the cascade. In both uses the model is a branching calculation that does not include the loss of the struck stream molecules and stops following molecules that have become slow relative to the stream; the sensitivity to the stopping rule was tested at Mach 8 and 12 (Appendix~\ref{app:labels}). With a forward-peaked
scattering law at the same viscosity the decay rate changes by at most 3\%, but fewer of the upstream-moving
molecules are recoiling stream molecules.

Fitted as $\ln\beta=a-bN^{p}$, the decay has $p=0.87\pm0.03$ for hard spheres, $0.73\pm0.02$ for $\omega=0.81$,
and $0.68\pm0.02$ for $\omega=1$ (mean over 40 independent runs and its 95\% interval; details and tests of the
fitting window and of the domain length are in Appendix~\ref{app:labels}). These are effective exponents over the resolved window of 2.7 decades of $\beta$. They increase with the hardness of the molecules, and the value for $\omega=1$ is consistent
with the $2/3$ of the relaxation models; they describe the resolved window, and the value for $\omega=0.81$ lies between the heuristic 0.76 and $2/3$. The data establish the dependence on $\omega$ and the slower decay relative to survival. The
precursor of hard spheres is much shorter than that of the softer molecules: at Mach 10, $\beta$ falls to
$10^{-5}$ within a column density of 13 for hard spheres, against 42 for $\omega=0.81$ and 90 for $\omega=1$. This
accounts for an observation of Yuan and Wu~\citeyearpar{YuanWu2022}: at Mach 5 the Shakhov model reproduces the
Boltzmann shock structure well for Maxwell molecules, but overpredicts the upstream temperature and heat flux
increasingly as $\omega$ decreases to the hard-sphere value. A relaxation model with a velocity-independent
collision frequency has the precursor of Maxwell molecules whatever its viscosity law, and, since it replaces every molecule that collides by one drawn from a target distribution fixed by a few local moments, it has no recoils.

\FloatBarrier
\section{Comparison with a many-kernel representation in physical variables}
\label{sec:baseline}

The preceding sections describe the distribution at each position by a few Gaussian populations and predict it across Mach number in scaled variables. Gaussians can also be used in a different way: as a compact representation of the whole phase-space function $f(x,\bm\xi)$, in which several hundred kernels act as basis functions without individual meaning. Such a representation of the same BGK shocks was studied by \citet{RoohiTransport}. Its results are summarised here because they are the reference against which the predictions of Sec.~\ref{sec:predictor} are judged; the model, its optimisation, the error measures, and the complete set of tests are reproduced in the supplementary material.

A positive mixture of $N=512$ diagonal Gaussians in the normalised coordinates $(x,\xi_x,\xi_y,\xi_z)$, with 4608 parameters fitted to $\log f$ of one DVM solution, reproduces that solution closely. For the Mach-5 shock the relative $L_2$ errors over the domain, averaged over three random initialisations, are 0.1--0.2\% in $\rho$, $u_x$, and $T$ and 1.6--1.8\% in the heat flux, the stress, and the third- and fourth-order nonequilibrium moments, the width of the density profile is within 2\% of the reference, and its position within 1.4\% of the shock thickness; multilinear grids with the same number of coefficients are in error by 79--104\% in the transport and higher moments.

\begin{table}
\caption{Many-kernel representation in physical variables \citep[results of][]{RoohiTransport}: leave-one-Mach-out errors in percent, mean $\pm$ sample standard deviation over three random initialisations. M6 moments are evaluated against the fine-grid reference. D2 and D3 denote global degree-2 and degree-3 maps; K3 and K4 denote three- and four-knot correspondence-preserving local maps.}
\label{tab:mach_holdout}
\centering
\scriptsize
\resizebox{\textwidth}{!}{%
\begin{adjustbox}{max width=\linewidth}
\begin{tabular}{llcccccccc}
\toprule
Case & map & $E_f$ & $E_\rho$ & $E_{u_x}$ & $E_T$ & $E_{q_x}$ & $E_{\sigma_{xx}}$ & $E_{M_{300}^{\rm neq}}$ & $E_{M_{400}^{\rm neq}}$\\
\midrule
M3 & global D2 & $11.18\pm1.38$ & $13.86\pm0.44$ & $3.07\pm0.31$ & $8.83\pm0.09$ & $23.16\pm3.70$ & $24.34\pm4.86$ & $31.11\pm3.53$ & $29.13\pm2.20$\\
M3 & global D3 & $11.04\pm0.40$ & $14.67\pm0.05$ & $3.75\pm0.21$ & $8.59\pm0.14$ & $36.38\pm4.69$ & $30.50\pm3.85$ & $41.81\pm5.42$ & $40.71\pm2.75$\\
M6 & global D2 & $41.90\pm6.24$ & $17.50\pm1.52$ & $4.85\pm0.80$ & $5.79\pm0.11$ & $54.24\pm3.39$ & $73.61\pm9.45$ & $62.28\pm8.71$ & $72.54\pm11.87$\\
M6 & local K3 & $21.85\pm4.06$ & $24.68\pm2.98$ & $5.58\pm1.78$ & $7.50\pm1.34$ & $66.24\pm12.75$ & $70.24\pm32.91$ & $77.17\pm7.04$ & $83.10\pm11.26$\\
M6 & global D3 & $42.86\pm5.40$ & $21.42\pm1.76$ & $6.85\pm0.61$ & $7.37\pm0.15$ & $38.30\pm7.72$ & $56.54\pm5.38$ & $46.61\pm7.80$ & $46.31\pm6.57$\\
M6 & local K4 & $11.45\pm0.94$ & $6.96\pm0.76$ & $3.05\pm0.04$ & $5.53\pm0.70$ & $39.50\pm3.52$ & $37.98\pm7.02$ & $39.05\pm5.87$ & $30.11\pm2.60$\\
\bottomrule
\end{tabular}
\end{adjustbox}%
}
\end{table}

The representation behaves differently over a range of shock strengths. With every kernel parameter a polynomial or piecewise-linear function of Mach number, fitted jointly to eight of the nine BGK shocks, a withheld shock is recovered with larger errors (table~\ref{tab:mach_holdout}): at Mach 3 with 11\% in the distribution and 23--42\% in the transport and higher moments, and at Mach 6, with the best of the tested models, with $11.5\pm0.9\%$ in the distribution and 30--40\% in the heat flux, the stress, and the fourth moment. Mach 12 lies outside the interval fitted with Mach numbers up to 8. In physical variables the distribution changes with Mach number mainly through the displacement and narrowing of its populations, which follow the Rankine--Hugoniot velocities and the downstream temperature, and the 5120 parameters of the correlated kernels have to be fitted to this dependence from eight solutions. In the scaled variables twelve numbers per position, combined linearly in $1/M^2$ from two Mach numbers, give the withheld Mach-6 shock to within 1.6\% in the distribution and 3.3\% in the heat flux, and the Mach-12 shock to within 4.7\% and 4.3\% (table~\ref{tab:predictor}). The two descriptions are complementary: the many-kernel mixture represents one given solution in full, and the scaled description carries over from one shock strength to another.

\FloatBarrier
\bigskip
\section{Discussion}

\subsection{Gas and model}

The principal reference data of this study are BGK solutions with a constant relaxation time. Sections~\ref{sec:models} and \ref{sec:dsmc} show what this choice affects. The adequacy of three populations, the streamwise reduced distribution at the shock centre ($n^*=0.5$), and the amount of upstream-moving mass inside the shock are common to the BGK, Shakhov, and ES-BGK models and to DSMC. The size of the departure from two Maxwellians, the separate upstream maximum, the peak temperature anisotropy, and the composition and decay of the precursor depend on the collision law, and the constant-relaxation-time BGK model differs most from the gas in each of these respects. Accuracies quoted for the BGK solutions therefore apply to that model; the hard-sphere and variable-hard-sphere distributions are closer to two Maxwellians and are represented more accurately by a given number of populations (table~\ref{tab:dsmc}). The same distinction applies to the populations themselves. Their number, their parameters, and their shares of the transport are reproducible properties of the distribution, but only the beam of a hard-sphere gas is a set of molecules with a common history (Sec.~\ref{sec:meaning}); the word population is used for the other two in the sense of an effective population throughout.

\subsection{Similarity, prediction, and the role of the Gaussian form}

The central result for prediction is the similarity of Sec.~\ref{sec:sim}: in the variables $(n^*,\hat\xi)$ and with temperatures in units of $T_2$, the distribution is nearly independent of Mach number above Mach 5 (figure~\ref{fig:params}). Section~\ref{sec:methods} separates this result from the three-population form in which it was first used. For strong shocks the scaled reduced distributions themselves, combined linearly in $1/M^2$, give the moments at equal $n^*$ more accurately than the populations do, and give a monotone profile in every test; the exception is the beam, whose width in the scaled velocity depends on Mach number, and a single Gaussian for the beam accounts for it (table~\ref{tab:methods}). The accuracy of the prediction therefore comes from the scaled variables, and what the Gaussian form adds is of a different kind: twelve numbers per position in place of a tabulated function, parameters that can be read as densities, velocities, and temperatures, a smoother dependence on Mach number when a source is a weak shock, and the separation of the one part of the distribution that does not scale. The accuracy of the three-population predictor is that of the three-population representation itself: for strong shocks it equals that of a direct fit of the target, and no profile is obtained where the populations of the sources overlap and their parameters do not correspond. In every method the positions follow from the moment equation~\eqref{eq:moment}, which is exact for the relaxation models and for Maxwell-type molecules and gives the hard-sphere shock width to within 5--8\%; this difference is common to the three methods and is the largest one of the hard-sphere predictions in physical space.

\subsection{Many kernels and few populations}

The 512-kernel mixture of Sec.~\ref{sec:baseline} and the few-population description serve different purposes. The mixture represents the whole phase-space function of one solution, including its variation in $x$, and reaches 1--2\% in the transport moments of a fitted state; it interpolates accurately between positions within one shock (supplementary material). The few-population description represents one position at a time with 4 to 16 numbers and reaches comparable accuracy in the heat flux and stress with three populations (figure~\ref{fig:map}). Between shock strengths the two behave differently. With kernel parameters that depend on Mach number in physical variables, the withheld Mach-3 and Mach-6 states have errors of 23--83\% in the transport and higher-order moments, and Mach 12 lies outside the fitted range (table~\ref{tab:mach_holdout}), whereas the predictions in scaled variables reach a few percent both between and beyond the source Mach numbers. The 512-kernel model differs from the predictions of Sec.~\ref{sec:predictor} also in its functional form, in the number of its parameters, and in the way its dependence on Mach number is fitted, so that the comparison is between the two descriptions as a whole; it suggests that compact representations that depend on Mach number be built in the scaled variables. On the upstream side at Mach 5 and above, a residual of 1--4\% remains with three or four Gaussians, in the region where the distribution next to the beam is singular in the strong-shock limit~\citep{Cercignani1999,Takata2000}.

\subsection{Scope}

The many-kernel representation was evaluated on BGK solutions. The DSMC gas is the variable-hard-sphere model with isotropic scattering; one test with the variable-soft-sphere law is reported in Appendix~\ref{app:labels}. The accuracies quoted for the few-population description refer to the reduced distributions $g$ and $h$ and to the moments tested; the transverse structure of the distribution is examined in Sec.~\ref{sec:meaning}. The predictions use kinetic solutions of the same gas at two Mach numbers and apply when both are strong shocks (Sec.~\ref{sec:predictor}). The exponents of the precursor are effective exponents over the 2.7 decades of the upstream-moving mass fraction that the simulations resolve (Sec.~\ref{sec:precgas}); the collision record and the cascade model on which the interpretation rests are summarised in Appendix~\ref{app:labels}. Gaussian kinetic solvers, in which the mixture parameters are obtained by minimising the residual of the kinetic equation and not by fitting its solution, are a natural extension.

\section{Conclusions}

The velocity distribution in a normal shock has been examined with discrete-velocity solutions of the BGK, Shakhov, and ES-BGK models up to Mach 12 and with DSMC of hard-sphere and variable-hard-sphere gases up to Mach 25, including runs in which every molecule carries a record of its collisions. Four conclusions follow.

First, the distribution is not a mixture of the two Rankine--Hugoniot Maxwellians, and the departure grows with Mach number: at Mach 12 it reaches 50\% of the streamwise reduced distribution for the BGK model and 14--30\% in DSMC, and at Mach 25 it is 16--37\% in DSMC. One further population is sufficient. Three anisotropic Gaussian populations, a beam, an intermediate population, and a hot population (figure~\ref{fig:overview}), reproduce the streamwise reduced distribution of the BGK model and of DSMC to within 4\% up to Mach 12, and of DSMC to within 6\% at Mach 25, and reduce the errors of two populations in the heat flux and the fourth-order moment by factors of up to 12. The collision records show what these populations are. For hard spheres the beam is the unscattered stream. The other two are effective populations: molecules that have collided once with the stream are assigned mostly to the hot population, so that the intermediate population describes the shape of the distribution and is not a class of molecules with a common history. Divided in the same way, the largest viscous stress of a hard-sphere shock is carried in nearly equal parts by the unscattered stream and by the scattered molecules, at every Mach number from 5 to 25, whereas for Maxwell-type molecules the unscattered stream carries a few percent of it, although the largest stress is nearly the same in the two gases. The heat flux carried by the unscattered stream is directed downstream, against the net heat flux, which is carried by the scattered molecules.

Second, the collision law matters most on the upstream side of the shock. At the centre ($n^*=0.5$) the streamwise reduced distributions of all models and of DSMC agree closely; the temperature anisotropy, which involves the transverse velocities, depends on the collision law throughout (table~\ref{tab:dsmc}). On the upstream side the hot population of the gas has the downstream velocity and is hotter than the downstream gas. In the BGK and ES-BGK models it is replaced there by a velocity-filtered group of upstream-moving molecules, which produces a separate maximum of the distribution that DSMC does not show. This maximum is therefore a property of those models and not of the gas. The BGK model with constant relaxation time also overestimates the peak temperature anisotropy, and the Shakhov model with the viscosity law of the gas is closer to DSMC.

Third, above Mach 5 the distribution in variables scaled with the Rankine--Hugoniot states is nearly independent of Mach number. Closed with one moment equation for the positions, this similarity turns two computed shocks into a prediction of a third: the Mach-12 shock is obtained from Mach 6 and 8 with three populations to within 5\% in the distribution, the heat flux, and the stress. At equal normalised density the scaled distributions combined directly give the heat flux and the stress of the Mach-25 shock from Mach 8 and 12 to within 0.3\%, which is within the numerical accuracy of the reference simulations; as functions of position the differences are 0.4--0.6\% for Maxwell-type molecules and 5\% for hard spheres. The positions follow from a moment equation that is exact for the relaxation models and for Maxwell-type molecules and gives the hard-sphere shock width to within 5--8\%, which accounts for the difference in physical space. The corrections to the strong-shock form are linear in $1/M^2$, as the kinetic equation in the scaled variables suggests. Comparing three ways of using the similarity shows where the accuracy comes from. The scaled distributions combined directly give the moments of strong shocks more accurately than the three populations do, and the one part of the distribution that does not scale, the beam, is represented by a single Gaussian. The accuracy is thus a consequence of the scaling; the Gaussian form contributes compactness and parameters that can be read as densities, velocities, and temperatures. The same comparison explains the behaviour of a many-kernel representation in physical variables \citep{RoohiTransport}, which reproduces a fitted shock closely and recovers withheld shock strengths with errors of tens of percent: in physical variables the dependence on Mach number has to be fitted from the available solutions, and in the scaled variables almost none remains.

Fourth, the upstream precursor of the gas does not consist of molecules that survive from the hot side. For the relaxation models the decay of the upstream-moving mass follows the survival law that leads to the two-thirds exponent of Lyubarskii and Narasimha. In DSMC the decay over the resolved range is a stretched exponential with effective exponents between 0.68 for Maxwell-type molecules and 0.87 for hard spheres, slower than survival predicts. The collision records give the reason: a collision in a gas does not remove a molecule, and where the upstream-moving fraction has fallen to $10^{-3}$ the most recent collision of 84--94\% of the upstream-moving molecules involved an unscattered molecule of the incoming stream. A linear cascade model in which both partners of every collision are followed, with a half-range Maxwellian of the downstream state as its only input, reproduces the decay rates to within 1.2\% without adjustable parameters (figure~\ref{fig:prediction}). The precursor is thus regenerated by collisions with the stream, and over the resolved range its decay rates are reproduced from the collision law and the downstream state alone, without data from the interior of the shock.

Together these results replace the two-Maxwellian picture of a strong shock by a description tied to the collision history of the molecules, and they indicate that reduced descriptions of shocks over a range of Mach numbers should be built in the Rankine--Hugoniot-scaled variables.

\appendix
\section{Details of the population fits}

The residual minimised at each position is
$\|g-\sum_kg_k\|_2^2/\|g\|_2^2+\|h-\sum_k2T_{\perp k}g_k\|_2^2/\|h\|_2^2$, with norms weighted by the
quadrature weights of the velocity grid. The parameters are $\ln n_k$, $u_k$, $\ln T_{\parallel k}$
and $\ln T_{\perp k}$, which keeps densities and temperatures positive. Positions are chosen
uniformly in $n^*$ between 0.002 and 0.998 (60 positions) and uniformly in $x$ over the same
interval (40 positions). The continuous $K=3$ branch has a residual larger than 1.5 times the best
fit at 7 of 81 positions for Mach 4, at 1 of 82 for Mach 5 and at none for the other Mach numbers.
The Shakhov distributions take small negative values, which a mixture of positive Gaussians cannot
represent; this contributes to their larger residuals.

\section{Collision record, cascade model, and statistics of the precursor}
\label{app:labels}

This appendix gives the methods behind the statements of Secs.~\ref{sec:meaning} and \ref{sec:precgas} that
rest on collision histories.

\emph{Collision record.} In a second series of DSMC runs (Mach 2, 3, 5, 8, and 12 with four runs per case, and Mach 25 with eight) each
simulated molecule carries two labels. They belong to the individual molecule, are updated at each of its
collisions, and are sampled with its velocity; they do not affect the dynamics: with the same random numbers
the sampled distributions are identical to those of the unlabelled code. Molecules created in the upstream
reservoir, and in the upstream cell that is refilled when the frame is shifted to keep the shock in place,
start as unscattered molecules of the stream; molecules created in the downstream reservoir or in the
refilled downstream cell start in the merged class defined below. The first label is a \emph{generation}. It
is 0 as long as the molecule belongs to the incoming stream and has collided only with other generation-0
molecules; in any other collision both partners receive the smaller of their two generations plus one, with
all values above 2 merged. The generation of a molecule is therefore fixed by its most recent collision
alone. Generation 1 means that this collision involved an unscattered molecule of the stream (either the
molecule itself, until then, or its partner); generation 2 that it involved a generation-1 molecule and no
unscattered one; the merged class that both partners were of generation 2 or higher. The label measures how
far, in collisions, a molecule is from the most recent contact with the unscattered stream. It is not a record
of the whole history: the sequence $(0,3)\to(1,1)$, $(1,1)\to(2,2)$, $(2,2)\to(3,3)$ returns a molecule to the
merged class after a collision with the stream, so that membership of the merged class does not mean that a
molecule has never collided with the stream. A flag for having \emph{ever} collided with the stream would not
discriminate in a steady shock, in which every molecule, apart from the few that enter from the downstream
reservoir, was itself once a molecule of the stream. The second label is a \emph{collision count} for
molecules of the stream: 0 for an unscattered molecule, 1 after its first collision with a scattered
molecule, and one more after every later collision. The reduced distribution, density, mean velocity, and
temperatures of every class are sampled in each cell. A label changes in every collision with a scattered
molecule, whatever the momentum exchanged. For hard spheres collisions at low relative speed are rare and a
change of label marks a real deflection; for Maxwell-type molecules they are as frequent as any other, so
that the label spreads through the stream without a corresponding change of velocities, and the mass
fractions of the classes then depend on the length of the domain. At Mach 8, halving the cell and the time
step changes the mass fractions of the generations by at most 0.005, using 50 molecules per cell by 0.012,
and halving the upstream domain by 0.002 for hard spheres and 0.010 for $\omega=0.81$.

\emph{Free-flight flag.} The generation does not say where the most recent collision took place. A third
label, used in a further series of runs at Mach 8 and 12 (four runs for each $\omega$), does: a molecule is flagged
when it crosses, moving upstream, the plane at which $\beta=10^{-2}$, and the flag is removed at its next collision,
whatever the partner. A flagged molecule has therefore flown from that plane to its present position without any
collision. This is the group that the survival estimate of Eq.~\eqref{eq:halfspace} takes to be the whole
precursor ahead of the plane.

\begin{table*}
\centering\small
\caption{Composition of the upstream-moving molecules in DSMC at the positions where $\beta=10^{-3}$ and $10^{-4}$, in percent: generations 1, 2, and the merged class; molecules that have flown without any collision from the plane where $\beta=10^{-2}$ (free-flight flag, mean of four runs and 95\% interval); and the same quantity in the linear cascade model driven by the DSMC flux through that plane (source molecules that have not collided).}
\label{tab:composition}
\begin{adjustbox}{max width=\linewidth}
\begin{tabular}{lccccccccc}
\toprule
 & & \multicolumn{5}{c}{$\beta=10^{-3}$} & \multicolumn{3}{c}{$\beta=10^{-4}$}\\
$\omega$ & $M$ & gen.~1 & gen.~2 & merged & free flight & model & gen.~1 & free flight & model\\
\midrule
0.5 & 8 & 94.3 & 4.1 & 1.6 & $10.6\pm0.4$ & 10.9 & 98.1 & $3.9\pm1.0$ & 3.9\\
0.5 & 12 & 94.2 & 4.0 & 1.7 & $10.9\pm0.7$ & 11.1 & 97.5 & $3.3\pm1.0$ & 3.9\\
0.81 & 8 & 91.4 & 5.4 & 3.2 & $10.7\pm0.8$ & 11.3 & 96.8 & $4.3\pm0.5$ & 4.5\\
0.81 & 12 & 91.9 & 5.0 & 3.2 & $11.0\pm1.2$ & 11.6 & 96.6 & $4.5\pm0.4$ & 4.7\\
1 & 8 & 84.0 & 9.1 & 6.9 & $11.4\pm1.1$ & 12.1 & 94.5 & $5.1\pm1.0$ & 5.3\\
1 & 12 & 84.0 & 9.0 & 7.0 & $11.5\pm0.5$ & 12.5 & 94.1 & $5.8\pm0.2$ & 5.5\\
\bottomrule
\end{tabular}
\end{adjustbox}
\end{table*}

\emph{Composition of the precursor.} Table~\ref{tab:composition} lists the composition of the upstream-moving
molecules at the positions where $\beta=10^{-3}$ and $10^{-4}$. Generation 1 holds 94\% of them for hard
spheres, 91--92\% for $\omega=0.81$, and 84\% for $\omega=1$ at the first position, and 94--98\% at the second.
By the definition of the label, the most recent collision of
these molecules involved an unscattered molecule of the stream, as one of its two partners (the molecule itself
may have been that partner); nothing is assumed about earlier collisions. The remaining
6--16\% had their most recent collision among scattered molecules, which make up 0.6\%, 1.2\%, and 3.4\% of the
gas at the first position; the merged class holds 1.6--1.7\%, 3.2\%, and 7\%.
The free-flight flag measures directly how many molecules arrive without collision: of the upstream-moving
molecules at the position where $\beta=10^{-3}$, 10.6--11.5\% have flown freely from the plane where
$\beta=10^{-2}$, and at the position where $\beta=10^{-4}$, 3.3--5.8\%; all the others have collided on
the way. The linear cascade model driven by the DSMC flux through the same plane (below) gives 10.9--12.5\%
and 3.9--5.5\% for the same quantity.
The share of stream molecules with a collision count of one is 0.47 and 0.46 for hard spheres, 0.45
and 0.46 for $\omega=0.81$, and 0.41 and 0.42 for $\omega=1$. These molecules move upstream at 0.3--0.4\,$u_1$ on
average. The kinematics explain the recoils: when a molecule moving straight upstream with speed $w$ meets a
stream molecule, the streamwise velocity of each partner after isotropic scattering is uniformly distributed
between $-w$ and $u_1$, so that each moves upstream with probability $w/(w+u_1)$. With the variable-soft-sphere
law at the same viscosity (deflection angle $\chi$ from $\cos\chi=2U^{1/\alpha}-1$, $U$ uniform; Mach 12,
$\omega=0.81$) the decay rate per unit $AN$ is 0.204 for $\alpha=1.4$ and 0.197 for $\alpha=2$, against 0.198 for
isotropic scattering, and the share of the merged class is 3.7\% and 3.2\%; the share of
stream molecules turned back by one collision falls from 0.46 to 0.30 and 0.18, because forward-peaked
scattering more often leaves the incident molecule moving upstream and the stream molecule moving downstream.

\emph{Linear cascade model.} Far enough upstream almost every molecule belongs to the stream and the others
collide only with stream molecules, so that their transport is linear. The model considers a uniform stream
of density $n_1$, velocity $u_1$, and temperature $T_1$ occupying $x<0$. Source molecules enter through $x=0$
with velocities drawn from the flux through that plane of the downstream Maxwellian (density $\rho_2$, mean
velocity $u_2$, temperature $T_2$), restricted to $\xi_x<0$; their streamwise speeds are sampled from a
Maxwellian three times hotter, with weights, so that large depths are reached. A followed molecule moves freely
and collides with the stream at the VHS rate $n_1A\,c_r^{2-2\omega}$: by null-collision sampling with a
majorant rate, a partner is drawn from the stream Maxwellian and accepted with probability proportional to
$c_r^{2-2\omega}$, so that accepted partners have the collision-rate-weighted distribution. The relative velocity
is turned to a random direction and both partners are followed from then on. The density of upstream-moving
molecules is accumulated as a function of depth from the time spent in each cell. Each run follows $10^8$
source molecules and about $10^9$ collisions, and four independent runs are made for each case. Nothing is
adjusted, and nothing from the DSMC shock enters except the Rankine--Hugoniot states.

In terms of the perturbation $\psi$ of the stream Maxwellian $M_1$, the model samples the gain of both
products of a collision and the loss of the incident molecule in the linearised equation
$\xi_x\partial_x\psi=Q(M_1,\psi)+Q(\psi,M_1)$. That equation contains one more term, the loss of the stream
molecule that was struck, which is a negative perturbation with the velocity distribution of the collision
partners. It is omitted, because it does not contribute to the quantity that is tallied: the struck molecules
have the velocities of the stream, of which the fraction with $\xi_x<0$ is $\beta_1\le5.4\times10^{-11}$ at Mach 5
and above, and their further collisions with the stream, at thermal relative speeds, cannot produce
upstream-moving molecules except from the same tail. The model is thus a branching process for the
upstream-moving perturbation and not a solution for the whole of $\psi$.

\begin{table}
\centering\small
\caption{Decay rate of the precursor: least-squares slope of $-\ln\beta$ against $AN$ over the cells with $10^{-4}<\beta<10^{-2}$, mean over independent runs and 95\% interval of the mean. DSMC shocks (eight runs) and the linear cascade model with a half-range Maxwellian source of the downstream state (four runs; tally cells of $1/1280$ of the domain). The last three columns give the means of the model with tally cells eight times larger and the termination constant $c=4$, 3, and 5. In the model the column density is the depth below the source plane.}
\label{tab:cascade}
\begin{adjustbox}{max width=\linewidth}
\begin{tabular}{lcccccc}
\toprule
 & & & & \multicolumn{3}{c}{larger tally cells}\\
$\omega$ & $M$ & DSMC & cascade model & $c=4$ & $c=3$ & $c=5$\\
\midrule
0.5 & 8 & $1.721\pm0.015$ & $1.726\pm0.003$ & 1.723 & 1.722 & 1.723\\
0.5 & 12 & $1.740\pm0.016$ & $1.737\pm0.001$ & 1.738 & 1.740 & 1.739\\
0.5 & 25 & $1.754\pm0.007$ & $1.747\pm0.002$ & 1.751 & -- & --\\
0.81 & 8 & $0.2526\pm0.0017$ & $0.2522\pm0.0002$ & 0.2527 & 0.2525 & 0.2520\\
0.81 & 12 & $0.1964\pm0.0012$ & $0.1985\pm0.0003$ & 0.1983 & 0.1983 & 0.1983\\
0.81 & 25 & $0.1257\pm0.0010$ & $0.1268\pm0.0001$ & 0.1266 & -- & --\\
1 & 8 & $0.07603\pm0.00064$ & $0.07647\pm0.00003$ & 0.07745 & 0.07748 & 0.07747\\
1 & 12 & $0.05135\pm0.00058$ & $0.05161\pm0.00006$ & 0.05218 & 0.05217 & 0.05217\\
1 & 25 & $0.02476\pm0.00022$ & $0.02504\pm0.00002$ & 0.02523 & -- & --\\
\bottomrule
\end{tabular}
\end{adjustbox}
\end{table}

A followed molecule is dropped when it returns through $x=0$, when it leaves the domain, or when its speed
relative to the stream has fallen below $u_1-c\sqrt{T_1}$ with $c=4$. The last rule is an approximation, since the stream Maxwellian has unbounded velocities. A molecule can
move upstream only if its speed relative to the stream exceeds $u_1$, and a collision between a molecule of
relative speed $w$ and a stream molecule of thermal speed $c'$ leaves both products with a relative speed of at
most $(w^2+c'^2)^{1/2}$; a dropped molecule could therefore produce an upstream-moving one in a single collision
only with a partner of thermal speed above $(2cu_1\sqrt{T_1}-c^2T_1)^{1/2}$, which is $6.0\sqrt{T_1}$ at Mach 5 and
$8.2\sqrt{T_1}$ at Mach 8. This bounds a single collision only; the cumulative effect of several collisions is
not bounded by the argument and was assessed numerically, by repeating the calculations at Mach 8 and 12 with
$c=3$ and $c=5$: the decay rates change by at most 0.26\% (from $0.2527\pm0.0002$ to $0.2520\pm0.0017$ at Mach 8, $\omega=0.81$, with $c=5$), and each of the twelve changes is smaller than the combined 95\% interval of the two values compared (table~\ref{tab:cascade}). The tally cells matter more for the soft molecules, whose fitted range spans few cells: with cells eight times smaller the rates change by at most 0.2\% for hard spheres and $\omega=0.81$ and decrease by 0.7--1.3\% for $\omega=1$; the smaller cells are used for the comparison with DSMC.

Table~\ref{tab:cascade} compares the decay rates. Because the decay is not exponential, a single rate
describes a window: each entry is the least-squares slope of $\ln\beta$ against $AN$ over all cells with
$10^{-4}<\beta<10^{-2}$, with equal weights, computed for each independent run; the mean over the runs and the
95\% interval of the mean are listed. For hard spheres at the three Mach numbers, for $\omega=0.81$ at Mach 8, and for $\omega=1$ at Mach 8 and 12
the model and the simulations agree within these intervals; for $\omega=0.81$ at Mach 12 and 25 and for
$\omega=1$ at Mach 25 the model is higher by 1.1\%, 0.9\%, and 1.2\%. A difference of this size is consistent
with the replacement of the source by a half-range Maxwellian. This is the independent test of the model: its source is fixed by the Rankine--Hugoniot states alone. When the model is driven instead by the
flux of upstream-moving molecules measured in the DSMC shock at the position where $\beta=10^{-2}$, it
reproduces $\beta$ over the next two decades with a root-mean-square difference of 0.01--0.02 in $\ln\beta$,
which compares the whole profile and not one rate; this is a test of the transport conditioned on the measured source, and the comparisons of
table~\ref{tab:composition} and figure~\ref{fig:mechanism} are of this second kind.

\begin{table}
\centering\small
\caption{Effective exponent $p$ of $\ln\beta=a-bN^{p}$ in DSMC for each Mach number. Mean over eight independent runs with the 95\% interval of the mean, and exponent fitted to the pooled data of the eight runs, for $10^{-2}>\beta>2\times10^{-5}$; means over the runs for the windows $10^{-2}$ to $2\times10^{-4}$ and $3\times10^{-3}$ to $2\times10^{-5}$.}
\label{tab:exponents}
\begin{adjustbox}{max width=\linewidth}
\begin{tabular}{lccccc}
\toprule
$\omega$ & $M$ & runs & pooled & upper window & lower window\\
\midrule
0.5 & 5 & $0.88\pm0.07$ & 0.90 & 0.89 & 0.87\\
 & 6 & $0.84\pm0.06$ & 0.88 & 0.88 & 0.79\\
 & 8 & $0.90\pm0.10$ & 0.92 & 0.87 & 0.90\\
 & 10 & $0.86\pm0.09$ & 0.88 & 0.94 & 0.81\\
 & 12 & $0.89\pm0.07$ & 0.90 & 0.88 & 0.86\\
 & 25 & $0.93\pm0.07$ & 0.96 & 0.90 & 0.93\\
0.81 & 5 & $0.70\pm0.05$ & 0.74 & 0.78 & 0.63\\
 & 6 & $0.73\pm0.04$ & 0.76 & 0.77 & 0.71\\
 & 8 & $0.73\pm0.07$ & 0.77 & 0.78 & 0.69\\
 & 10 & $0.74\pm0.06$ & 0.76 & 0.72 & 0.73\\
 & 12 & $0.73\pm0.09$ & 0.75 & 0.75 & 0.71\\
 & 25 & $0.79\pm0.08$ & 0.82 & 0.73 & 0.81\\
1 & 5 & $0.72\pm0.06$ & 0.74 & 0.68 & 0.72\\
 & 6 & $0.66\pm0.05$ & 0.69 & 0.66 & 0.61\\
 & 8 & $0.68\pm0.03$ & 0.67 & 0.68 & 0.65\\
 & 10 & $0.69\pm0.07$ & 0.71 & 0.68 & 0.66\\
 & 12 & $0.68\pm0.06$ & 0.67 & 0.71 & 0.66\\
 & 25 & $0.62\pm0.07$ & 0.67 & 0.70 & 0.56\\
\bottomrule
\end{tabular}
\end{adjustbox}
\end{table}

\emph{Form of the decay.} The fit $\ln\beta=a-bN^{p}$ with $p$ free was made separately for each of the eight
independent runs of every case, for $10^{-2}>\beta>2\times10^{-5}$ (2.7 decades). The exponent $p$ is an
effective exponent over this resolved window. Table~\ref{tab:exponents} lists, for every Mach number, the mean
over the eight runs with the 95\% interval of that mean, the exponent fitted to the pooled data of the eight
runs, and the means for two other windows. For a single Mach number the interval is $\pm0.03$ to $\pm0.10$, and
none of the three gases shows a trend with Mach number that exceeds it. Over the five Mach numbers between 5 and 12 (40 runs for
each gas) the mean exponent and the 95\% interval of the mean are $0.87\pm0.03$ for hard spheres, $0.73\pm0.02$
for $\omega=0.81$, and $0.68\pm0.02$ for $\omega=1$; this interval measures the scatter between runs and not the
bias of the window, which is shown by the last two columns: the other windows change the means by up to 0.04. At Mach 25 the exponents of the eight runs are $0.93\pm0.07$, $0.79\pm0.08$, and $0.62\pm0.07$, which differ from these means by less than the interval of a single Mach number.
Within this range of $\beta$ a pure exponential ($p=1$) fits the hard-sphere
data worse than $p=0.87$, and for $\omega=0.81$ the difference between the fitted 0.73 and $2/3$ is about three times the 95\% interval but of the same size as the dependence on the window. The cascade model, which reaches
$\beta=10^{-7}$, gives $p=0.89$--0.92 for hard spheres, 0.79 for $\omega=0.81$, and 0.72 for $\omega=1$, with a
local decay rate that keeps decreasing slowly; the asymptotic law lies beyond the range of both.
All DSMC precursor results use the total mass fraction $\beta$; at Mach 5 and above the contribution of the
upstream equilibrium, $\beta_1\le5.4\times10^{-11}$, is far below the window.

\emph{Length of the domain.} At Mach 8 the simulations were repeated with upstream domains of 120\,$\ell_1$ for
$\omega=0.81$ and 180\,$\ell_1$ for $\omega=1$ in place of 60\,$\ell_1$ (eight runs each). For $\omega=0.81$,
$\ln\beta$ at a given column density changes by 0.008 in the root mean square above $\beta=2\times10^{-4}$ and by
0.03 down to $2\times10^{-5}$, where the statistical noise dominates, and the exponent fitted to the pooled
data is 0.77 with either domain. For $\omega=1$, whose precursor extends beyond the standard domain, the change is 0.02 above $\beta=2\times10^{-4}$, and the exponent rises from 0.67 to 0.70.

\FloatBarrier

\begin{bmhead}[Supplementary material.]
The supplementary material contains the complete account of the many-kernel phase-space representation summarised in Sec.~\ref{sec:baseline} (model and optimisation, evaluation procedures, complete results, conservative reconstruction, computational cost).
\end{bmhead}

\begin{bmhead}[Funding.]
This research received no specific grant from any funding agency, commercial or not-for-profit sectors.
\end{bmhead}

\begin{bmhead}[Declaration of interests.]
The author reports no conflict of interest.
\end{bmhead}

\begin{bmhead}[Data availability statement.]
The codes and processed data that support the findings of this study are openly available at \url{https://github.com/Ehsan-Roohi/shock-velocity-distribution-populations}: the DSMC and cascade codes with the collision record and the free-flight flag, the reduced BGK solver with the recomputed Mach-8 solution and its resolution tests, the population fits, the three prediction methods with their positivity checks, and the per-run statistics of the precursor. The code of the many-kernel representation is available at \url{https://github.com/Ehsan-Roohi/KineticGaussian}. The raw DSMC output, the complete DVM phase-space arrays, and the trained models and evaluation files of the many-kernel study are available from the author on reasonable request because of their size.
\end{bmhead}

\begin{bmhead}[Declaration of AI use.]
The author used ChatGPT (OpenAI) and Claude (Anthropic) as assistants for code development and debugging, data analysis, drafting and editing of the text, and critical review of the manuscript. The author checked the computations, results, and references, and takes full responsibility for the content of the paper.
\end{bmhead}

\end{document}


\maketitle
\noindent This document reproduces the study of the many-kernel (512-Gaussian) phase-space representation of BGK normal shocks \citep{RoohiTransport} that the main paper summarises in its section on that representation and uses as a reference. Sections, tables, figures, and equations of this document carry the prefix S; other numbers refer to the main paper.

\section{Overview: a many-kernel phase-space representation and its range of validity}
\label{sup:overview}

The main paper describes the distribution at each position by a few Gaussian populations and predicts it across Mach number in scaled variables. Gaussians can also be used in a different way: as a compact representation of the whole phase-space function $f(x,\bm\xi)$, in which several hundred kernels act as basis functions without individual meaning. This document applies such a representation to the same BGK shocks and determines where it is accurate and where it stops being valid. It is the reference against which the few-population description and the predictions of Sec.~8 of the main paper are judged. The model and its optimisation, the error measures, and the complete set of tests are given in Secs.~\ref{app:model}--\ref{sec:shock-results}.

\subsection{Representation of a fitted shock}

The distribution is written as a positive mixture of $N=512$ Gaussians in the normalised coordinates $(x,\xi_x,\xi_y,\xi_z)$, Eq.~\eqref{eq:phase_gaussian}. With diagonal kernels the model has 4608 parameters, which replace $2.26\times10^8$ stored values of the Mach-5 DVM solution. The parameters are fitted to $\log f$ of one DVM solution with a Huber loss, and every moment reported below is recomputed from the fitted mixture with the DVM quadrature. Errors are relative $L_2$ errors over the stored positions, $E_\phi=\|\hat\phi-\phi\|_2/\|\phi\|_2$. All accuracies quoted in this section refer to the velocity domain and quadrature of the DVM solution on which the mixture is fitted and evaluated, and to the moments listed; the tails of the mixture beyond that domain are not constrained by the fit (Secs.~\ref{sec:conservative} and \ref{app:controls}).

\begin{figure}[!htbp]
\centering
\includegraphics[width=\textwidth]{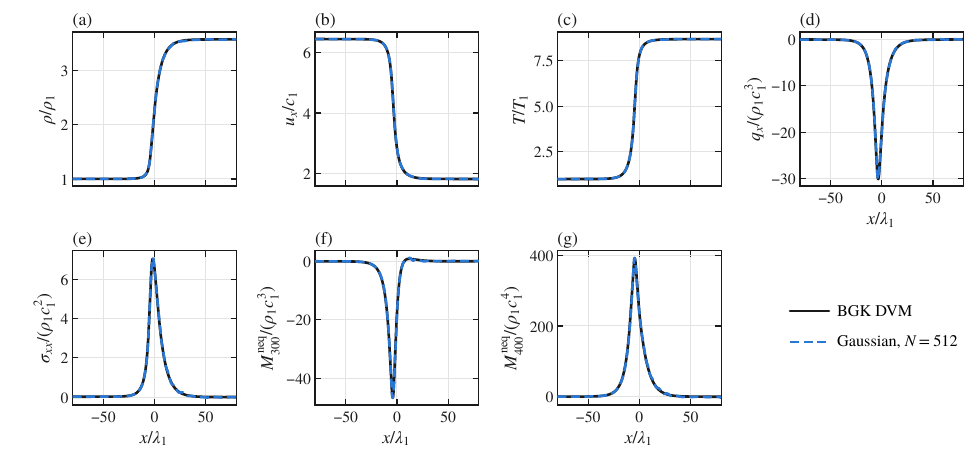}
\caption{Mach-5 normal shock reconstructed by the diagonal $N=512$ log-density model (seed 1234): (a)--(c) low-order fields, (d, e) heat flux and normal stress, and (f, g) third- and fourth-order nonequilibrium moments, compared with the BGK DVM. All seven moments enter the error calculation.}
\label{fig:m5}
\end{figure}

For the Mach-5 shock (figure~\ref{fig:m5}) the errors, averaged over three random initialisations, are 0.23\%, 0.09\%, and 0.12\% in $\rho$, $u_x$, and $T$, and $1.7\pm0.8\%$, $1.6\pm0.2\%$, $1.8\pm0.1\%$, and $1.8\pm0.2\%$ in $q_x$, $\sigma_{xx}$, $M_{300}^{\mathrm{neq}}$, and $M_{400}^{\mathrm{neq}}$ (table~\ref{tab:m5_seeds}). For the Mach-3 shock they are below 0.15\% and 1.5--2.1\% (figure~\ref{fig:m3}). These norms extend over the whole domain, of which the shock layer ($0.01<n^*<0.99$) is about a quarter, and they do not measure the position or the thickness of the reconstructed shock. Restricted to the layer and normalised by the jumps across the shock, the root-mean-square errors of $\rho$, $u_x$, and $T$ at Mach 5 are 0.40\%, 0.16\%, and 0.16\% (largest error of the density 1.5\% of its jump); the width of the density profile between $n^*=0.2$ and 0.8 is within 2\% of the reference and the position $n^*=0.5$ within 0.16\,$\lambda_1$, 1.4\% of the shock thickness. Three further tests, reported in Sec.~\ref{sec:shock-results}, complete these numbers. Fitting $\log f$ instead of sampled moments lowers the stress error tenfold, from 17.8\% to 1.7\%, whereas a correlated $x$--$\xi_x$ kernel gives no measurable gain (tables~\ref{tab:m3_ablation} and \ref{tab:m5_seeds}). Spatial grid points withheld from the fit, every fifth one, have the same errors as the fitted ones to within 1.3\% of the error itself (table~\ref{tab:holdout}), so the mixture interpolates in $x$. Finally, the heat-flux error varies from 1.1\% to 3.6\% between random initialisations, and choosing the saved model with the smallest loss does not reduce this variation (table~\ref{tab:checkpoint}).

The accuracy is a property of the Gaussian localisation and not of the number of coefficients. Multilinear grids that store $\log f$ with the same number of coefficients, including one whose $x$ nodes are placed with knowledge of the reference gradients, give errors of 79--104\% in the transport and higher moments (figure~\ref{fig:shock_baseline}, tables~\ref{tab:shock_baseline} and \ref{tab:grid_layouts}). A two-Maxwellian description and a 13-moment description with the same or smaller storage lose the third- and fourth-order moments (table~\ref{tab:codecs}). These comparisons concern regular grids; sparse-grid, tensor, and low-rank compression are outside their scope.

The fitted mixture does not conserve the fluxes of mass, momentum, and energy exactly. Section~\ref{sec:conservative} constructs a positive continuous-velocity reconstruction that enforces the Rankine--Hugoniot fluxes simultaneously under analytic integration and under the DVM quadrature. For the fitted Mach-5 shock it gives a heat-flux error of 1.53\%. Exact fluxes do not by themselves ensure accurate transport: without the moment term the same constraints leave a heat-flux error of 11\%.

\subsection{Dependence on Mach number in physical variables}

The representation behaves differently over a range of shock strengths. Every kernel parameter was made a function of Mach number, either a polynomial of degree 2 or 3 or a piecewise-linear function with three or four knots, and the resulting models with 15,360 or 20,480 coefficients were fitted jointly to eight of the nine BGK shocks (Sec.~\ref{app:model}). The kernels are shared between Mach numbers, so that the dependence on Mach number is fitted to the data and not obtained by interpolating between separately fitted mixtures, whose kernels do not correspond.

On the shocks used for training the distribution error is 1.3--1.9\% on average (figure~\ref{fig:full_mach}a). For a withheld shock it is much larger (table~\ref{tab:mach_holdout}). The withheld Mach-3 shock is recovered with an error of 11\% in the distribution and 23--42\% in the transport and higher moments. For the withheld Mach-6 shock the polynomial models give 42--43\% in the distribution; the four-knot local model reduces this to $11.5\pm0.9\%$, but its heat-flux, stress, and fourth-moment errors remain 30--40\%. Mach 12 lies outside the interval fitted with Mach numbers up to 8. The polynomial evaluated there gives distribution errors of several hundred percent and more (Sec.~\ref{sec:parametric}) and a density many orders of magnitude above the reference (figure~\ref{fig:full_mach}c), and the range check of Eq.~\eqref{eq:domain_guard} excludes it.

Sections~7 and 8 of the main paper showed why this happens and what removes it. In physical variables the distribution changes with Mach number mainly through the displacement and narrowing of its populations, which follow the Rankine--Hugoniot velocities and the downstream temperature; 5120 kernel parameters have to be fitted to this dependence from eight solutions. In the variables scaled with the Rankine--Hugoniot states the three populations are nearly independent of Mach number, and twelve numbers per position, combined linearly in $1/M^2$ from two Mach numbers, give the withheld Mach-6 shock to within 1.6\% in the distribution and 3.3\% in the heat flux, and the Mach-12 shock, which lies outside the range of the 512-kernel model, to within 4.7\% and 4.3\% (table~8 of the main paper); the scaled distributions combined directly, without any Gaussian, are more accurate still in the moments (table~12 of the main paper). The two descriptions are therefore complementary. The 512-kernel mixture represents a given solution in full, with its dependence on $x$ and its tails, to 1--2\% in the transport moments. The three populations describe a given solution to within 5\% of the peak heat flux (Sec.~4 of the main paper), and their parameters carry over from one shock strength to another.

\begin{table}
\caption{Leave-one-Mach-out errors in percent, mean $\pm$ sample standard deviation over three seeds. M6 moments are evaluated against the grid-refined fine reference. D2 and D3 denote global degree-2 and degree-3 maps; K3 and K4 denote three- and four-knot correspondence-preserving local maps.}
\label{tab:mach_holdout}
\centering
\scriptsize
\resizebox{\textwidth}{!}{%
\begin{adjustbox}{max width=\linewidth}
\begin{tabular}{llcccccccc}
\toprule
Case & map & $E_f$ & $E_\rho$ & $E_{u_x}$ & $E_T$ & $E_{q_x}$ & $E_{\sigma_{xx}}$ & $E_{M_{300}^{\rm neq}}$ & $E_{M_{400}^{\rm neq}}$\\
\midrule
M3 & global D2 & $11.18\pm1.38$ & $13.86\pm0.44$ & $3.07\pm0.31$ & $8.83\pm0.09$ & $23.16\pm3.70$ & $24.34\pm4.86$ & $31.11\pm3.53$ & $29.13\pm2.20$\\
M3 & global D3 & $11.04\pm0.40$ & $14.67\pm0.05$ & $3.75\pm0.21$ & $8.59\pm0.14$ & $36.38\pm4.69$ & $30.50\pm3.85$ & $41.81\pm5.42$ & $40.71\pm2.75$\\
M6 & global D2 & $41.90\pm6.24$ & $17.50\pm1.52$ & $4.85\pm0.80$ & $5.79\pm0.11$ & $54.24\pm3.39$ & $73.61\pm9.45$ & $62.28\pm8.71$ & $72.54\pm11.87$\\
M6 & local K3 & $21.85\pm4.06$ & $24.68\pm2.98$ & $5.58\pm1.78$ & $7.50\pm1.34$ & $66.24\pm12.75$ & $70.24\pm32.91$ & $77.17\pm7.04$ & $83.10\pm11.26$\\
M6 & global D3 & $42.86\pm5.40$ & $21.42\pm1.76$ & $6.85\pm0.61$ & $7.37\pm0.15$ & $38.30\pm7.72$ & $56.54\pm5.38$ & $46.61\pm7.80$ & $46.31\pm6.57$\\
M6 & local K4 & $11.45\pm0.94$ & $6.96\pm0.76$ & $3.05\pm0.04$ & $5.53\pm0.70$ & $39.50\pm3.52$ & $37.98\pm7.02$ & $39.05\pm5.87$ & $30.11\pm2.60$\\
\bottomrule
\end{tabular}
\end{adjustbox}%
}
\end{table}

\FloatBarrier
\bigskip
\section{The 512-kernel phase-space representation: model and optimisation}
\label{app:model}

\subsection{Positive phase-space model for shocks}
\label{sec:phase-model}

The shock VDF is represented in normalised phase space $\bm z=(\tilde x,\tilde\xi_x,\tilde\xi_y,\tilde\xi_z)$ by
\begin{equation}
\log \hat f_\theta(\bm z)=\log\sum_{m=1}^{N}
\exp\left[a_m-\frac12Q_m(\bm z)\right].
\label{eq:phase_gaussian}
\end{equation}
The index $m$ labels a kernel, $N$ is the number of kernels, $a_m$ is its learned log-amplitude, and $Q_m$ is a positive quadratic form. Because Eq.~\eqref{eq:phase_gaussian} is a log-sum-exp of positive Gaussian contributions, $\hat f_\theta$ is positive by construction. The diagonal form uses
\begin{equation}
Q_m(\bm z)=\sum_{\ell=1}^{4}
\left(\frac{z_\ell-\mu_{m\ell}}{s_{m\ell}}\right)^2,
\label{eq:diag_q}
\end{equation}
where $\mu_{m\ell}$ and $s_{m\ell}>0$ are the centre and width of kernel $m$ in coordinate $\ell$. Equation~\eqref{eq:diag_q} restricts each kernel to axis-aligned localisation in the four normalised phase coordinates. Each diagonal kernel has one amplitude, four centres, and four widths, so the parameter count is $9N$. For $N=512$, the model has 4608 parameters.

A correlated block between physical position and streamwise velocity was tested through
\begin{equation}
\begin{aligned}
Q_m^{x\xi_x}&=\frac{\alpha_m^2-2r_m\alpha_m\beta_m+\beta_m^2}{1-r_m^2},\\
\alpha_m&=\frac{\tilde x-\mu_{mx}}{s_{mx}},\qquad
\beta_m=\frac{\tilde\xi_x-\mu_{m\xi_x}}{s_{m\xi_x}},
\end{aligned}
\label{eq:xvx}
\end{equation}
where $r_m\in(-1,1)$ is a learned correlation coefficient. Equation~\eqref{eq:xvx} adds exactly one off-diagonal degree of freedom, allowing a kernel to align with a tilted ridge in $(x,\xi_x)$ while the transverse $\xi_y$ and $\xi_z$ terms remain diagonal. This model has $10N$ parameters. Across the three tested seeds, however, it provides no clear performance advantage over the diagonal form at $N=512$ once the objective is fixed.

The main objective is a Huber loss in log-density,
\begin{equation}
\mathcal L_f=\frac{1}{B}\sum_{b=1}^{B}
H_\delta\!\left(\log\hat f_\theta(\bm z_b)-\log[f(\bm z_b)+\epsilon]\right),
\label{eq:logloss}
\end{equation}
where $B$ is the minibatch size in sampled phase points, $H_\delta$ is the Huber penalty with threshold $\delta$, and $\epsilon$ is the log floor. The pointwise log error is clipped at 20 during optimisation. By operating on log density, Eq.~\eqref{eq:logloss} gives low-probability regions of the VDF a meaningful influence without allowing numerical zeros to dominate; the Huber form also limits the influence of isolated large residuals. For comparison, a sampled moment-informed objective adds normalised errors in selected moments at sampled spatial and velocity points. All reported moments, however, are recomputed after training by Eqs.~(2.5)--(2.10) of the main paper; no reported shock moment is taken directly from a training target.

\subsection{Mach-conditioned phase-space model}

For the nine-Mach study, the correlated phase-space mixture is conditioned on a normalised Mach coordinate $\widetilde M$. Every unconstrained kernel parameter is represented by a degree-$d$ polynomial in the Legendre basis,
\begin{equation}
\begin{aligned}
\bm p_m(\widetilde M)&=\sum_{k=0}^{d}\bm p_{mk}P_k(\widetilde M),\\
\widetilde M&=2\frac{M-M_{\min}}{M_{\max}-M_{\min}}-1,
\end{aligned}
\label{eq:mach_conditioning}
\end{equation}
where $P_k$ is the Legendre polynomial of degree $k$; the bounded centre, width, and correlation transforms are applied afterward. Equation~\eqref{eq:mach_conditioning} therefore gives every kernel parameter one shared global polynomial dependence on normalised Mach number while retaining positivity after the bounded transforms. Conditioning all ten parameters of each $x$--$\xi_x$ kernel gives $10N(d+1)$ coefficients: 15,360 for $d=2$ and 20,480 for $d=3$ at $N=512$. In float32 these shared models occupy 61,440 and 81,920 bytes. Depending on the excluded case, one model replaces 5.81--9.90 GB of eight-state phase-space data, a nominal compression ratio of approximately $7.1\times10^4$--$1.61\times10^5$.

The coefficients are fitted jointly to all Mach numbers and are not obtained by interpolating between separately fitted mixtures. A Gaussian mixture is unchanged when its kernels are renumbered, so that kernel $m$ of one fit does not correspond to kernel $m$ of another, and interpolation between separate fits is not defined. In the joint fit each kernel keeps its index at all Mach numbers, and its parameters vary continuously with $M$.

For M6 the global polynomial is also replaced by a piecewise-linear basis with the same kernels,
\begin{equation}
\bm p_m(M)=\sum_{j=1}^{K}h_j(M)\bm p_{mj},
\qquad \sum_{j=1}^{K}h_j(M)=1,
\label{eq:local_mach_map}
\end{equation}
where $h_j$ is piecewise linear between the knots and is nonzero on the two intervals adjacent to knot $j$. The three-knot map has knots at $M=1.5$, 5, and 12 and has $3\times10N=15{,}360$ coefficients, the same number as for degree 2. The four-knot map has knots at $M=1.5$, 4, 8, and 12 and has $4\times10N=20{,}480$ coefficients, the same number as for degree 3. M6 is not used in the fit and lies between the knots at Mach 4 (M4) and Mach 8 (M8). Because the kernels are fitted jointly, each kernel keeps its index across the knots, as in the polynomial model. With equal numbers of coefficients, the comparison shows the effect of a local in place of a global dependence on Mach number.

The fitted Mach interval for either basis is computed from the fitted cases alone. The conditional representation therefore carries the range check
\begin{equation}
\mathcal A(M)=\mathbf 1\!\left(|\widetilde M|\le1\right),
\label{eq:domain_guard}
\end{equation}
where $\mathcal A=1$ inside the fitted interval and $\mathcal A=0$ outside it. The withheld M3 and M6 states satisfy Eq.~\eqref{eq:domain_guard}. When M12 is withheld, $M_{\max}=8$ and $\widetilde M=2.2308$, so that Mach 12 lies outside the interval and the model is not evaluated there. The local basis is likewise not used beyond its end knots. The polynomial evaluated at M12 without this check is reported with the results to show what the check excludes; it is not a prediction of the model.

\subsection{Optimisation settings}
\label{sec:optimization}

Shock coordinates are affinely normalised by their midpoint and half-width to approximately $[-1,1]^4$. Kernel centres are parameterised as $1.15\tanh(\cdot)$, and logarithmic widths are bounded to $[-5.6,-1.25]$ through a sigmoid map. Widths are initialised at log-scale $-2.65$ and log-amplitudes at $-7.0$ with small Gaussian perturbations. Centres are initialised from $5\times10^5$ representative phase points drawn with the density-weighted sampling that is also used in the fit.

Each optimisation step samples 24 physical locations and 384 velocity nodes per location. The M3 sampler uses mass exponent 0.55 and a 15\% uniform-velocity component; the M5 sampler uses exponent 0.45 and a 25\% uniform component, which places more samples in the tails. The Huber threshold is $\delta=1$, the pointwise log-error clip is 20, and the log floor is $10^{-35}$. Sample weights are proportional to the square root of local sampled density and clipped to $[0.25,4]$. Models are trained for 80,000 AdamW steps with default $\beta_1=0.9$, $\beta_2=0.999$, zero weight decay, gradient-norm clipping at 1, and a base learning rate of $5\times10^{-4}$. Centre and width learning rates are multiplied by 0.5 and 0.25, respectively; the model is saved every 2000 steps. Table~\ref{tab:training} summarises these settings. The shock models were fitted on NVIDIA Tesla M40 24-GB GPUs.

The conditional models use the same $24\times384$ phase-point minibatch, mass exponent 0.55, 15\% uniform-velocity sampling, and $10^{-35}$ floor. At each step one of the eight training Mach cases is selected, its physical locations and velocities are sampled, the selected global or local basis generates the kernel parameters, and the log-density loss is evaluated. Every tenth step, a moment penalty of weight 0.01 is computed from 12 physical locations and 1536 velocity samples for $\rho$, $u_x$, $T$, $q_x$, $\sigma_{xx}$, and $M_{300}^{\mathrm{neq}}$. The 18 global-polynomial models and six local-M6 models are each trained for 160,000 steps with a 1000-step warmup, base learning rate $5\times10^{-4}$, final learning-rate ratio 0.05, and gradient clipping at 1. Seeds 1234, 2026, and 3407 are used for every comparison. Table~\ref{tab:conditional_protocol} lists the withheld cases, bases, parameter counts, and settings. Evaluation uses 50,000 sampled phase points and full spatial moment profiles; all six local runs finished normally and use the model of the last step, as for the global models.

The fitting procedure is therefore as follows. The coordinates are normalised with the fitted cases alone; the kernel centres are initialised from the sampled phase points; steps on the log-density loss alternate with the moment penalty; the model is saved at fixed intervals; the model of the last step is used, without reference to the errors of the withheld cases; and all moments are then recomputed with the DVM quadrature. The settings are stored in the configuration files of the code repository (\url{https://github.com/Ehsan-Roohi/KineticGaussian}).

\begin{table}
\caption{Principal optimisation settings. For shocks, ``uniform fraction'' denotes the fraction of velocity samples drawn uniformly rather than from the mass-weighted distribution.}
\label{tab:training}
\centering
\small
\begin{tabular*}{\textwidth}{@{\extracolsep{\fill}}lcccccc@{}}
\toprule
Shock model & $N$ & steps & base rate & $x\times\xi$ samples/step & centre pool & uniform fraction\\
\midrule
M3 diagonal & 512 & 80,000 & $5\times10^{-4}$ & $24\times384$ & $5\times10^5$ & 0.15\\
M5 diagonal & 512 & 80,000 & $5\times10^{-4}$ & $24\times384$ & $5\times10^5$ & 0.25\\
M5 $x$--$\xi_x$ & 512 & 80,000 & $5\times10^{-4}$ & $24\times384$ & $5\times10^5$ & 0.25\\
\bottomrule
\end{tabular*}
\vspace{5pt}

\end{table}

\begin{table}
\caption{Tests of the Mach-conditioned model. Each row uses three seeds, producing 18 global-polynomial runs and six local-M6 runs with the same parameter counts.}
\label{tab:conditional_protocol}
\centering
\small
\begin{adjustbox}{max width=\linewidth}
\begin{tabular}{lccccc}
\toprule
Withheld & role & train range & degree or knots & parameters & steps\\
\midrule
M3 & interpolation & $[1.5,12]$ & 2 / 3 & 15,360 / 20,480 & 160,000\\
M6 & interpolation & $[1.5,12]$ & 2 / 3 & 15,360 / 20,480 & 160,000\\
M12 & out-of-domain test & $[1.5,8]$ & 2 / 3 & 15,360 / 20,480 & 160,000\\
M6 local & interpolation & $[1.5,12]$ & 3 / 4 & 15,360 / 20,480 & 160,000\\
\bottomrule
\end{tabular}
\end{adjustbox}
\end{table}

Table~\ref{tab:conditional_protocol} separates three distinct tests with the same three seeds and 160,000 steps. M3 and M6 are bracketed interpolation cases, M12 is an out-of-domain test because removing it truncates the training interval at M8, and the local-M6 row keeps the two global parameter counts while changing only the Mach-coordinate basis. In the M6 comparison only the locality of the basis therefore changes, and not the storage or the optimisation.

\FloatBarrier
\section{Evaluation and comparison procedures of the 512-kernel study}
\label{app:protocols}

\subsection{Error norms and storage count}

For a reference profile or field $g$ and reconstruction $\hat g$, the principal metric is
\begin{equation}
E_g=\frac{\|\hat g-g\|_2}{\|g\|_2}.
\label{eq:error}
\end{equation}
For shock profiles, the norm is taken over the spatial points after the VDF has been integrated by the DVM quadrature. Nominal compression is defined as the number of stored reference scalar values divided by the number of representation parameters,
\begin{equation}
C_{\mathrm{nom}}=\frac{N_{\mathrm{raw}}}{P}.
\label{eq:compression}
\end{equation}
Equation~\eqref{eq:compression} counts the coefficients of grid and Gaussian representations in the same way. The ratio is nominal, since stored DVM data could also be reduced by lossless compression, symmetry, or reduced velocity coordinates. The comparison made here is at equal numbers of floating-point coefficients.

\subsection{Random initialisation and choice of the saved model}

Each M5 model was fitted from three independent random
initialisations. For a paired comparison, the same random seeds
(1234, 2026, and 3407) were used for the diagonal and correlated
models. The seeds control Gaussian-centre initialisation, stochastic
phase-space sampling, and minibatch construction. Results are reported for the model at the last step (80,000) for every seed. Table~\ref{tab:checkpoint} compares this model with the saved model of smallest recorded training loss. The smallest minibatch loss does not always give the smallest moment error, in particular for $q_x$. The model of the last step is therefore used throughout, so that the choice of model does not depend on the reported moments.

\subsection{Withheld spatial points: procedure}

To test whether the representation interpolates between the stored spatial points, every fifth interior M5 point is excluded from the fit and from the initialisation of the centres. This leaves 1760 fitted and 440 withheld points. Errors are evaluated separately on the two sets, over the whole domain and in the central region $|x/\lambda_1|\le10$ (labelled shock layer in table~\ref{tab:holdout} and figure~\ref{fig:holdout}; it is not the layer $0.01<n^*<0.99$ of Sec.~\ref{sup:overview}).

\subsection{Leave-one-Mach-out tests}

The parametric test is stricter than the test with withheld points because an entire solution is removed. For each withheld state in $\{\mathrm{M3},\mathrm{M6},\mathrm{M12}\}$, separate degree-2 and degree-3 models are trained from three seeds on the other eight Mach cases. A comparison of bases with equal parameter counts repeats the M6 exclusion with three- and four-knot local maps over the same three seeds. The reported distribution metric is a relative $L_2$ error over 50,000 phase points sampled from the common evaluation list; the weighted log-density root-mean-square error is recorded separately. Moment errors use Eq.~\eqref{eq:error} over the complete stored spatial profile. Training-case statistics include only cases actually seen by a given model. Data of the withheld state are never used for choosing the saved model or for the coordinate normalisation. M12 evaluations are not clipped to the training interval, so loss of physical validity is visible in the reported profiles.

\subsection{Equal-storage shock grids}

For comparison, $\log f$ is stored on coarse four-dimensional grids in $(x,\xi_x,\xi_y,\xi_z)$ and reconstructed by multilinear interpolation before the same DVM moment quadrature is applied. Table~\ref{tab:grid_layouts} gives the tested layouts at two storage sizes: 4608 values, matching the diagonal $N=512$ model, and 5120 values, matching the correlated model. The layouts are a uniform grid with nearly equal numbers of nodes in the four coordinates; a uniform grid with more nodes in $x$ and $\xi_x$; and a grid of the second kind with nonuniform $x$ nodes. The $x$ nodes of the third grid are placed with the gradients and transport profiles of the reference solution, which favours the grid in the comparison.

\section{Complete results of the 512-kernel representation}
\label{sec:shock-results}

\subsection{Mach-3 reconstruction and the role of the objective}

\begin{figure}[!htbp]
\centering
\includegraphics[width=\textwidth]{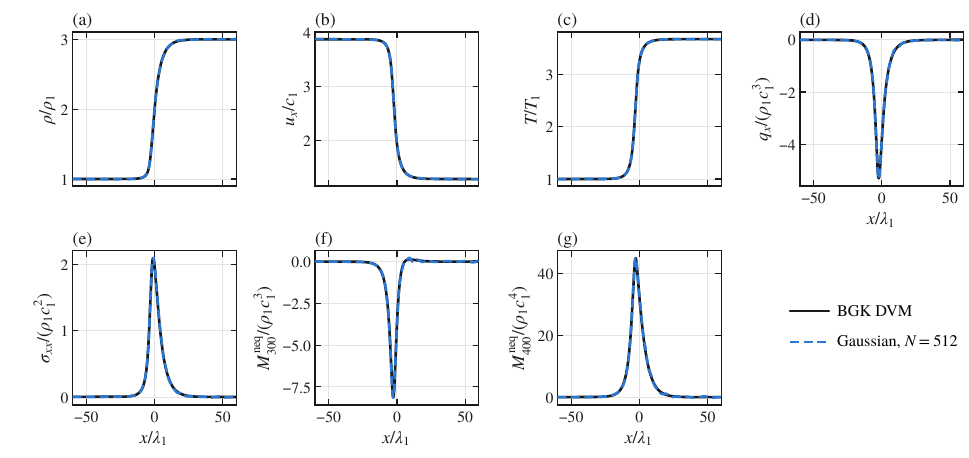}
\caption{Mach-3 normal shock reconstructed by the diagonal $N=512$ positive phase-space model trained on $\log f$: (a)--(c) low-order fields, (d, e) heat flux and normal stress, and (f, g) third- and fourth-order nonequilibrium moments, compared with the BGK DVM.}
\label{fig:m3}
\end{figure}

Figure~\ref{fig:m3} compares the M3 DVM profiles with a diagonal $N=512$ Gaussian model trained only on $\log f$.

The low-order profiles are nearly indistinguishable from the reference: $E_\rho=0.142\%$, $E_{u_x}=0.0656\%$, and $E_T=0.0648\%$. The localised heat-flux minimum, normal-stress maximum, negative third moment, and fourth-order deviation are preserved with errors of 1.48\%, 1.68\%, 2.03\%, and 2.07\%, respectively. The fourth-order moment is a sensitive test because its equilibrium part has been subtracted.

Table~\ref{tab:m3_ablation} shows that the objective changes the errors more than the kernel form or the number of kernels does. With the same diagonal $N=512$ model, sampled moment-informed training gives stress and heat-flux errors of 17.84\% and 7.97\%, compared with 1.68\% and 1.48\% for log-density training. Adding an $x$--$\xi_x$ block under the sampled-moment objective does not reduce those two errors, and the tested $N=1024$ variant increases $q_x$ and $M_{300}^{\mathrm{neq}}$ errors. Log-density fitting is therefore used. The comparison does not separate the contributions of phase-space coverage, stochastic moment quadrature, and optimisation to this difference.

\begin{table}
\caption{M3 relative errors in percent for the comparison of objectives and model sizes. The log-density row uses the model at the last step. A dash indicates that the fourth-order deviation was not stored for the moment-loss runs.}
\label{tab:m3_ablation}
\centering
\begin{adjustbox}{max width=\linewidth}
\begin{tabular}{lccccccc}
\toprule
Configuration & $E_\rho$ & $E_{u_x}$ & $E_T$ & $E_{q_x}$ & $E_{\sigma_{xx}}$ & $E_{M_{300}^{\rm neq}}$ & $E_{M_{400}^{\rm neq}}$\\
\midrule
Diagonal 512, $\log f$ & 0.142 & 0.0656 & 0.0648 & 1.475 & 1.681 & 2.030 & 2.066\\
Diagonal 512, moment loss & 0.935 & 0.306 & 0.514 & 7.969 & 17.843 & 8.235 & --\\
$x$--$\xi_x$ 512, moment loss & 0.934 & 0.300 & 0.476 & 8.058 & 17.266 & 8.644 & --\\
$x$--$\xi_x$ 1024, moment loss & 0.940 & 0.280 & 0.541 & 10.423 & 17.762 & 10.515 & --\\
\bottomrule
\end{tabular}
\end{adjustbox}
\end{table}

\subsection{Mach-5 reconstruction}

The M5 shock has larger ranges of the moments. Figure~\ref{fig:m5} uses the diagonal $N=512$ model and seed 1234. The representative errors are $0.231\%$, $0.0752\%$, and $0.0900\%$ for $\rho$, $u_x$, and $T$, and 1.12\%, 1.66\%, 1.71\%, and 1.80\% for $q_x$, $\sigma_{xx}$, $M_{300}^{\mathrm{neq}}$, and $M_{400}^{\mathrm{neq}}$. The plotted fourth-order peak is roughly two orders of magnitude larger than in the M3 case, and the error remains 1.80\% for this moment.

The comparison of objectives was made at Mach 3 (table~\ref{tab:m3_ablation}).

\subsection{Kernel form, random initialisation, and choice of the saved model}

\begin{figure}[!htbp]
\centering
\includegraphics[width=\textwidth]{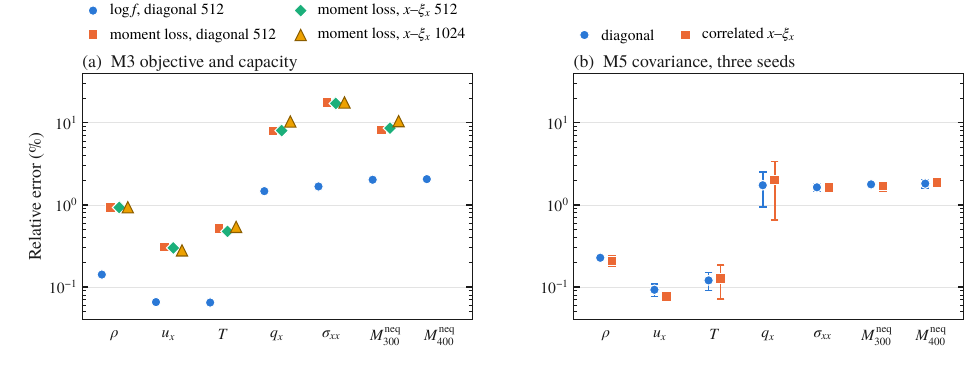}
\caption{Comparison of objectives and architectures. (a) For M3, changing the objective from sampled moment supervision to direct log-density fitting reduces the displayed transport and higher-moment errors by factors of approximately 4--11. The fourth-order deviation was not stored for the moment-loss runs, and the tested $N=1024$ moment-loss run does not reduce the errors. (b) For M5 at $N=512$, the diagonal and correlated $x$--$\xi_x$ models are compared across three seeds; error bars are one sample standard deviation. Values are those of tables~\ref{tab:m3_ablation} and \ref{tab:m5_seeds}.}
\label{fig:ablation}
\end{figure}

Figure~\ref{fig:ablation} shows the M3 comparison of objectives and the M5 comparison of kernel forms. For M5, the diagonal and $x$--$\xi_x$ models are each evaluated over three seeds. Their mean nonequilibrium errors differ by less than one sample standard deviation, as listed in table~\ref{tab:m5_seeds}. The correlated model has a slightly smaller mean stress error, while the diagonal model has slightly smaller mean $q_x$ and $M_{400}^{\mathrm{neq}}$ errors. With three initialisations the two kernel forms are not distinguished. The diagonal model is used for the representative profile because it has 10\% fewer parameters and comparable errors in this sample.

\begin{table}
\caption{M5 errors of the models of the last step in percent over seeds 1234, 2026, and 3407. Values are mean $\pm$ sample standard deviation. The low-order and the nonequilibrium moments are listed separately.}
\label{tab:m5_seeds}
\centering
\small
\begin{adjustbox}{max width=\linewidth}
\begin{tabular}{lccc}
\toprule
Architecture & $E_\rho$ & $E_{u_x}$ & $E_T$\\
\midrule
Diagonal & $0.2275\pm0.0077$ & $0.0927\pm0.0156$ & $0.1206\pm0.0299$\\
$x$--$\xi_x$ & $0.2081\pm0.0293$ & $0.0770\pm0.0062$ & $0.1275\pm0.0566$\\
\bottomrule
\end{tabular}
\end{adjustbox}
\vspace{5pt}

\begin{adjustbox}{max width=\linewidth}
\begin{tabular}{lcccc}
\toprule
Architecture & $E_{q_x}$ & $E_{\sigma_{xx}}$ & $E_{M_{300}^{\rm neq}}$ & $E_{M_{400}^{\rm neq}}$\\
\midrule
Diagonal & $1.740\pm0.796$ & $1.642\pm0.165$ & $1.784\pm0.085$ & $1.823\pm0.221$\\
$x$--$\xi_x$ & $2.023\pm1.361$ & $1.634\pm0.090$ & $1.704\pm0.236$ & $1.871\pm0.208$\\
\bottomrule
\end{tabular}
\end{adjustbox}
\end{table}

\begin{figure}[!htbp]
\centering
\includegraphics[width=\textwidth]{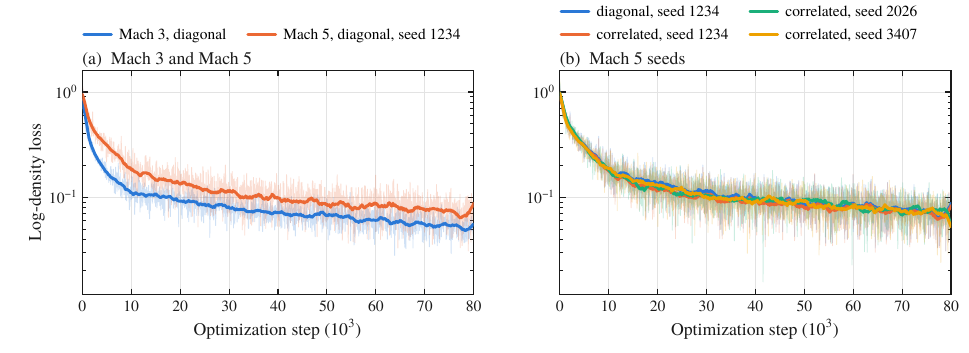}
\caption{Shock optimisation histories. (a) M3 and M5 diagonal models. (b) M5 diagonal and correlated runs with different seeds. Thin lines are the recorded minibatch Huber losses and thick lines their 2000-step moving averages in $\log_{10}$. The loss remains noisy after the rapid initial decay; the model at the last step is used for every seed instead of a selection by a reported moment.}
\label{fig:training}
\end{figure}

The heat-flux error has the largest sample standard deviation in table~\ref{tab:m5_seeds}. One seed gives $E_{q_x}=3.59\%$ for the correlated model and 2.64\% for the diagonal model, whereas the other seeds are close to 1.2--1.7\%. The variation is not consistently removed by choosing the saved model with the smallest stored minibatch loss; for some runs that choice increases the error of $q_x$. Figure~\ref{fig:training} shows that the loss remains noisy late in the optimisation. The model at the last step is therefore used for every seed instead of selecting by a reported moment.

\subsection{Withheld spatial points: results}

\begin{figure}[!htbp]
\centering
\includegraphics[width=\textwidth]{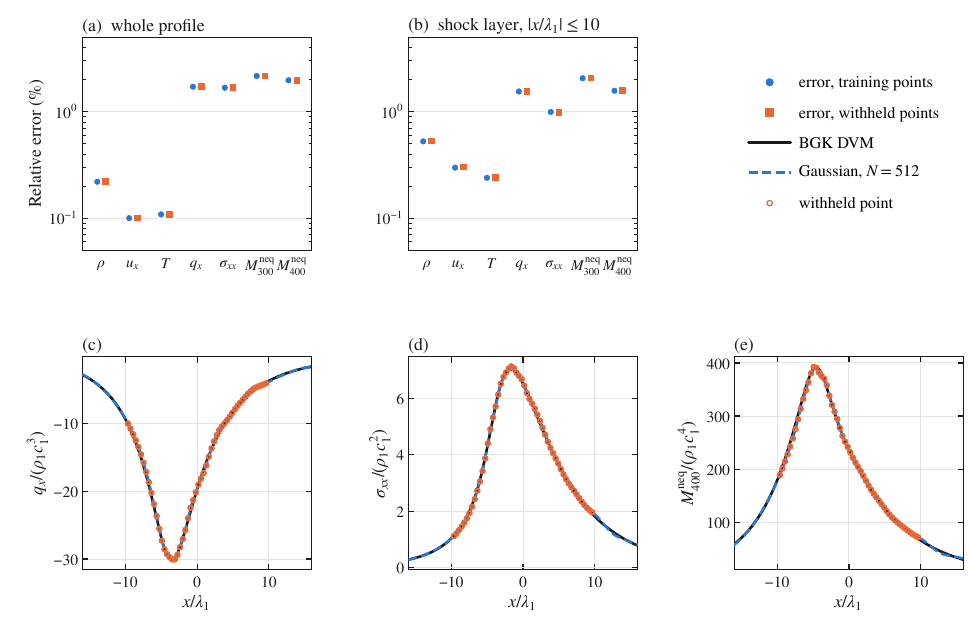}
\caption{Interpolation at withheld points for the M5 shock. Every fifth interior point (440 of 2200) is excluded from training and centre initialisation. (a, b) Relative errors on training and withheld points over the whole profile and within the shock layer, $|x/\lambda_1|\le10$ (table~\ref{tab:holdout}). (c)--(e) Sensitive profiles in the shock layer; open circles mark the withheld points, which retain the accuracy of the training set.}
\label{fig:holdout}
\end{figure}

The test with withheld points examines interpolation between stored spatial points within one fixed M5 state. Global errors at training and withheld points agree to within one unit in the fourth significant digit. In the shock layer, table~\ref{tab:holdout} and figure~\ref{fig:holdout} show that the difference between training and withheld points is at most 1.3\% of the corresponding training error. The test therefore shows spatial interpolation within the M5 solution; other shock strengths are tested in Sec.~\ref{sec:parametric}.

\begin{table}
\caption{M5 errors at training and withheld spatial points in percent. Every fifth interior spatial point is excluded from the fit and from the initialisation of the centres. The rows labelled shock layer use $|x/\lambda_1|\le10$.}
\label{tab:holdout}
\centering
\begin{adjustbox}{max width=\linewidth}
\begin{tabular}{llccccccc}
\toprule
Region & points & $E_\rho$ & $E_{u_x}$ & $E_T$ & $E_{q_x}$ & $E_{\sigma_{xx}}$ & $E_{M_{300}^{\rm neq}}$ & $E_{M_{400}^{\rm neq}}$\\
\midrule
Global & training & 0.2200 & 0.1002 & 0.1086 & 1.7181 & 1.6811 & 2.1706 & 1.9767\\
Global & withheld & 0.2199 & 0.1002 & 0.1086 & 1.7181 & 1.6811 & 2.1706 & 1.9767\\
Shock layer & training & 0.5272 & 0.2982 & 0.2396 & 1.5507 & 0.9957 & 2.0657 & 1.5752\\
Shock layer & withheld & 0.5318 & 0.3019 & 0.2413 & 1.5502 & 0.9859 & 2.0680 & 1.5726\\
\bottomrule
\end{tabular}
\end{adjustbox}
\end{table}

\subsection{Equal-storage regular-grid comparison}
\label{sec:shock-storage}

\begin{figure}[!htbp]
\centering
\includegraphics[width=\textwidth]{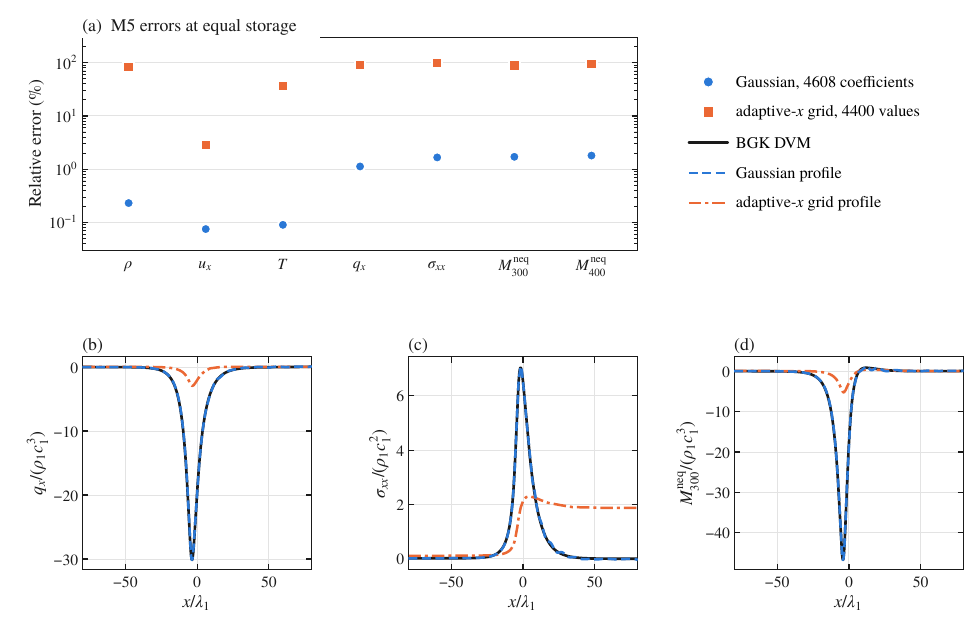}
\caption{M5 equal-storage comparison. (a) Relative errors of the diagonal Gaussian model and the best tested regular grid, whose $x$ nodes are placed with the reference solution. (b)--(d) Heat-flux, stress, and third-order profiles. With this 4400-value grid, adaptive physical-node placement does not prevent large moment errors (table~\ref{tab:shock_baseline}).}
\label{fig:shock_baseline}
\end{figure}

The last test compares the Gaussian mixture with $\log f$ stored on a coarse grid with the same number of coefficients. With the storage of the diagonal model, a four-dimensional regular grid has 4608 values to distribute among one physical and three velocity coordinates. The grid with reference-adapted $x$ nodes stores 4400 values and concentrates its spatial nodes near the shock; it gives 84.3\% density error, 36.6\% temperature error, and 89--98\% errors in the transport and high-order moments. The corresponding Gaussian errors are 0.075--0.231\% for low-order fields and 1.12--1.80\% for nonequilibrium quantities. Figure~\ref{fig:shock_baseline} compares the profiles, table~\ref{tab:shock_baseline} reports the main set of moments, and table~\ref{tab:grid_layouts} gives both coefficient counts and all three grid layouts.

\begin{table}
\caption{M5 equal-storage errors in percent. The Gaussian has 4608 parameters and nominal compression $4.91\times10^4$; the regular grid with reference-adapted $x$ nodes stores 4400 values and has nominal compression $5.14\times10^4$.}
\label{tab:shock_baseline}
\centering
\begin{adjustbox}{max width=\linewidth}
\begin{tabular}{lccccccc}
\toprule
Method & $E_\rho$ & $E_{u_x}$ & $E_T$ & $E_{q_x}$ & $E_{\sigma_{xx}}$ & $E_{M_{300}^{\rm neq}}$ & $E_{M_{400}^{\rm neq}}$\\
\midrule
Diagonal Gaussian & 0.231 & 0.075 & 0.090 & 1.125 & 1.663 & 1.708 & 1.804\\
Adaptive regular grid & 84.285 & 2.884 & 36.586 & 90.440 & 97.884 & 88.712 & 95.761\\
\bottomrule
\end{tabular}
\end{adjustbox}
\end{table}

\begin{table}
\caption{Regular-grid layouts at both storage sizes. Errors are percentages. ``Balanced'' distributes nodes nearly uniformly; ``physics uniform'' favours $x$ and $\xi_x$; ``adaptive $x$'' places physical nodes using reference gradients and transport profiles.}
\label{tab:grid_layouts}
\centering
\small
\begin{adjustbox}{max width=\linewidth}
\begin{tabular}{rrlrrrrr}
\toprule
Size & stored & layout & $E_{q_x}$ & $E_{\sigma_{xx}}$ & $E_{M_{300}^{\rm neq}}$ & $E_{M_{400}^{\rm neq}}$ & max.\\
\midrule
4608 & 4608 & balanced & 98.776 & 97.471 & 102.515 & 97.827 & 102.515\\
4608 & 4400 & physics uniform & 89.064 & 102.957 & 87.445 & 97.126 & 102.957\\
4608 & 4400 & adaptive $x$ & 90.440 & 97.884 & 88.712 & 95.761 & 97.884\\
5120 & 5120 & balanced & 100.658 & 97.944 & 103.745 & 98.361 & 103.745\\
5120 & 5000 & physics uniform & 80.461 & 95.219 & 80.124 & 92.106 & 95.219\\
5120 & 5000 & adaptive $x$ & 79.464 & 92.792 & 79.238 & 91.093 & 92.792\\
\bottomrule
\end{tabular}
\end{adjustbox}
\end{table}

At equal storage the Gaussian mixture is more accurate than all three multilinear grids. Tensor decompositions, reduced velocity coordinates, and nonlinear maps are other representations with which it could be compared at equal numbers of coefficients.

\subsection{Parametric shock strength and the range check}
\label{sec:parametric}

\begin{figure}[!htbp]
\centering
\includegraphics[width=\textwidth]{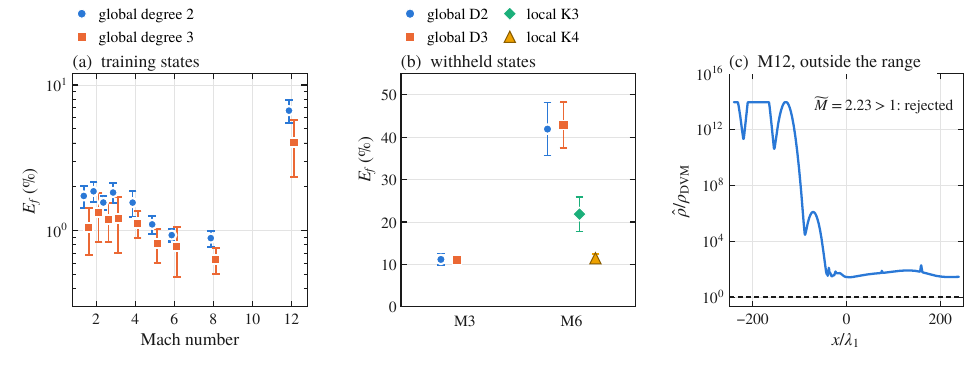}
\caption{Mach-conditioned model: training, withheld states, and range test. (a) Sampled distribution errors of the training states over the nine-Mach database. (b) Errors of the withheld M3 and M6 states of the global degree-2 and degree-3 maps (D2, D3) and, at M6, of the three- and four-knot local maps (K3, K4). (c) Degree-2 density ratio, without the range check, for the withheld M12 evaluation, which Eq.~\eqref{eq:domain_guard} rejects beyond the M8 training boundary; the dashed line marks unity, and the corresponding sampled errors without the range check are given in the text. Error bars are one sample standard deviation over three seeds. Moment-resolved in-range errors are reported in table~\ref{tab:mach_holdout}.}
\label{fig:full_mach}
\end{figure}

Figure~\ref{fig:full_mach}(a) summarises the 18 completed global conditional runs. Averaged over the cases used in training, the sampled distribution error is 1.87\% for degree 2 and 1.27\% for degree 3. The error is 0.89\% and 0.63\% at M8 and 6.67\% and 4.04\% at the trained M12 endpoint. Withheld complete solutions in figure~\ref{fig:full_mach}(b) provide the corresponding interpolation test.

Table~\ref{tab:mach_holdout} reports the bracketed M3 and M6 cases. At M3, both global degrees give approximately 11\% sampled distribution error, 13.9--14.7\% density error, and 23--42\% transport and higher-moment errors. At M6, the global maps give approximately 42\% sampled distribution error. These errors are much larger than at the fitted Mach numbers, most of all in the transport moments; raising the polynomial degree lowers some errors and raises others.

Table~\ref{tab:mach_holdout} also shows the effect of the basis at M6. The four-knot map reduces the distribution error from $42.86\pm5.40\%$ for global degree 3 to $11.45\pm0.94\%$ with the same 20,480 parameters, a 73.3\% reduction. Against the grid-refined fine M6 reference, errors fall from 21.42\% to 6.96\% in density, 6.85\% to 3.05\% in velocity, 7.37\% to 5.53\% in temperature, 56.54\% to 37.98\% in stress, and 46.31\% to 30.11\% in the fourth-order nonequilibrium moment. The mean heat-flux error changes from 38.30\% to 39.50\%, an increase of 1.2 percentage points, smaller than the reported sample standard deviations. The K3 run has an intermediate distribution error but larger moment errors; K3 and K4 differ in both the placement of the knots and the number of coefficients. A local dependence on Mach number therefore reduces the errors of the distribution and of most moments, while the heat-flux error is unchanged and the transport moments keep errors of 30--40\%.

When M12 is withheld it lies outside the fitted interval. Its normalised coordinate is 2.2308, so that Eq.~\eqref{eq:domain_guard} excludes it. Evaluated without the range check, the polynomial of degree 2 gives sampled distribution errors of 204\%, 454\%, and $1.81\times10^6$\%, while degree 3 gives $1.97\times10^{15}$--$3.09\times10^{16}$\%. Figure~\ref{fig:full_mach}(c) shows a representative density ratio for degree 2. A polynomial evaluated beyond its normalised training interval can grow rapidly, while positivity constrains only the sign of $f$ and does not bound its integrated moments. This evaluation shows the consequence of omitting the range check; it is not a prediction of the model.

Replacing the archived-grid M6 reference moments by the grid-refined fine solution changes the local-K4 aggregate errors by less than 0.21 percentage points. At M12 the largest change between the archived-grid and refined references is 6.46\% for stress, small relative to the values obtained without the range check. The two tests are complementary: the withheld states give the accuracy within the range, and the range check excludes Mach numbers beyond the fitted interval.

\FloatBarrier
\section{Conservative continuous-velocity reconstruction of shock solutions}
\label{sec:conservative}
The errors of Sec.~\ref{sec:shock-results} measure the reconstruction of macroscopic fields under the DVM quadrature; they are not errors of the conserved fluxes. This section asks whether a reconstructed shock distribution can carry the Rankine--Hugoniot (RH) fluxes exactly, both under the DVM quadrature and under analytic integration over all velocities, and what such constraints imply for the accuracy of the transport moments. The construction is applied to the fitted models of Sec.~\ref{sec:optimization} without refitting them, and the additional numbers that it stores are counted separately from the comparison of Sec.~\ref{sec:shock-storage}. Variants of the construction are described in Sec.~\ref{app:controls}.

For a distribution $f$, the longitudinal fluxes of mass, momentum, and energy are
\begin{equation}
\bm F[f]=\int f\left(\xi_x,\ \xi_x^2,\ \tfrac12\xi_x|\bm\xi|^2\right)^T\,\mathrm d\bm\xi,
\label{conservative:eq:flux}
\end{equation}
and their quadrature counterparts replace the integral by the DVM quadrature. Both are compared with the upstream RH fluxes $\bm F_*=(J,P,E)^T$.

\subsection{Reconstruction}
\label{conservative:sec:models}
\emph{Fitted state.} For the separately fitted Mach-5 model (512 Gaussians, 4608 coefficients), the kernels are converted to physical units and averaged over the reflections $\xi_y\mapsto-\xi_y$ and $\xi_z\mapsto-\xi_z$, which keeps the mixture positive. At a given $x$ it is a positive combination of velocity Gaussians with masses $n_j^0$.

\emph{Withheld Mach number.} For a target Mach number $M$ between training neighbours $M_L$ and $M_H$, let $b=(M-M_L)/(M_H-M_L)$. Each neighbour model (the conditional model of Sec.~\ref{sec:parametric}, 20,480 coefficients) receives one positive amplitude multiplier per component, calibrated against the DVM moments of its own training state. Seven DVM moment profiles of each neighbour are stored, together with the coordinates, at 129 uniformly spaced positions ($2\times8\times129=2064$ values in total) and are interpolated to give target moments: $\rho$, $u_x$, and $T$ are interpolated in RH-normalised form, and $q_x$, $\sigma_{xx}$, $M_{300}^{\rm neq}$, and $M_{400}^{\rm neq}$, normalised by the upstream velocity to the powers 3, 2, 3, and 4, are interpolated linearly in $b$. A positive prior distribution is formed from the two neighbour mixtures by shifting each to the target bulk velocity, dilating it by $\sqrt{T_t/T_i}$, rescaling its density, and mixing with weights $1-b$ and $b$; this gives at most 1024 velocity Gaussians. No field of the withheld state enters. The stored numbers occupy 106,624 bytes, against 81,920 bytes for the conditional coefficients alone.

\emph{Constrained correction.} Let $n_j$ be the analytic masses of the components and $\bm h(n)=(\rho,u_x,T,q_x,\sigma_{xx},M_{300}^{\rm neq},M_{400}^{\rm neq})$ the central moments of the mixture, computed analytically. The reconstructed masses minimise
\begin{equation}
 \mathcal J(n)=\sum_j\left[n_j\log\frac{n_j}{n_j^0}-n_j+n_j^0\right]
 +\frac{\alpha}{2}\left\|S^{-1}[\bm h(n)-\bm h_t]\right\|^2
 \quad\text{subject to}\quad A^Tn=\bm1_6,\quad n_j>0,
 \label{conservative:eq:nonlinear}
\end{equation}
where $\bm h_t$ contains the target moments (for the fitted state, the quadrature moments of the fitted Gaussian), $S$ is a diagonal matrix of moment scales, $\alpha=10^6$, and the six columns of $A$ are the three analytic and three quadrature fluxes of Eq.~\eqref{conservative:eq:flux} for each Gaussian, normalised by the RH fluxes. Both integration conventions therefore reproduce the RH fluxes exactly. The first term keeps the masses close to the prior; it is not a maximum-entropy closure \citep{ConservativeLevermore}. The problem is solved by successive linearisation of $\bm h(n)$ with a line search. An accepted solution must be finite, have positive density and temperature, and satisfy all six flux constraints to $2\times10^{-10}$.

\subsection{Results}
\begin{figure}[!htbp]
\centering
\includegraphics[width=.86\textwidth]{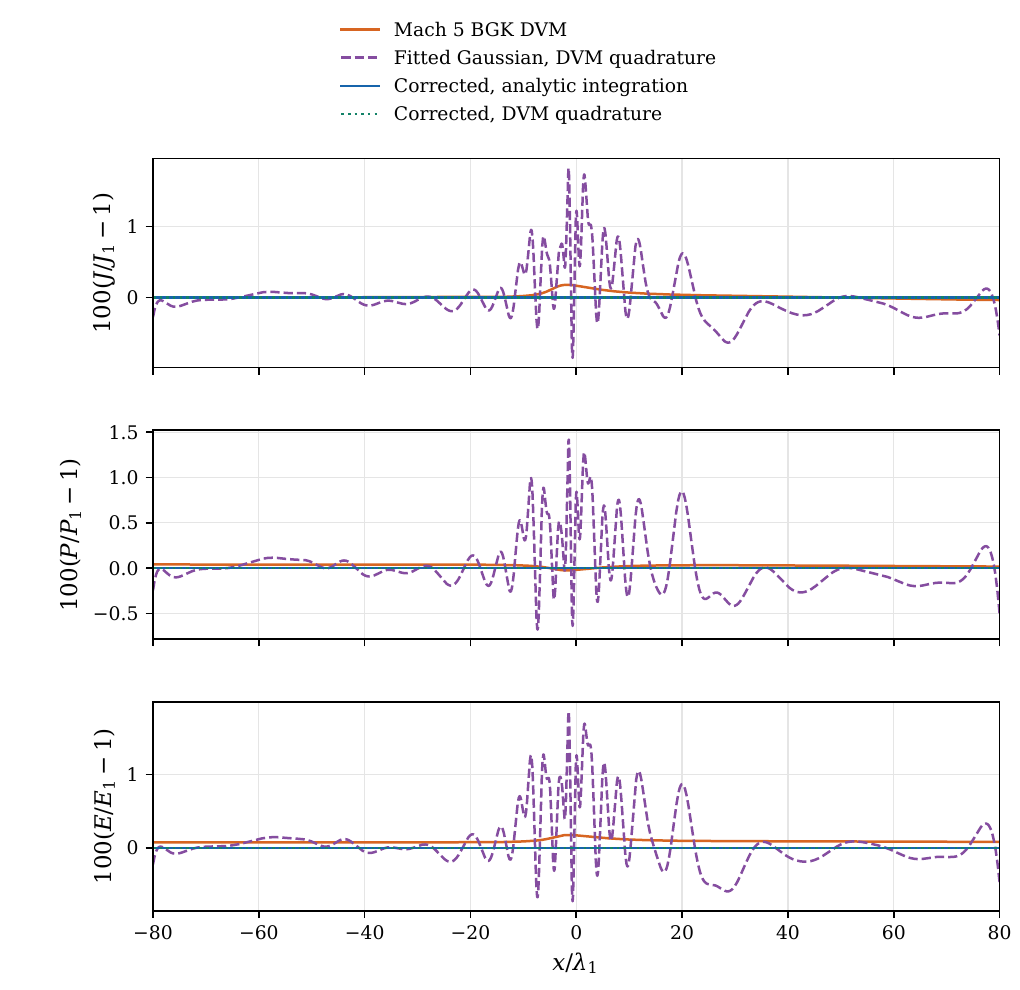}
\caption{Mach-5 mass, momentum, and energy fluxes relative to their upstream values: BGK DVM, the fitted Gaussian under the DVM quadrature before the correction, and the conservative reconstruction under analytic and DVM integration.}
\label{conservative:fig:M5flux}
\end{figure}

\emph{Fitted Mach-5 shock.} Before the correction, the quadrature energy flux of the fitted Gaussian varies across the profile by 2.586\% (figure~\ref{conservative:fig:M5flux}). Projection onto the three quadrature flux constraints alone removes this difference but raises the heat-flux error from 1.125\% to 1.649\%, and analytic integration of distributions corrected in this way shows continuous-flux discrepancies as large as 37.2\%: exact balance under the quadrature does not imply exact continuous balance. Imposing both sets of flux constraints without the moment term is feasible but distorts the distribution, and the quadrature heat-flux error rises to about 11.37\%. With the moment term the reconstruction satisfies all six constraints and has analytic heat-flux and fourth-order errors of \FiveQ\% and \FiveMfour\% (table~\ref{conservative:tab:analytic}).

\begin{figure}[!htbp]
\centering
\includegraphics[width=.88\textwidth]{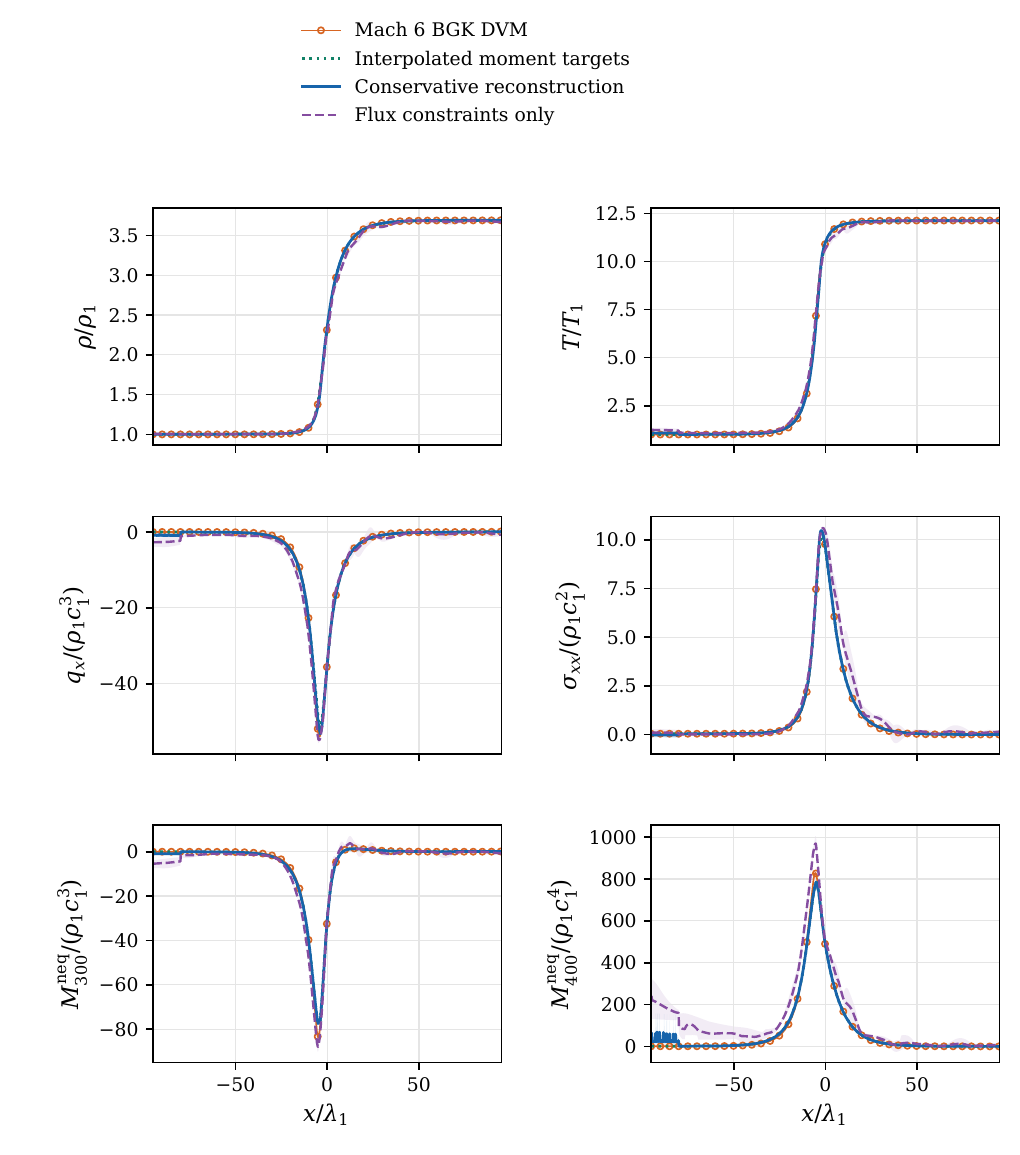}
\caption{Withheld Mach 6: BGK DVM, the moment targets interpolated from the Mach-5 and Mach-8 training states, the conservative reconstruction with the moment term, and the same prior with flux constraints only. Lines are means over three random initialisations; shading spans their range.}
\label{conservative:fig:M6}
\end{figure}

\emph{Withheld Mach numbers.} Mach 3 is reconstructed from the Mach-2.5 and Mach-4 neighbours and Mach 6 from Mach 5 and 8, each with three seeds; Mach 4 is reconstructed from the separately fitted Mach-3 and Mach-5 models. The mean analytic heat-flux errors are \ThreeQ\% at Mach 3 and \SixQ\% at Mach 6, and \FourQ\% at Mach 4 (table~\ref{conservative:tab:analytic} and figure~\ref{conservative:fig:M6}). Two points bear on the interpretation. First, the method uses stored moment profiles of the neighbouring states, including the high-order ones, and the result is dominated by these targets: the Mach-3 errors are identical for the three seeds to the digits shown. The result shows that interpolated moments can be converted into a positive distribution with exact fluxes; the dependence of the distribution on Mach number is the subject of Sec.~7 of the main paper. Second, the method was developed with these same states.

\emph{Numerical checks.} All 16,800 positions of the eight profiles satisfy both flux conventions within the threshold; 16,687 reach a relative Karush--Kuhn--Tucker residual below $10^{-6}$, 105 reach the maximum number of updates, and eight stagnate. Changing the number of updates from 20 to 80 changes no moment by more than $2.2\times10^{-6}$ of its scale. The 129-point tables reproduce the training profiles to within 0.73--0.94\% in heat flux.

\emph{Accuracy of the distribution.} Conservation and accurate moments do not ensure an accurate distribution. The sampled distribution errors of the reconstructed states are 5.2\% for the fitted Mach-5 shock, 2.1--2.3\% at Mach 3, 3.3\% at Mach 4, and 6.7--11.3\% at Mach 6, and the errors restricted to the tails ($|\bm\xi-\bm u|^2>9T$) are 7--61\%. At Mach 6 the analytic and quadrature fourth moments of the same distribution differ by 5.95--11.11\% although the three fluxes agree exactly, so agreement of the fluxes does not establish that the quadrature resolves high moments.

\begin{figure}[!htbp]
\centering
\includegraphics[width=.90\textwidth]{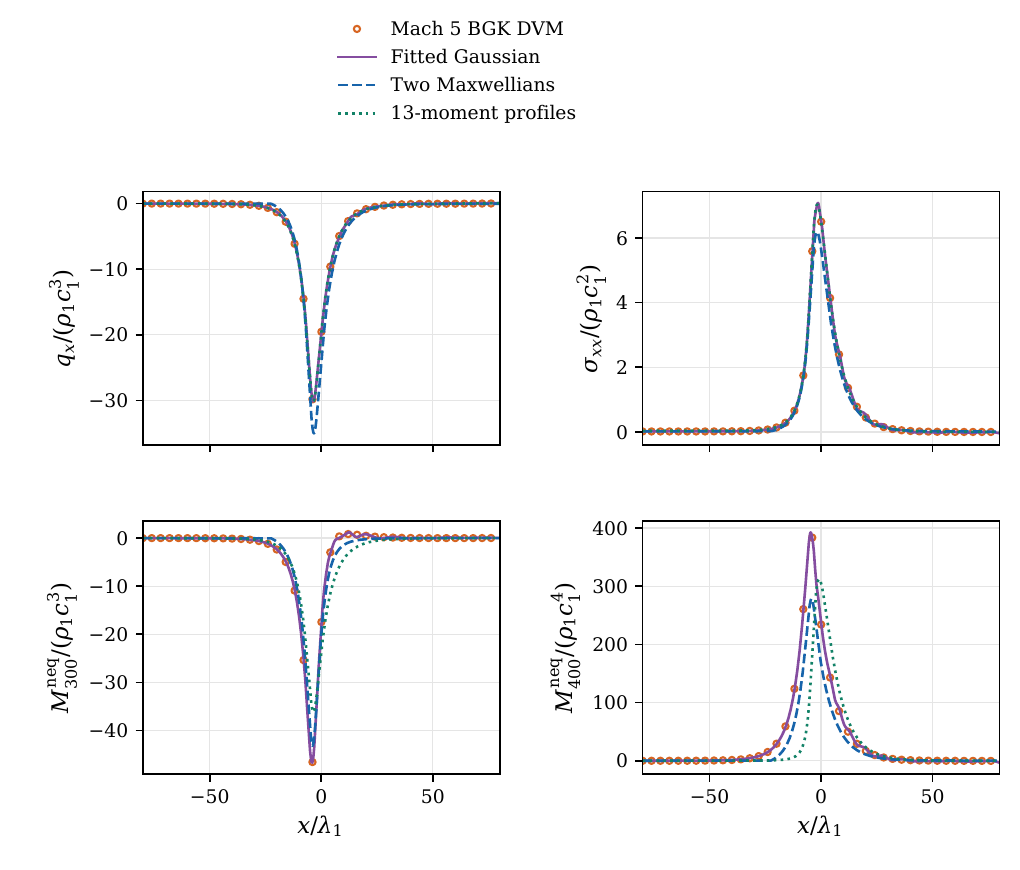}
\caption{Mach-5 transport and regenerated higher moments for the descriptions of table~\ref{tab:codecs}.}
\label{conservative:fig:M5codecs}
\end{figure}

\emph{Comparison at equal storage.} Table~\ref{tab:codecs} and figure~\ref{conservative:fig:M5codecs} compare the fitted Gaussian with two physically motivated descriptions that use the same or fewer stored coefficients: a positive mixture of the two RH Maxwellians with weights derived from the density profile, in the spirit of \citet{MottSmith1951}, and a Grad 13-moment distribution built from five stored profiles, which is not a solution of the regularised 13-moment equations \citep{ConservativeR13}. The two-Maxwellian description remains positive but misses transport and fourth-order structure (errors of 12--33\%). The 13-moment description reproduces the stress and heat flux that it stores but has errors of 33--63\% in the regenerated third and fourth moments and a negative quadrature mass of 0.07\% (Mach 3) and 0.64\% (Mach 5) of the positive mass.

\begin{table}
\centering
\small\setlength{\tabcolsep}{4pt}
\caption{Analytic central-moment errors (\%) of the conservative reconstruction over all positions of the corresponding DVM solution. M5 is the fitted state; M3, M4, and M6 are withheld, with one row per seed.}
\label{conservative:tab:analytic}
\begin{adjustbox}{max width=\linewidth}
\begin{tabular}{lrrrrrrr}
\toprule
Case/seed & $\rho$ & $u_x$ & $T$ & $q_x$ & $\sigma_{xx}$ & $M_{300}^{\rm neq}$ & $M_{400}^{\rm neq}$\\
\midrule
M5/1234 & 0.130 & 0.087 & 0.204 & 1.529 & 1.713 & 1.719 & 1.846\\
M3/1234 & 0.392 & 0.428 & 0.602 & 3.110 & 2.274 & 6.021 & 4.155\\
M3/2026 & 0.392 & 0.428 & 0.602 & 3.110 & 2.274 & 6.021 & 4.155\\
M3/3407 & 0.392 & 0.428 & 0.602 & 3.110 & 2.274 & 6.021 & 4.155\\
M6/1234 & 0.317 & 0.465 & 0.874 & 4.431 & 3.086 & 6.079 & 5.882\\
M6/2026 & 0.318 & 0.482 & 0.874 & 4.365 & 2.097 & 6.252 & 11.013\\
M6/3407 & 0.316 & 0.463 & 0.823 & 3.534 & 2.079 & 5.710 & 4.919\\
M4/1234 & 0.392 & 0.493 & 0.804 & 3.677 & 2.376 & 6.703 & 5.225\\
\bottomrule
\end{tabular}
\end{adjustbox}
\end{table}

\begin{table}
\centering
\small\setlength{\tabcolsep}{4pt}
\caption{Descriptions with equal or smaller storage, evaluated with the DVM quadrature. Errors are in percent. The 13-moment description stores five profiles and is not a solution of the regularised 13-moment equations.}
\label{tab:codecs}
\begin{adjustbox}{max width=\linewidth}
\begin{tabular}{llrrrrr}
\toprule
Mach & Description & Stored coefficients & $q_x$ & $\sigma_{xx}$ & $M_{300}^{\rm neq}$ & $M_{400}^{\rm neq}$\\
\midrule
3 & Gaussian & 4608 & 1.475 & 1.681 & 2.030 & 2.066\\
3 & Two Maxwellians & 1808 & 22.559 & 15.333 & 17.093 & 27.045\\
3 & 13-moment profiles & 4608 & 0.061 & 0.304 & 33.106 & 51.316\\
5 & Gaussian & 4608 & 1.125 & 1.665 & 1.708 & 1.807\\
5 & Two Maxwellians & 2208 & 17.128 & 12.417 & 15.516 & 33.310\\
5 & 13-moment profiles & 4608 & 0.166 & 0.729 & 33.568 & 62.572\\
\bottomrule
\end{tabular}
\end{adjustbox}
\end{table}

\FloatBarrier
\subsection{Why exact fluxes do not fix the heat flux}
At fixed fluxes $J$, $P$, and $E$ and exact transverse symmetry,
\begin{equation}
 q_x=E-Pu_x+\frac{J}{2}u_x^2-\frac{3J}{2}T ,
 \label{conservative:eq:qx}
\end{equation}
so that two conservative states differ by
\begin{equation}
 \delta q_x=(Ju_x-P)\,\delta u_x+\frac{J}{2}(\delta u_x)^2-\frac{3J}{2}\,\delta T .
 \label{conservative:eq:dqx}
\end{equation}
Small, weakly correlated errors in the bulk velocity and temperature can thus produce substantial relative heat-flux errors. Conservation constrains the relation between moments but does not determine the transport state, and the fourth-order moment carries information that is absent from all three fluxes. This is why the moment term is needed in addition to the flux constraints. The reconstruction is a post-processing step; its objective can have several local minima, and it has been tested at the Mach numbers used in its development.

\FloatBarrier
\section{Variants of the conservative reconstruction}
\label{app:controls}
Several variants of the reconstruction of Sec.~\ref{sec:conservative} were also tested. They show what exact constraints determine and what they leave open.

\emph{Normalisation with the moments of the source Gaussian.} Replacing the tabulated bulk moments of each neighbour by moments integrated analytically from its own Gaussian makes the density, velocity, and temperature identities exact and removes a jump at the edge of the neighbour's spatial domain (up to 2.51\% of the training scale in the fourth-order moment at Mach 6). It also shows a property of the source models. At an upstream Mach-6 position, $x/\lambda_1\simeq-59.6$, the neighbouring Mach-8 model has an analytic temperature of 3.95 over all velocities but 1.14 within the velocity box of the DVM: about 0.265\% of its mass lies outside the box and carries 71\% of its thermal energy (64--80\% for the three seeds). The log-density fit does not constrain the Gaussian tails outside the sampled velocities. With this normalisation the dense distribution error at that position rises from 0.66\% to 135\%.

\emph{Function-space regularisation.} Adding to Eq.~\eqref{conservative:eq:nonlinear} a weighted $L_2$ distance between the reconstructed and prior distributions, computed from Gaussian overlap integrals, and penalties on the difference between analytic and quadrature low-order moments reduces the dense distribution error at the withheld Mach numbers (from 51--82\% to 26--37\% at Mach 6) but raises it for the fitted Mach-5 state from 0.30\% to 4.99\%; 478 of the 512 evaluated positions reach stationarity.

\emph{Tail filtering and bulk-guided fits.} Multiplying each source mixture by a Gaussian factor that suppresses velocities outside the DVM box, followed by an affine restoration of its bulk moments, was scanned over 38 filter strengths for nine source evaluations. No candidate satisfied both the bound of $10^{-3}$ on the content outside the box and a 1\% limit on the change of the resolved distribution; the smallest changes were 1.0--2.4\% for the Mach-5 sources and 9.0--16.3\% for the Mach-8 sources. A least-squares fit of the component masses to the prior, with the analytic fluxes, density, and energy as exact constraints, passed all numerical checks at five selected positions but increased the distribution error relative to the unmodified prior, from 0.15\% to 1.2\% for the fitted Mach-5 state and from 0.28--0.70\% to 3.7--4.6\% at the withheld Mach 6, and the tail-restricted error from 0.25--1.8\% to 14--17\%.

These variants show that positivity, exact fluxes, exact bulk moments, and bounds on the unresolved tails do not by themselves give an accurate resolved distribution; an accurate initial mixture is required. They also show that the uniform velocity grids of the archived high-Mach references affect such tests: the upstream Maxwellian integrated on the archived Mach-6 grid has a temperature of 0.950 instead of unity, against 0.9999974 on the GL3 grid (see Sec.~2.4.2 of the main paper).

\emph{Other models.} Shock profiles from the regularised 13- and 26-moment equations \citep{ConservativeR13,GuEmerson2009} can be compared with the present results only if the collision operator, Prandtl number, viscosity law, boundary states, and length convention are matched. Such a comparison is outside the scope of this study.

\FloatBarrier
\section{Computational cost of the 512-kernel representation}
\label{app:cost}

Table~\ref{tab:cost} reports the measured M5 training times. The diagonal seed-1234 run required 2.56 h on an NVIDIA Tesla M40 24-GB GPU, while the three correlated runs averaged $2.68\pm0.59$ h. These times are those of the present implementation. The 4608 diagonal parameters represent $2.261\times10^8$ M5 phase-space values, giving the nominal coefficient ratio in table~\ref{tab:shock_baseline}. Moment evaluation still requires the DVM velocity quadrature.

\begin{table}
\caption{Measured M5 training cost. The correlated value is mean $\pm$ sample standard deviation over three seeds; the diagonal timing is the representative seed-1234 run used in figure~\ref{fig:m5}.}
\label{tab:cost}
\centering
\small
\setlength{\tabcolsep}{4pt}
\begin{adjustbox}{max width=\linewidth}
\begin{tabular}{@{}lccc@{}}
\toprule
Model & runs & parameters & time (h)\\
\midrule
Diagonal $N=512$ & 1 & 4608 & 2.56\\
$x$--$\xi_x$, $N=512$ & 3 & 5120 & $2.68\pm0.59$\\
\bottomrule
\end{tabular}
\end{adjustbox}
\end{table}

The conditional parameter counts are given in table~\ref{tab:conditional_protocol}: a map with 15,360 parameters occupies 61,440 bytes in single precision and one with 20,480 parameters 81,920 bytes, whether the basis is global or local. These numbers are the storage of the representation and do not include the cost of computing the eight BGK-DVM solutions.

\section{Last step compared with the smallest training loss}

Table~\ref{tab:checkpoint} compares the two choices for all three correlated seeds and all three diagonal seeds. Only one of the six models of minimum loss has a lower heat-flux error than the model at the last step. The smallest recorded minibatch loss therefore does not identify a better model. Using the last step keeps the choice of model independent of the moments that are reported.

\begin{table}
\caption{M5 heat-flux error for the saved model with the smallest recorded training loss (best) and the model at the last step, 80,000 (final). Values are percentages.}
\label{tab:checkpoint}
\centering
\begin{adjustbox}{max width=\linewidth}
\begin{tabular}{lrrr}
\toprule
Model & seed & best & final\\
\midrule
$x$--$\xi_x$ & 1234 & 1.564 & 1.243\\
$x$--$\xi_x$ & 2026 & 5.442 & 3.595\\
$x$--$\xi_x$ & 3407 & 1.141 & 1.232\\
Diagonal & 1234 & 1.139 & 1.125\\
Diagonal & 2026 & 2.713 & 2.639\\
Diagonal & 3407 & 1.626 & 1.456\\
\bottomrule
\end{tabular}
\end{adjustbox}
\end{table}

\FloatBarrier

\FloatBarrier
\section{Reproducibility details}

The code repository of this study (\url{https://github.com/Ehsan-Roohi/KineticGaussian}) contains the codes and configurations for the principal shock representation cases and for the equal-storage grids. The stored prediction profiles, the tables of seeds and saved models, the 18 evaluations of the global models and the six of the local M6 models, the grid-refined M6 and M12 moment profiles, and the files that document the DVM solutions are available from the author on request. Shock coordinates are normalised by the midpoint and half-width of each stored coordinate. The bounded centre and scale parameterisations, initialisation values, optimiser settings, sampling fractions, saving interval, and seeds stated in Sec.~\ref{sec:optimization} are included. The adaptive-$x$ grid in table~\ref{tab:grid_layouts} places nodes by weighted quantiles of a function of $|\partial_x\rho|$, $|q_x|$, and $|\sigma_{xx}|$. With these files the principal conditional, equal-storage, and grid-refinement results can be traced to stored data.

\section{Remarks on the many-kernel representation}

\subsection{What controls the accuracy of the 512-kernel representation?}

Within the M3 comparison in table~\ref{tab:m3_ablation}, changing from the sampled-moment objective to direct log-density fitting lowers every reported error, including a reduction of the stress error by more than a factor of ten. Within the three-seed M5 comparison in table~\ref{tab:m5_seeds}, the mean differences between diagonal and $x$--$\xi_x$ kernels are smaller than one sample standard deviation for each nonequilibrium moment. Log-density training and the simpler diagonal form are therefore used. At the M5 4608-coefficient storage, all tested multilinear four-dimensional grids give transport and higher-moment errors above 87\%, whereas the Gaussian errors are 1.12--1.80\% (table~\ref{tab:shock_baseline}); a two-Maxwellian description and a 13-moment description with the same or smaller storage lose the third- and fourth-order moments (table~\ref{tab:codecs}). Storage counts are nominal coefficient counts; sparse-grid, tensor, and low-rank representations remain useful comparisons.

\subsection{Conservation and accuracy of the distribution}

Section~\ref{sec:conservative} separates three properties. Exact quadrature fluxes do not imply exact continuous fluxes: corrections under the quadrature alone leave analytic flux discrepancies as large as 37.2\%. Exact fluxes in both conventions do not imply accurate transport: without the moment term, the quadrature heat-flux error of the fitted Mach-5 reconstruction rises to about 11.37\%, and Eq.~\eqref{conservative:eq:dqx} shows how small, weakly correlated velocity and temperature errors are amplified in $q_x$ at fixed fluxes. Accurate moments do not imply an accurate distribution, because the sampled, dense, and tail errors of the distribution remain larger than the moment errors. A conservative reconstruction is therefore to be used together with separate tests of the moments, the tails, and the integration convention.